\documentclass{aa}  

\usepackage{graphicx}
\usepackage{natbib}
\bibpunct{(}{)}{;}{a}{}{,} 

\usepackage{nameref}
\usepackage{longtable}
\usepackage{filecontents,catchfile}
\def\toprule{\hline}
\def\midrule{\hline}
\def\bottomrule{\hline}

\usepackage{xcolor}
\usepackage{hyperref}
\usepackage{indentfirst} 
\usepackage{afterpage}
\usepackage{orcidlink}
\usepackage{caption}

\usepackage{txfonts}
\begin{document}

   \title{Discovery of young, oxygen-rich supernova remnants in PHANGS-MUSE galaxies}

   \author{T. Kravtsov\inst{1,2}\orcidlink{0000-0003-0955-9102} \and
            J.~P. Anderson\inst{1,3}\orcidlink{0000-0003-0227-3451}\and
            H. Kuncarayakti\inst{2,4}\orcidlink{0000-0002-1132-1366}\and
            K. Maeda\inst{5}\and
            S. Mattila\inst{2,6}\orcidlink{0000-0001-7497-2994}
          }

   \institute{European Southern Observatory, Alonso de C\'ordova 3107, Casilla 19, Santiago, Chile\\
              \email{thtkra@utu.fi}
         \and
             Department of Physics \& Astronomy, University of Turku, Vesilinnantie 5, Turku, FI-20014, Finland
        \and
            Millennium Institute of Astrophysics MAS, Nuncio Monsenor Sotero Sanz 100, Off. 104, Providencia, Santiago, Chile
        \and
            Finnish Centre for Astronomy with ESO (FINCA), FI-20014 University of Turku, Finland
        \and
            Department of Astronomy, Kyoto University, Kitashirakawa-Oiwake-cho, Sakyo-ku, Kyoto 606-8502, Japan
        \and
            School of Sciences, European University Cyprus, Diogenes Street, Engomi, 1516 Nicosia, Cyprus
        }

   \date{Received --- 00, 2023; accepted -- 00, 2023}

   \titlerunning{Discovery of seven extragalactic O-rich SNRs}
    \authorrunning{Kravtsov et al.}

 
  \abstract
   {Supernova remnants (SNRs) are the late stages of supernovae before their merging into the surrounding medium. Oxygen-rich supernova remnants represent a rare subtype with strong visible light oxygen emission.}
   {We present a new method to detect SNRs exploiting the capabilities of modern visible-light integral-field units based on the shapes of the SNR emission lines.}
   {We search for unresolved shocked regions with broadened emission lines using the medium-resolution integral-field spectrograph MUSE on the Very Large Telescope. The spectral resolving power allows shocked emission sources to be differentiated from photoionised sources based on the linewidths.}
   {We find 307 supernova remnants, including seven O-rich SNRs. For all O-rich SNRs, we observe the [O\,III]$\lambda\lambda$4959,5007 emission doublet. In addition, we observe emissions from [O\,I]$\lambda\lambda$6300,6364, [O\,II]$\lambda\lambda$7320,7330, H$\alpha$+[N\,II]$\lambda$6583 and [S\,II]$\lambda\lambda$6717,6731 to varying degrees. The linewidths for the O-rich SNRs are generally broader than the rest of the SNRs in the sample of this article. The oxygen emission complexes are reminiscient of SNR 4449-1 and some long-lasting SNe. For the O-rich SNRs, we also search for counterparts in archival data of other telescopes; we detect X-ray and mid-IR counterparts for a number of remnants.}
   {We have shown efficacy of the method to detect SNRs presented in this article. In addition, the method is also effective in detecting the rare O-rich SNRs, doubling the sample size in the literature. The origin of O-rich SNRs and their link to specific SN types or environments is still unclear, but further work into this new sample will unquestionably help us shed light on these rare remnants.}


   \maketitle

   
%

\section{Introduction}
Supernova remnants (SNRs) are the last evolutionary phase of supernovae (SNe) before they merge into the surrounding interstellar medium (ISM). They exhibit complex interactions between the ejecta and any significant circumstellar medium (CSM) together with the interstellar medium (ISM). In the case of core-collapse SNe, additional luminosity may arise from the compact central object - a neutron star or a black hole - due to magnetic breaking or accretion. These interactions give rise to signatures of shocked gases, most notably of hydrogen, nitrogen and sulphur. The timescale of this phase is from thousands of years to a million years \citep{micelotta18}.

Oxygen-rich supernova remnants (O-rich SNRs) are objects with strong obsered oxygen features. The prototypical O-rich SNR is Cassiopeia A, which has knots dominated by [O\,III] $\lambda \lambda$4959,5007 \citep{Milisavljevic12}. However, to date very few O-rich remnants are known. In the Milky Way (MW) and the Magellanic Clouds there are eight such objects. These are Cassiopeia A, Puppis A and G292+1.8 in the MW, 0540-69.3 and LMC N132D in the Large Magellanic Cloud (LMC), 1E 0102.2-7219, B0049-73.6 and B0103-72.6 in the Small Magellanic Cloud (SMC) \citep{Vogt2010, Schenck2014}. The latter two have no published optical-wavelength spectra. However, both have X-ray observations that show enhanced oxygen abundances. This low number of O-rich SNRs is in stark contrast to the around 300 SNRs discovered in the MW \footnote{https://www.mrao.cam.ac.uk/surveys/snrs/}. From the literature, only one object designated as an O-rich SNR outside the MW system is found; SNR 4449-1 in NGC 4449. SNR 4449-1 was originally detected in the radio \citep{seaquist78}, then subsequently in the optical \citep{balick78} and in X-rays \citep{blair83}. The optical spectra examined in \cite{balick78} showed a similarity between the SNR and the O-rich knots of Cas A and N132D, a remnant in the LMC.
   
Some long-lasting SNe show similar spectral features to SNR 4449-1. These include SN 1957D in M83, SN 1979C in M100 and SN 1995N in MCG-02-38-017 \citep{long12, Milisavljevic08, Wesson2023}. SN1980K also has strong oxygen emission lines \citep{Milisavljevic12}. A sample of late-time observations of SNe can be found in \cite{Milisavljevic12}. \cite{blair15} also discovered an ejecta-dominated SNR in M83 labeled B12-174a with broad H$\alpha$, [S\,II]$\lambda \lambda$6717,6731 and weaker O-lines. More recently, \cite{Caldwell23} reported a discovery of the SNR labeled WB92-26 with strong N and O-lines that they dubbed a `N-rich SNR'. 
   
There is an apparent connection between dust formation and O-rich remnants. Cas A has clear infrared (IR) excesses implying the presence of dust. For example, Cas A has been estimated to host 0.5 M$_{\odot}$ of silicate dust, with carbon dust formation being quenched due to the O-rich nature of the remnant \citep{delooze17}. Based on modeling of optical emission line profiles, \cite{niculescuduvaz22} estimate dust masses for SN 1979C to be $\approx$ 0.3-0.65\,M$_{\odot}$. SNR 4449-1 and other O-rich SNRs have also been detected at X-ray wavelengths. \cite{Patnaude2003} report a total luminosity of 2.4$\times$ 10$^{38}$ erg/s for SNR 4449-1, an order of magnitude more luminous than Cas A, and comparable to the 6.5$\times$ 10$^{38}$ erg/s luminosity of SN 1979C, which has been argued to host a newly formed black hole \citep{Patnaude11}.

The nature of O-rich SNRs and their connection to specific SNe is still unclear. While Cas A was spectroscopically identified as a type IIb \citep{Krause2008}, SN1979C was classified as an extremely bright type II with a fast declining light curve \citep{deVaucouleurs1981}. Given the low number of known O-rich SNRs (nine in total), any additional discoveries and their subsequent characterisation can significantly aid in understanding these remnants and their progenitor properites. In addition, it appears that only a few SNe have been recovered years after the explosion. In these long-lasting SNe, strong oxygen and H$\alpha$ features are often observed \citep{Milisavljevic12, kuncarayakti16}, which could indicate a shared, but uncommon, evolutionary stage for certain CCSNe. 

In this paper, we introduce a new methodology for finding SNRs, report the discovery of new O-rich SNRs and characterise our sample of seven extragalactic O-rich remnants. We additionally compile a sample of "recovered SNe" that we argue provide a strong link to O-rich remnants, perhaps indicating a common origin. In Section 2 we present our observations and data reductions. In Section 3 we describe our methods for finding SNRs with MUSE and search for counterparts in data obtained with other facilities. In section 4 we show our results, then discussion and conclusions are presented in Sections 5 and 6 respectively.

\section{Observations}
The project presented in this paper was motivated by an initial search for shock-ionised regions in an optical-wavelength spectroscopic datacube of Messier 61 (NGC 4303). This led to the serendipitious discovery of two objects that appeared consistent with O-rich SNRs, particularly SNR 4449-1. Following this initial discovery, we initiated a systematic search for O-rich SNRs. The discovery was based on VLT Multi Unit Spectroscopic Explorer (MUSE, \citealt{bacon10}) observations of NGC 4303, data which is part of the PHANGS-MUSE ESO program. The Physics at High Angular resolution in Nearby GalaxieS (PHANGS)  survey uses Atacama Large Millimeter/submillimeter Array (ALMA, \citealt{Leroy21}), Hubble Space Telescope (HST, \citealt{Lee22}), James Webb Space Telescope (JWST, \citealt{Lee23}) and MUSE \citep{Emsellem22} (at the Very Large Telescope) data to study small-scale physics of gas and star formation. These high-resolution observations allow for a wide range of studies as a spin-off from the original survey aims. We limit our search to this dataset, because it has some of the best spatial coverage of nearby galaxies. These data also guarantees HST observations for most of the galaxies (16 out of 19) and gives a reasonable coverage of JWST images - observations that are still ongoing. The proximity of these galaxies also gives a good archival Chandra X-ray coverage.

The MUSE data used is from ESO Program ID 1100.B-0651 \citep{Emsellem22}. The observations were made in wide-field mode (WFM) both with and without adaptive optics (AO) between November 2017 and February 2021. The data were retrieved from ESO Phase 3 archive\footnote{https://archive.eso.org/wdb/wdb/adp/phase3\_spectral/form} and were reduced by the PHANGS-MUSE collaboration. For the four host galaxies of O-rich SNRs we searched archives of HST, JWST and Chandra for supporting material. The datasets used are described in Appendix C. We downloaded the Level 3 reduced data from Mikulski Archive for Space Telescope. The HST observations are partially based on PHANGS-HST observations \citep{Lee22}. For HST observations we used DOLPHOT to measure the magnitudes of our targets. With JWST data we employed jdaviz\footnote{https://doi.org/10.5281/zenodo.5513927} and CIAO for the Chandra datasets\citep{jdadf, 2006SPIE.6270E..1VF}.

In addition to the archival data for our O-rich SNRs, we also download reduced spectra of WB92-26 observed with MMT/Hectospec in October 2006, November 2006, and Octboer 2007 were downloaded from the MMT/Hectospec M31 archive\footnote{https://oirsa.cfa.harvard.edu/signature\_program/}. 
Reduced spectra of SNR 4449-1 observed with HST FOS in January 1993 were downloaded from MAST. Reduced spectrum of SN 1979C observed with Gemini-N GMOS in May 2017 was downloaded from WISeREP\footnote{www.wiserep.org}. Spectra of SNR 0540-69.3 was extracted from a MUSE observation that was part of the ESO program ID 0102.D-0769 observed between January and March 2013. The data was downloaded from ESO Phase 3 archive.

\section{Methods}
In this section we first review the traditional, historic methods of detecting extragalactic SNRs. We then describe our new methodology that makes use of the large field of view (FOV) integral field unit (IFU) capabilities of the MUSE instrument. We define the steps needed to go from a reduced MUSE datacube to a list of unresolved broad-line regions and finally to a list of SNR candidates. At the end of the section we present our methodology as applied to one galaxy in the dataset.

\subsection{Identification of SNRs}
A classic method to find SNRs is to measure the line ratio of [S\,II]/H$\alpha$, with a ratio higher than 0.4 being interpreted to be a likely SNR \citep{Smith1993, Matonick97}. However, it has been noted that this method might not be exhaustive and could omit some SNRs, e.g. coinciding HII regions producing strong H$\alpha$ emission that could skew the line ratio \citep{blair97}. Also, \cite{Fernandes21} reported several SNRs with [S\,II]/H$\alpha$ ratio lower than 0.4. This line-ratio method has been often used as it enables searches for SNRs over large FOVs (e.g. the full extent of large - on the sky - nearby galaxies) through the use of narrow-band filters. So while emission line ratios can be used to differentiate between different emission mechanisms, such as photoionization or shock excitation, the measured ratios can differ significantly from the empirical values. Thus, other detection methods should be investigated in order to recover a more complete picture of SNR populations.

With new technologies (such as IFUs) it is now possible to obtain additional, spatially resolved spectral information for all regions of galaxies with a single pointing. With MUSE, we can observe a whole galaxy in spectral space at high enough resolution in order to allow us to develop and test new tools to discover SNRs. We expect SNRs to have ejecta velocities of over 200 km/s up to 100 kyr after the SN explosion (\citealt{micelotta18} see their figure 1), which is in a stark contrast to the typical thermal velocities of HII regions of 10-20 km/s \citep{Arenas18}. Planetary nebulae have similar velocities of few tens of km/s caused by the expansion of the nebula\citep{Schonberner18}. The broader nature of emission lines produced in SNRs can therefore be used to separate and identify such objects from other emission-line sources. With IFU data, we can use a combination of line ratios and velocities to isolate and identify SNRs.

Following the availability of MUSE IFU data and the possibility to use velocity information to find SNRs, we proceeded to search for broadened emission lines using narrow-line off center images. The emission lines used in this survey generally do not have strong absorption lines associated with them in the stellar continua. Therefore, a simple linear approximation for the continuum subtraction was considered sufficient to detect broad-line emissions without the need to consider contamination from absorption lines. These line maps then show the presence of broadened emission lines. Sources for these regions include: shocked sources such as SNRs, but additionally HII regions with high S/N producing large full width at zero intensity (FWZI), emission-line stars and late-type stars with a complex continuum. These latter stars can be in the host galaxy or appear as foreground stars. In section \ref{separating} we outline how we differentiate between these different sources.

The broadened-emission-line maps produced above contain both unresolved point sources and extended emission regions. The remnants are a few to tens of parsecs in size and appear as unresolved sources at the distances in our MUSE dataset, between 9-20 Mpc, or 0.37"-0.85" for an SNR with diameter of 37 pc (i.e. something the size of the Cygnus Loop \citealt{Fesen18}). Extended emission sources are large HII regions or emission from diffuse ISM. To identify unresolved point sources, we process the maps through SExtractor. The result of this method is presented in Fig. \ref{sources_m61} for NGC 4303. We then extract the point source spectra from the original MUSE datacube based on the resulting coordinates and remove the background continuum with an annular ring around the source. Shock-powered region sources are generally easily identifiable by eye due to distinctly broadened emission lines and enhanced [S\,II]$\lambda\lambda$6717,6731 emission, but we also measure the line ratios of H$\alpha$, [N\,II]$\lambda$6583 and [S\,II]$\lambda\lambda$6717,6731 in order to confirm the identified shocked regions. In addition, we require all of the aforementioned lines to be detected in order to be considered a regular SNR; for O-rich SNRs we forfeit this requirement. We note that broad sources with log(H$\alpha$/[S\,II]) < 0.4 are most likely SNRs, but also that this cutoff is empricial, as discussed previously. Thus we inspect sources near this cutoff more closely.

   \begin{figure*}
   \centering
   \includegraphics[width=19cm]{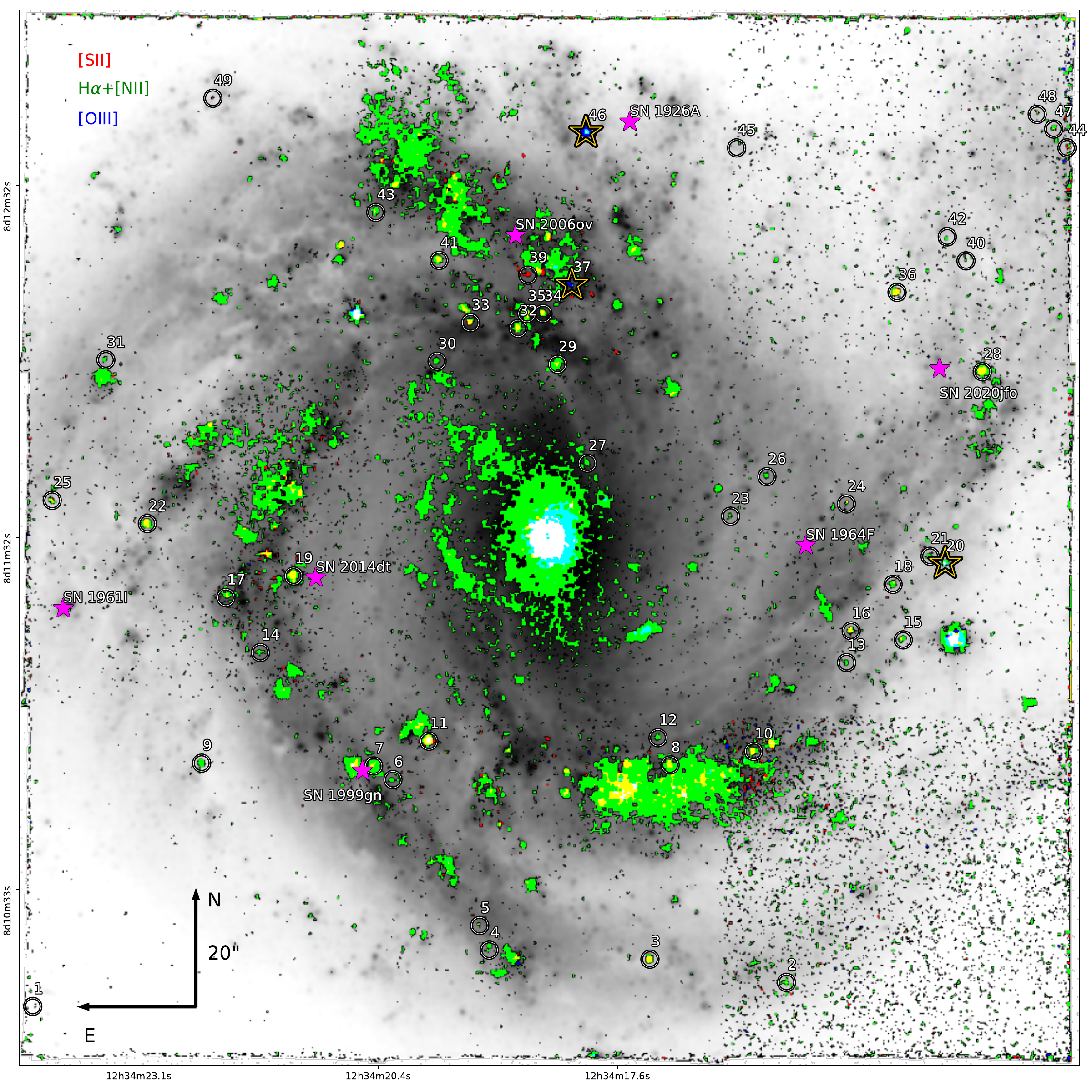}
        \caption{Colour image of M61 compiled from three narrow-line maps used in this survey; red, green and blue colours correspond to [SII], H$\alpha$ and [OIII], respectively. The background image is the collapsed cube of the MUSE observation. White circles and numbers are the SNRs in the galaxy confirmed by this survey. Magenta stars indicate the locations of previously observed SNe in the field of view. Gold stars show the locations of the three O-rich SNRs in the galaxy.}
         \label{sources_m61}
   \end{figure*}

         \label{sources_m61}

\subsection{Application to NGC 4303} \label{separating}

\cite{martinez-rodriguez24} recently showed the effectiveness of using integral field spectroscopy to search for SNRs, using a conceptually similar technique to that which we have presented, i.e. identifying broadened emission regions. The authors noted a lack of O-rich events in their sample. However, the total number of SNRs they discovered was much less than through our methods: 19 in their work compared with a total of more than 300 here, even though Martinez-Rodriguez et al. searched for SNRs in a much larger galaxy sample (around a 1000 galaxies compared to the sample of 19 used here). The differences between the detection rates of these studies are likely related to the discovery techniques. While both search for broadened emission, \cite{martinez-rodriguez24} only select the broadest (>400km/s) H$\alpha$ components and therefore youngest SNR candidates. They therefore ignore the possibility of detecting SNRs with broadened metal lines.

In this subsection we outline our methodology for finding new SNR candidates within MUSE datacubes of nearby galaxies using NGC 4303 as an example case. The datacube of this galaxy is a 3 x 3 MUSE mosaic centered on the galaxy cenntre (leading to a total field size of 3' x 3'). Most of the extent of NGC 4303 was thus observed, however the outer spiral arm regions are outside of the FOV. Even though the galaxy is quite face-on, there is an inclination of 27$^\circ$ \citep{tschoke2000} which causes, along with the galaxy rotation, redshift and blueshift of the emission lines that is detectable within the MUSE data. To account for this shift, we inspect each cube, spaxel by spaxel, identifying H$\alpha$ emission and fitting a Gaussian to it. By comparing the shift of the Gaussian mean with the Gaussian mean of the center of the galaxy, we produce a redshift map of the galaxy, describing the effect of the galaxy inclination. Using this redshift map we remove the galaxy rotational effects by estimating the spectral bin drift. In NGC 4303, the rotation causes a drift of $\approx$2\,\,Å and in other galaxies this can be larger.

\begin{table}[h]
    \centering
    \resizebox{\columnwidth}{!}{\begin{tabular}{l|c|c|c|c|c}
    Line	&	Central	&	Slice&	Continuum& Continuum &  Continuum \\
    name & wavelength & bandwidth & (blue) & (red) & bandwidths \\
    \hline
    		&Å&Å&Å&Å&Å\\
    \hline
    [OIII] & 4995 & 14 & 4897 & 5100 & 11\\
    H$\alpha$ & 6555 & 6.25 & 6530 & 6630 & 11\\
    NII & 6588 & 6.25 & 6530 & 6630 & 11\\
    SII & 6724 & 6.25 & 6630 & 6730 & 11\\
    \end{tabular}}
    \caption{Emission lines used in our survey, their sampled wavelengths and bandwidths.}
    \label{tab:lines_used}
\end{table}

\begin{figure}[h]
  \centering
  \includegraphics[width=\hsize]{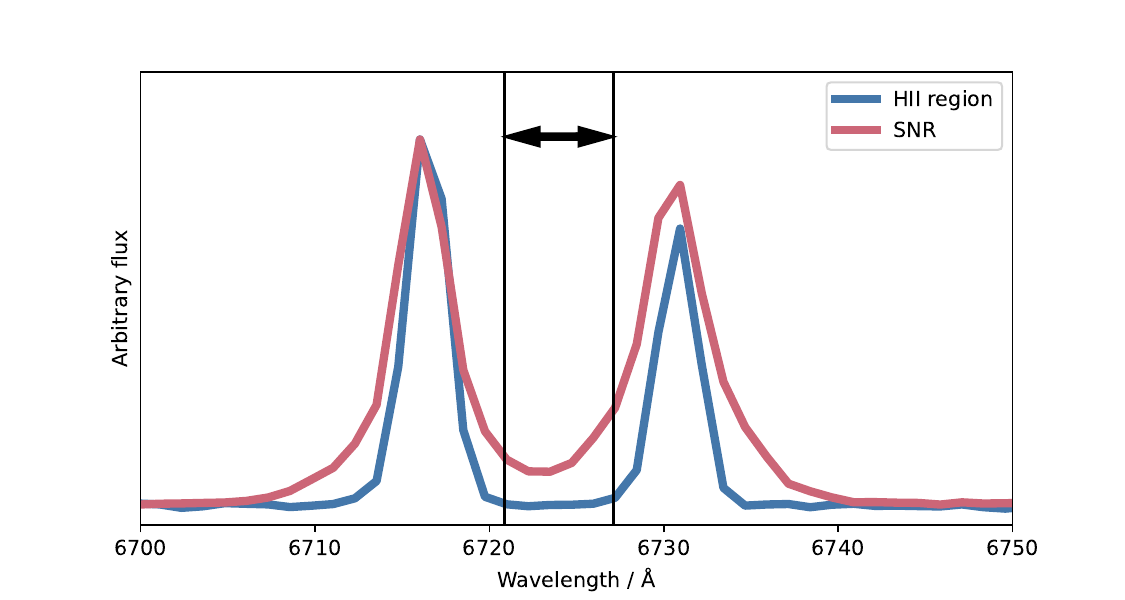}
     \caption{Comparison of [SII]$\lambda\lambda$6717,6731 emission lines from a SNR and an HII region, scaled to similar peak fluxes. The area bounded by black vertical lines (black arrows) is extracted and continuum subtracted. The difference in flux extracted between the two sources is apparent.}
        \label{sii_comparison}
  \end{figure}

We now have our MUSE spectra of NGC 4303 with all spaxels wavelength corrected to a common reference - the galaxy centre. With this in hand, we proceed to extract off-centered linemaps from the cube. Lines used and their respective bandwidths are listed in Table \ref{tab:lines_used}. For example, in the case of the [N\,II]$\lambda$6583 emission line, we extract a linemap from the red side of the emission line. This linemap is centered at 6588\,\,Å, so we select the MUSE spectral bin that includes this wavelength as our central point. From this spectral bin, we select a wavelength range that is 6.25\,\,Å wide, or five spectral bins of 1.25\,\,Å, thus from 6584.88\,\,Å to 6591.13\,\,Å. This range covers the wing of an emission line with a FWHM of about 350 km/s. This methodology is depicted in Fig. \ref{sii_comparison}, as applied to the [S\,II]$\lambda\lambda$6717,6731 doublet. The figure shows how the method applies to a shocked ionised SNR region and also a photoionised HII region, reiterating the point above - SNRs have broader lines the other emission-line sources. The final linemap is produced by summing the flux in the defined wavelength range. In the case of the [O\,III]$\lambda\lambda$4959,5007 doublet, we do not expect to observe any emission lines between them. Therefore, we opted to use a wider bandwith for the linemap in order to increase the S/N of the oxygen emission without the risk of being contaminated other emission lines.

Next, we estimate the continuum level to be extracted from the line maps. We do this using regions offset from the emission lines. In the case of [N\,II]$\lambda$6583, we center them to wavelengths 6530\,\,Å and 6630\,\,Å. For all lines, we use regions offset by +/- 14\,\,Å from the emission line centre to estimate the continuum at the emission-line wavelength. The continuum tends to be either very weak compared to the emission lines or almost linear in such a narrow passband, so we can fit a simple linear continuum model to them. We subtract this model continuum from the spectrum.

Following the above steps, we are now left with maps showing sources of broadened emission lines - unbroadened sources (e.g. most HII regions) will have zero flux following the continuum subtraction. Visually inspecting the newly produced continuum-subtracted linemap we can see some extended emission regions related to strong star-formation (for example the northern quadrant of NGC 4303 in Fig. \ref{sources_m61}). These are caused by high S/N of the emission lines. While the FWHM of an HII region   is effectively constant and dominated by instrument broadening, bright HII regions might still have observable wings that bleed into our linemap. We also see bright unresolved point sources in the maps, which are our points of interest.

For the next step we opt to use SExtractor to extract the coordinates of the point sources from the line maps; at this point we are only interested in the centroid of the point source. We iteratively change the source extraction parameters until no obvious point source is missed in visual inspection. SExtractor produces a catalog of identified sources. Using this catalogue we then extract the 1D spectra of these point sources from the original MUSE datacube. One spectrum is extracted using an aperture with a width of the seeing at the time of the observation, 0.8" or 4 pixels in the case of NGC 4303. A second spectrum is extraced using an annular ring around the first extraction region with a size of 0.2". The second spectrum is used to continuum-subtract the first. We are then left with continuum-subtracted 1D spectra of our point sources and SNR candidates. At these scales the diffuse emission from nearby HII regions and the stellar population is near-homogeneous. By scaling the amount of flux extracted by the ratio of spaxels in both the source aperture and the annulus, we can remove the HII contaminations with a direct subtraction.

With a list of SNR candidate spectra in hand, we now visually inspect each one to identify their nature as true SNRs, contaminant HII regions or other sources. We fit Gaussian lines to visible strong lines, specifically H$\alpha$, [N\,II]$\lambda$6583 and [S\,II]$\lambda\lambda$6717,6731, to measure their FWHM. Spectra that have self-consistent broadened emission lines, i.e. the detected lines of a given source all have similar FWHMs and are higher than 3\,Å (135 km/s), are deemed to be originating from shocked regions and thus be SNRs. We also qualitatively note that these broadened emission sources tend to have stronger [N\,II]$\lambda$6583 and [S\,II]$\lambda\lambda$6717,6731 emission compared to H$\alpha$, indicating a shocked source. HII regions contaminating this sample are identified by having emission line widths close to the instrument broadening, indicating photoionization sources. Star contaminants are identified either from their continua, e.g. M-type stars have metal lines such as TiO, or by the shape of the emission line; emission-line stars have strong hydrogen emission lines that have a Lorentzian profile or they show narrow CaII$\lambda\lambda\lambda$8498,8542,8662 emission lines. This process is then repeated for all catalogues of every emission line used to produce the maps (i.e. [O\,III]5007, H$\alpha$, [N\,II]$\lambda$6583 and [S\,II]$\lambda\lambda$6717,6731). In some cases the same SNR can be identified several times from different line maps. The final step is to then cross-match all the detected SNRs from all line maps based on their respective centroid coordinates. Our final product is then a full list of SNR candidates in NGC 4303. We then repeat these steps for all galaxies within our sample.

We found a total of 307 SNRs in our survey across all galaxies in our sample. Out of these, 35 SNRs had detected oxygen emission, with at minimum a detection in the [O\,III] linemap. This is to be expected as SNRs sometimes exhibit cooling lines such as the [O\,III]$\lambda\lambda$4959,5007 doublet from the SNR interior \citep{Makarenko23}. During the last visual inspection we also took note of the oxygen emission lines of [O\,III]$\lambda\lambda$4959,5007, [O\,I]$\lambda\lambda$6300,6364 and [O\,II]$\lambda\lambda$7320,7330. From this subset, we noted that seven remnants in particular exhibit unusually strong and broad oxygen lines when compared to H$\alpha$, [N\,II]$\lambda$6583 and [S\,II]$\lambda\lambda$6717,6731. We label these events as `O-rich SNRs', and we characterise and discuss them in detail in Chapter \ref{results}.

\subsection{Biases and restrictions}
Before the SNR indentification process outlined in the previous subsection, we also need to consider the completeness of our methodology and what types of false positives we may detect. We assume that the interesting sources are point sources in the MUSE data. This is mostly fine for compact shocked sources such as young SNRs, but we also have a bias for detecting these. There might be objects with narrower emission lines that are closer to the MUSE spectral resolution, e.g. older SNRs, or there may be resolved objects such as the SNR Cygnus Loop. These latter objects might instead show up as arcs with relatively narrow emission lines, but with high peculiar velocity shift at different position angles of the arc.

We also expect to observe objects, such as emission-line stars, P-Cygni stars and Wolf-Rayet stars. All of these objects have strong broad H$\alpha$ emission lines, with widths larger than a few hundred km/s. However, these will have very different emission profiles from shocked sources; emission-line stars generally have lorentzian emission lines with narrow NIR CaII emission lines, while P-Cygni stars show typical blueshifted absorption features. Wolf-Rayet stars have a distinct "red bump" at $\sim$5808\,\,Å which is observable with MUSE \citep{gomezgonzales21}.

The extracted SNR spectra is shock-excited, meaning that the Case B condition for the estimation of extinction is not fulfilled. We therefore estimate extinction based on the Balmer decrement and adopt the fluxes for H$\alpha$ and H$\beta$ from \cite{Emsellem22} for the environments of each individual SNR. We calculate extinction assuming reddening law value R$_V$ = 3.1 and adopt the model from \cite{Cardelli89}. This may cause inaccuracies as the colour excess is measured on larger scales and does not probe the intrinsic SNR extinction.

\subsection{Counterparts at other wavelengths} \label{counterparts}
After the identification of O-rich SNRs in our sample, we searched for counterparts in the archival data of other facilities. SN1979C, another long-observed SN and SNR 4449-1 have been detected in X-rays \citep{Patnaude11, Patnaude2003}, IR \citep{Tinyanont16} and radio \citep{weiler81, Lacey07, bietenholz10}. Depending on the age of the remnants and the depth of archival data, we thus expect to find counterparts to these sources as well. We performed archival searches for HST, JWST and Chandra data using the Mikulski Archive for Space Telescopes (MAST)\footnote{mast.stsci.edu} and Chandra Data Archive (CDA)\footnote{https://cxc.cfa.harvard.edu/cda/}.

For HST, we find that all O-rich SNR host galaxies have been observed with Wide Field Camera 3 with at least filters F275W, F336W, F438W, F555W and F814W, mostly due to PHANGS-HST and the Legacy ExtraGalactic UV Survey (LEGUS).

With JWST the observations for PHANGS-JWST are still ongoing as of writing; NGC 4303 has only partial coverage with the Near Infrared Camera (NIRCam) and does not include the positions of the O-rich SNRs in the galaxy. In the other cases NIRCam observations include filters F200W, F300M, F335M and F360M. JWST's Mid-Infrared Instrument (MIRI) has observations with filters F770W, F1000W, F1130W and F2100W. In Chandra data, there are usually multiple epochs of data. We measure both the stacked flux and generate a lightcurve based on them. The former assumes a constant flux between the epochs and gives a more accurate flux measurement, while the latter assumes an evolving lightcurve. We use the coordinates from MUSE data and attempt to visually identify point sources in the datasets of other instruments. Absolute coordinate solutions in the publicly available PHANGS-MUSE dataset have been validated against Gaia DR1 \citep{Emsellem22} and we expect World Coordinate System coordinates in the space telescope archival data to be accurate down to the image quality of MUSE observations, i.e. less than 0.8". 
\\

In this section we have outlined our methodology for discovering new SNRs in our dataset, using NGC 4303 as an example. During the search, we discovered a total of seven unresolved sources, that have broad emission oxygen emission lines. In the following section we define our list of O-rich SNRs and proceed to characterise each event in terms of its observed features at optical wavelengths, compare them to a recovered SN and the prototypical O-rich remnant SNR 4449-1. We will also discuss the counterparts detected at other wavelengths.

\section{Results} \label{results}
\begin{figure}[h]
   \centering
   \includegraphics[width=\hsize]{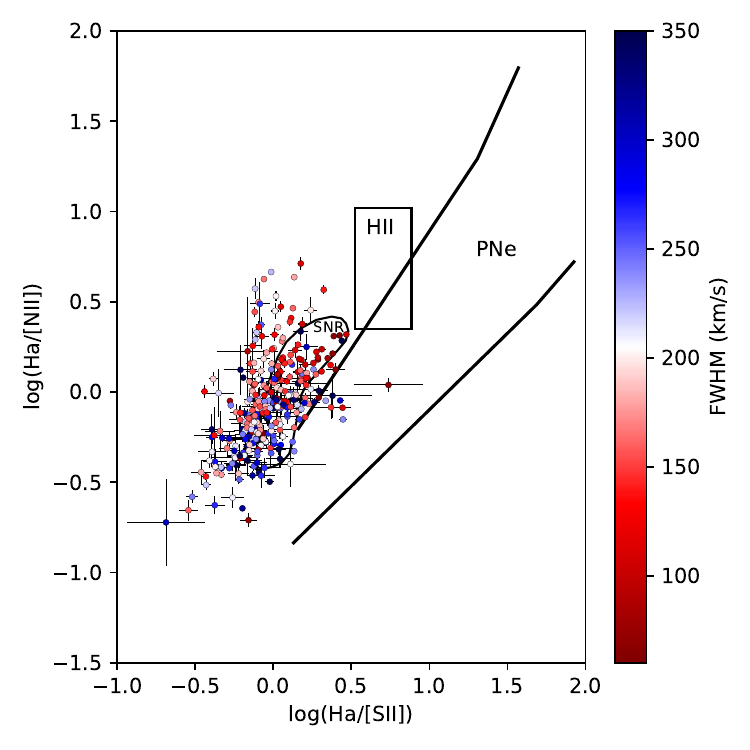}
    \caption{Our sample of broadened emission lines that have detectable H$\alpha$, [N\,II]$\lambda$6583 and [S\,II]$\lambda\lambda$6717,6731 emission lines and that we identify as SNRs. Colour of the points represent the FWHM of the [N\,II]$\lambda$6583 with the instrumental broadening removed. The different regions in the figure have been adopted from \citealt{meaburn10}. Most of our sample aligns with the SNR region with log(H$\alpha$/[S\,II]) < 0.5.}
      \label{dpuser_plot}
         \end{figure}

\begin{figure*}
        \centering
        \includegraphics[width=19cm]{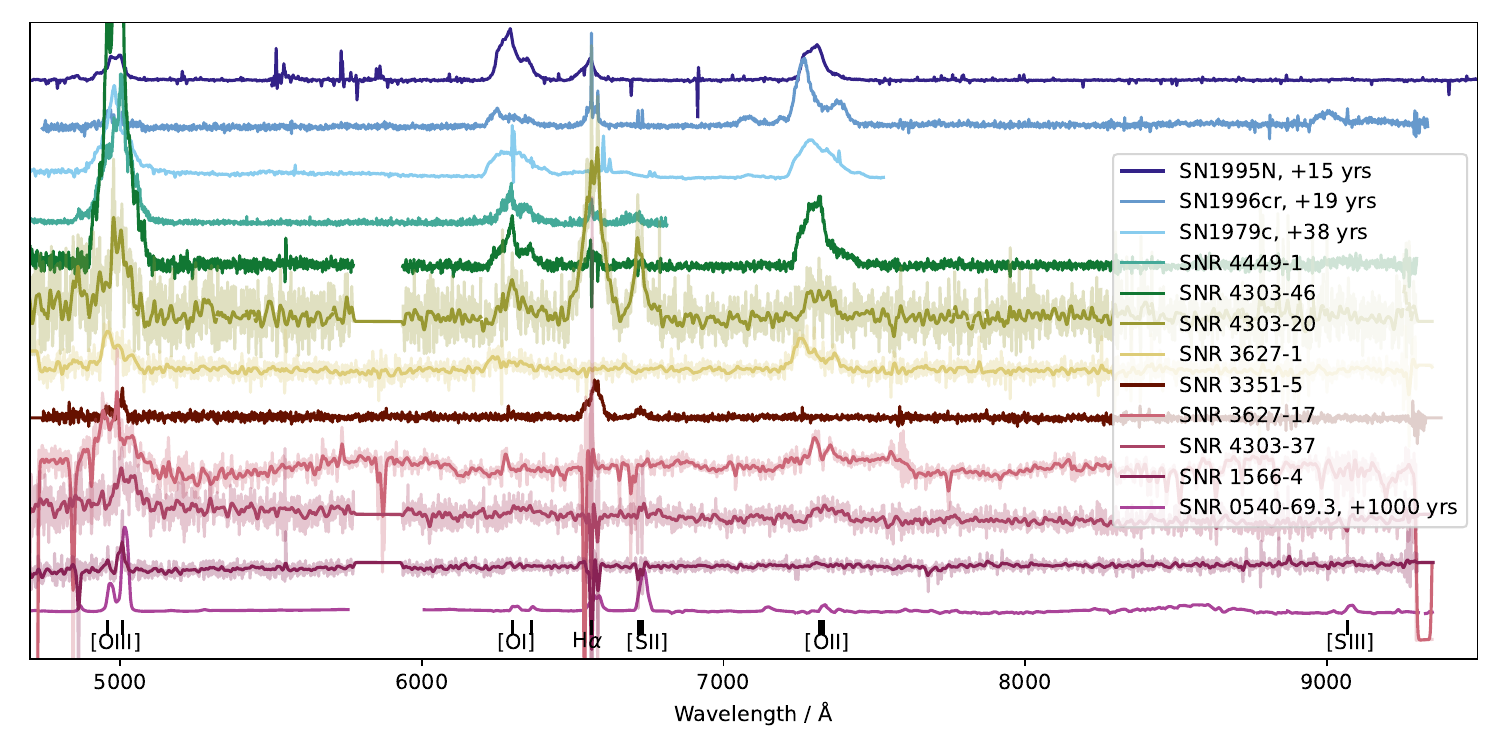}
        \caption{Optical wavelength spectra of all seven O-rich SNRs discovered by our work. We addditionally display the spectra of three SNe at 15 to 40 years post explosion, SNR 4449-1, and also include a spectrum of SNR 0540-69.3, an old O-rich SNR in the LMC.}
        \label{fig:source_specs}
   \end{figure*}

Through the methodology outlined in the previous section, we have found a total of 307 SNR candidates in the PHANGS-MUSE sample. All detected SNRs are listed in \nameref{snr-list}. We leave a thorough analysis of this full sample for a future publication. For the newly discovered SNRs, we use a naming convention of SNR XXXX-YY, where XXXX is the galaxy number in the NGC/IC catalogues and YY is the running number of the SNR based on the declination from South to North. The SNRs are presented in Fig. \ref{dpuser_plot} on an emission-line diagnostic plot. The SNR population on average exhibit [S\,II]$\lambda\lambda$6717,6731 and [N\,II]$\lambda$6583 fluxes comparable to the H$\alpha$ flux. We note a trend in the FWHM of the SNRs, where the linewidth tends to narrow as the SNR gets closer to the photoionized zone in the figure. The outlier remnant with log(H$\alpha$/[S\,II]) > 0.5 is the O-rich SNR 3627-17. For the rest of the O-rich SNRs, the emission lines are either too blended to measure accurately or too weak to detect. A total of seven SNRs (just 2\% of the whole sample) exhibit optical emission line features reminiscient of young O-rich SNRs (see Fig. \ref{fig:source_specs}).

We will discuss each O-rich remnant separately in the succeeding sections. In general, the remnants are heterogenous (as will be outlined below), showing different strength H$\alpha$ and [S\,II] emission, but all of them have at least a detectable broad [O\,III] emission and sometimes also [OII] and [OI] emission lines (defining them as `O-rich'). We note that the spectrum of SNR 4449-1 is reminiscient of the O-rich SNRs presented in this paper to varying degrees. Similar features have also been observed in old SNe/young SNRs, including SN1957D, SN1979C, SN1980K, SN1996cr and SN1995N \citep{Knox12, Fesen99, Bauer08, Wesson2023}. We will compare all of our sources to both SNR 4449-1 and SN 1979C in the figures. The two sources are similar to each other and they are also most reminiscient to our O-rich SNRs. The spectra are scaled in order to highlight the spectral features and their similarities. We will describe the spectral features, possible counterpart detections and how they compare to other remnants and long-lasting SNe.


\begin{figure}[h]
   \centering
   \includegraphics[width=\hsize]{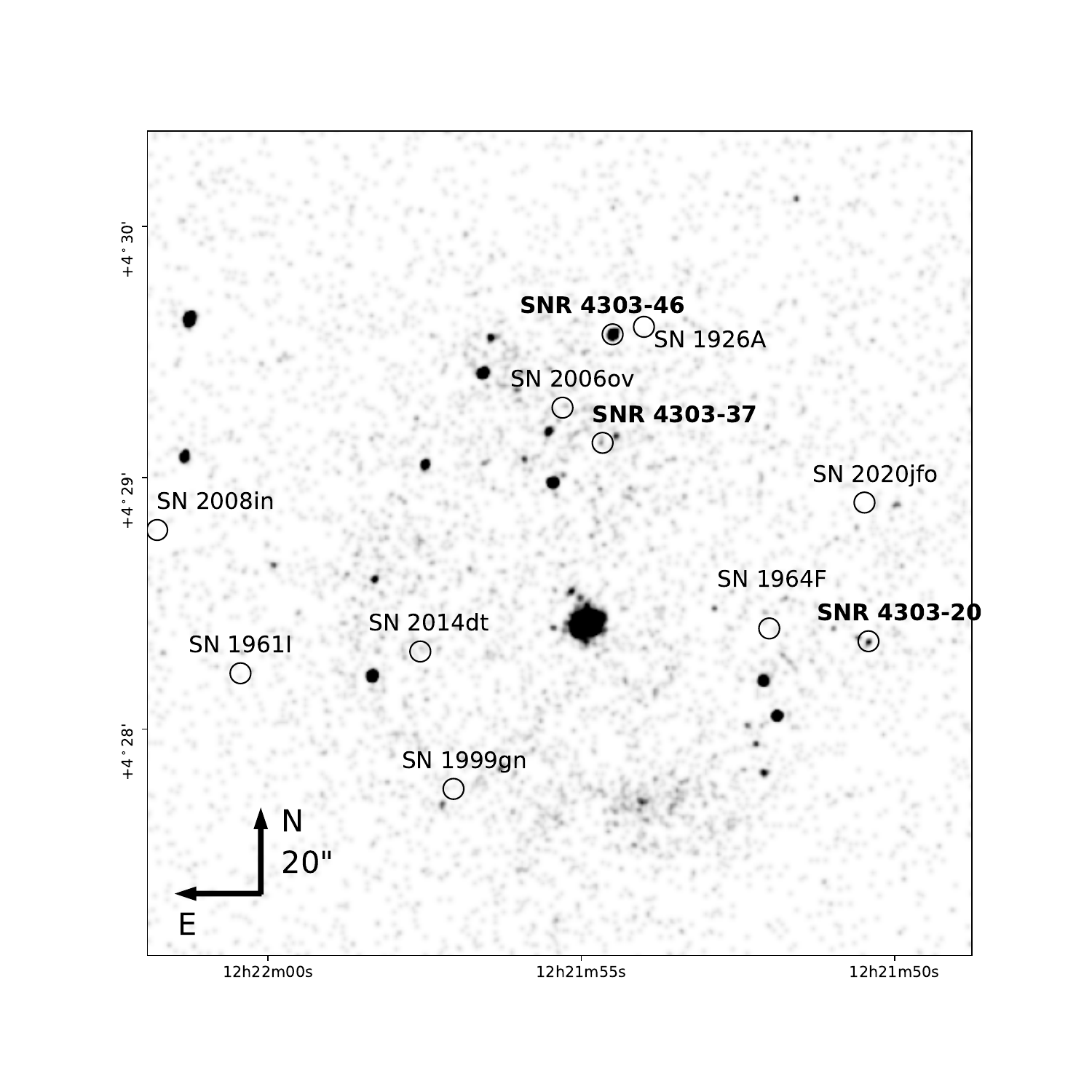}
      \caption{X-ray image of NGC 4303 obtained with Chandra, observed in August 2001. None of the historical SNe have an associated source in X-ray, but both SNR 4303-46 and SNR 4303-20 are detected. SNR 4303-20 is considerably dimmer compared to SNR 4303-46, while SNR 4303-37 has very faint detection.}
         \label{xray_source_specs}
   \end{figure}

\begin{figure}[h]
   \centering
   \includegraphics[width=\hsize]{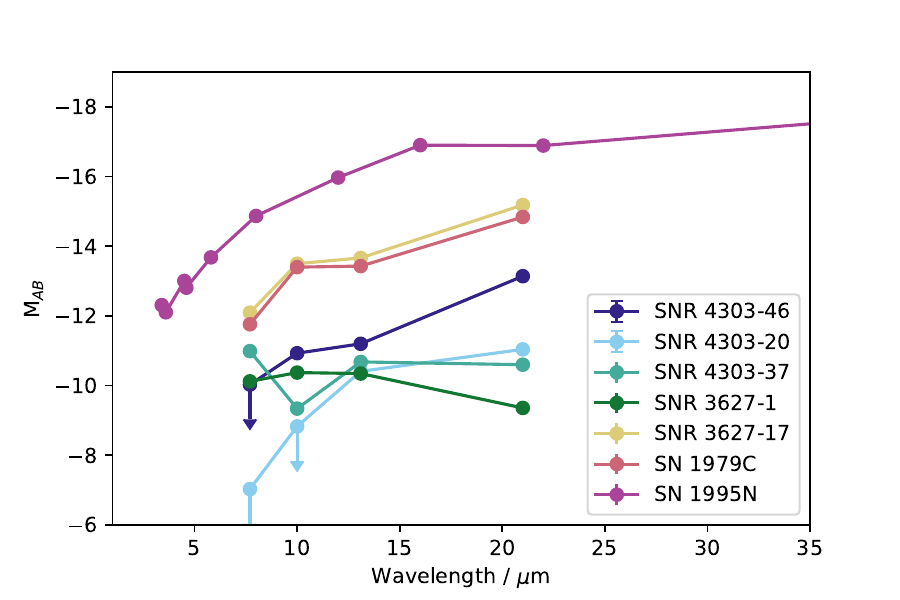}
      \caption{JWST observations of the new O-rich SNRs compared to SN 1979C with JWST and SN 1995N from \cite{Wesson2023}.
         \label{jwst_data}}
    \end{figure}

Searching the space-telescope archives MAST and CDA, we find that most of the sources have a Chandra counterpart, i.e. an unresolved source within 1" of the object, (see example of NGC 4303 in Fig. \ref{xray_source_specs}). The luminosities for all sources range in the order of few $10^{37}-10^{38}$ erg/s for these objects based on measurements outlined in section \ref{counterparts} (see Table \ref{tab:xray}). This is near the limit of ultra-luminous X-ray sources and around the Eddington limit of neutron star mass objects. These are also comparable to SN 1979C; L$_x$ = 6.5$\times$ 10$^{38}$ erg/s and SN 1957D; L$_x = 1.7\times 10^{37}$ erg/s \citep{long12}. HST counterparts are generally only detected in F555W and F814W, with weaker detection or non-detections in other filters such as the near-ultraviolet filter F275W. This seems to agree with previous observations of SNR 4449-1 where the dominant emission features are in visible/NIR oxygen lines (see fig. 1 in \citealt{blair83}). In JWST we observe bright point sources at these remnant locations. The detections are presented in Fig. \ref{jwst_data}. Specifically SNR 4303-46 and SNR 3627-17 show very similar features compared to SN 1979C.

Given the similarity between our discovered O-rich SNRs and old SNe/young SNRs, we searched for known SNe at the locations of our SNRs. However, no matches were found. SNR 4303-46 is near to SN1926A, but the projected separation is over 5", corresponding to a distance of 350 pc (see Figs. \ref{sources_m61} and \ref{xray_source_specs}). We discuss this possible connection below.

Now, we discuss each identified O-rich SNR in detail, characterising its optical-wavelength properties and noting detections at other wavelengths. A summary of the observed properties of our sample is presented in Appendix A.

\subsection{SNR 4303-46}
\begin{figure}[h]
   \centering
   \includegraphics[width=\columnwidth]{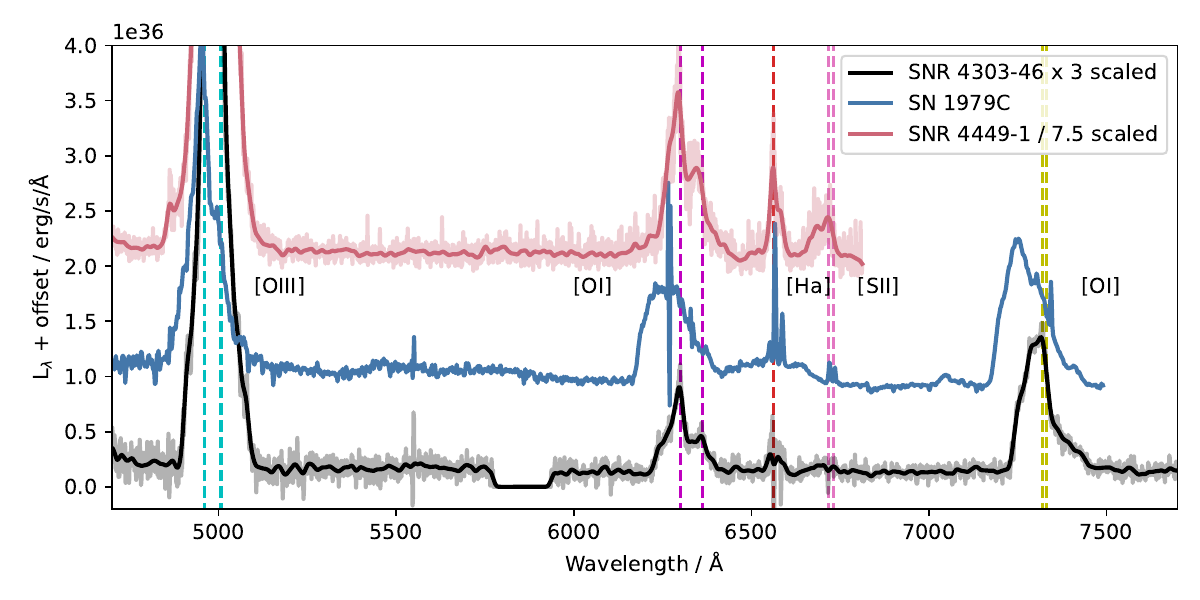}
      \caption{Comparison of SNR 4303-46, SNR 4449-1 and SN 1979C. Solid lines are smoothed spectra, shaded represents the original spectrum. Vertical lines indicate different emission lines; [O\,III] (cyan), [O\,I] (magenta), H$\alpha$ (red), [S\,II] (pink) and [O\,II] (yellow).}
         \label{snr4303-53}
   \end{figure}

SNR 4303-46 was detected in the off-center linemaps in [O\,III] and H$\alpha$, being the brightest point source in the [O\,III] linemap of NGC 4303 (see Fig. \ref{sources_m61}). The MUSE spectrum (from 4700\,\,Å to 9400\,\,Å) is presented in Fig. \ref{snr4303-53}. The most striking and dominant component observed is the [O\,III]$\lambda\lambda$4959,5007 emission feature, which is brighter than the rest of the spectrum, defining it as an O-rich SNR. SNR 4303-46 also displays strong, broad lines of [O\,I]$\lambda\lambda$6300,6364 and [O\,II]$\lambda\lambda$7320,7330. A comparatively weak H$\alpha$+[N\,II]$\lambda$6583 is also visible, along with a marginal [S\,II]$\lambda\lambda$6717,6731 detection.

The FWHM of the [O\,III]$\lambda\lambda$4959,5007 complex is over 4000 km/s and is asymmetrical. We measure a total flux of 1.8$\pm$0.3$\times$ 10$^{38}$ erg/s. Due to the high S/N of the spectrum we can see that the emission feature is a blend of several [O\,III] doublets with different velocities with respect to the assumed rest wavelength. Overall this is very similar to both SNR 4449-1 and SN 1979C, especially to the former.

\begin{figure}[h]
   \centering
   \includegraphics[width=\hsize]{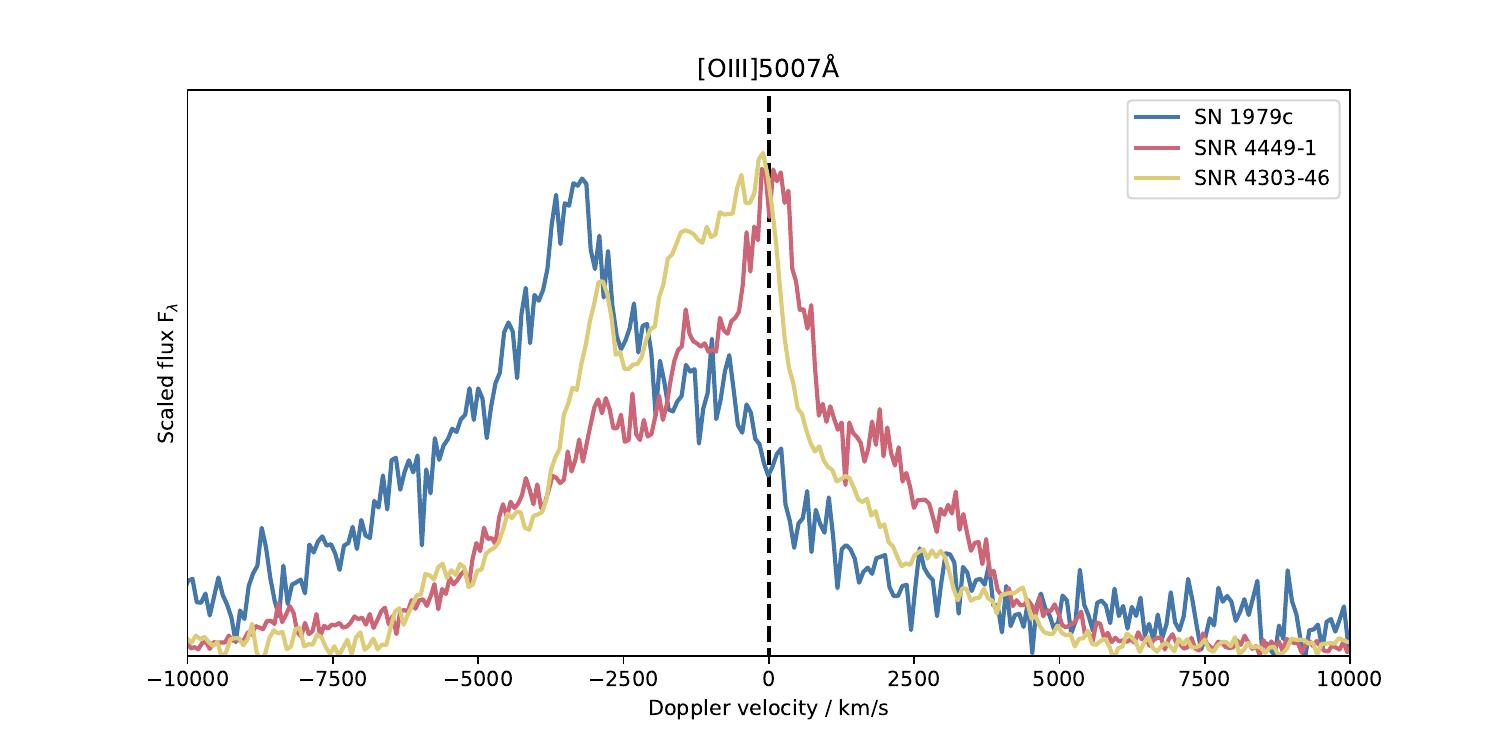}
   \includegraphics[width=\hsize]{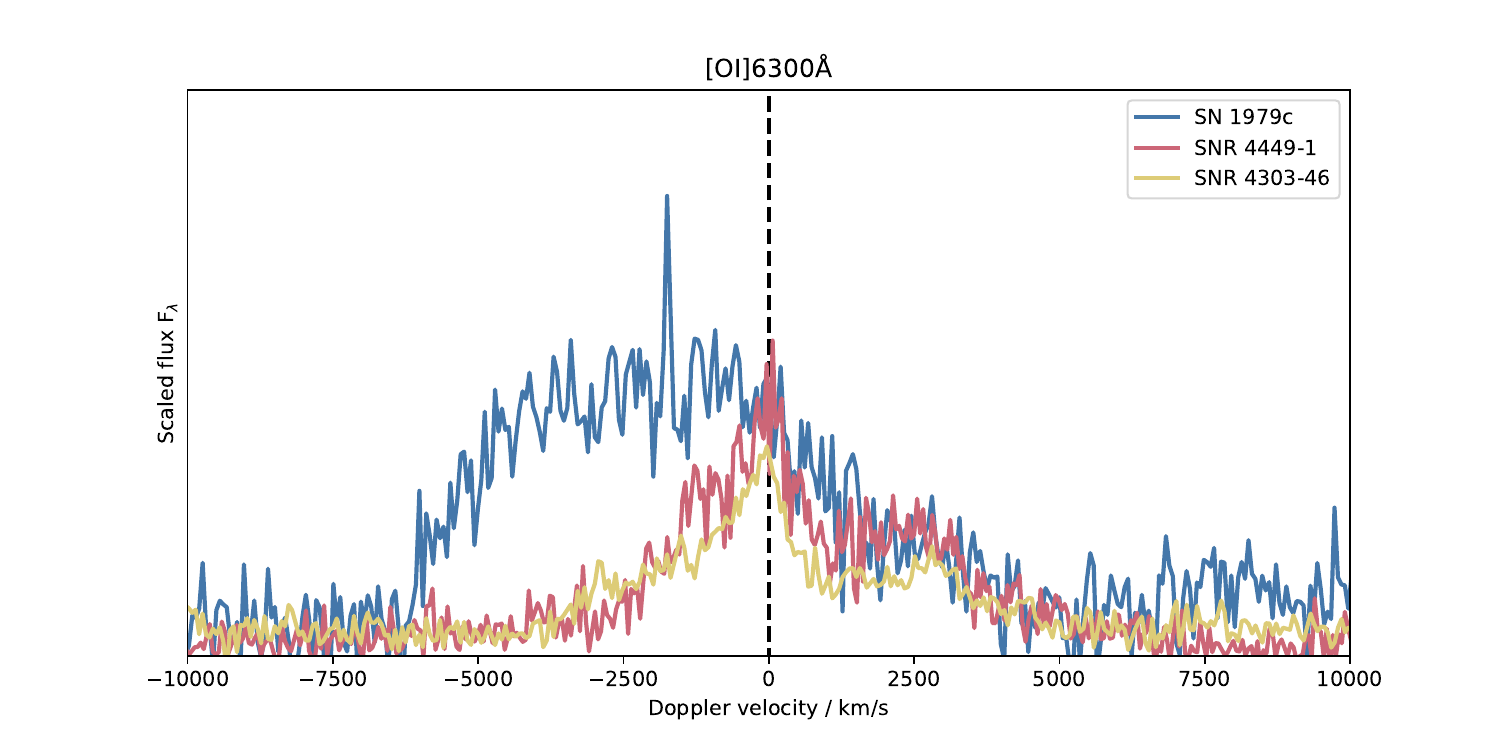}
   \includegraphics[width=\hsize]{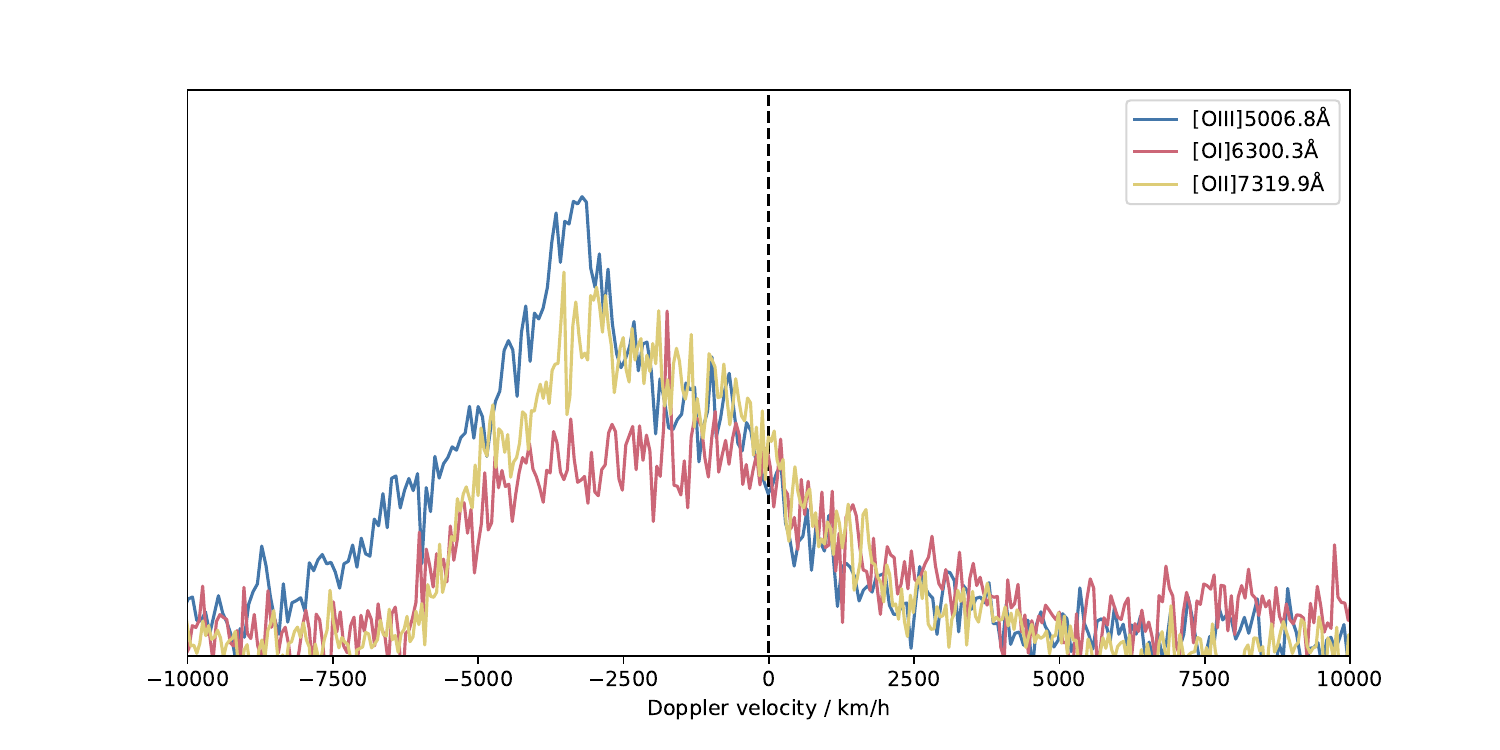}
   \includegraphics[width=\hsize]{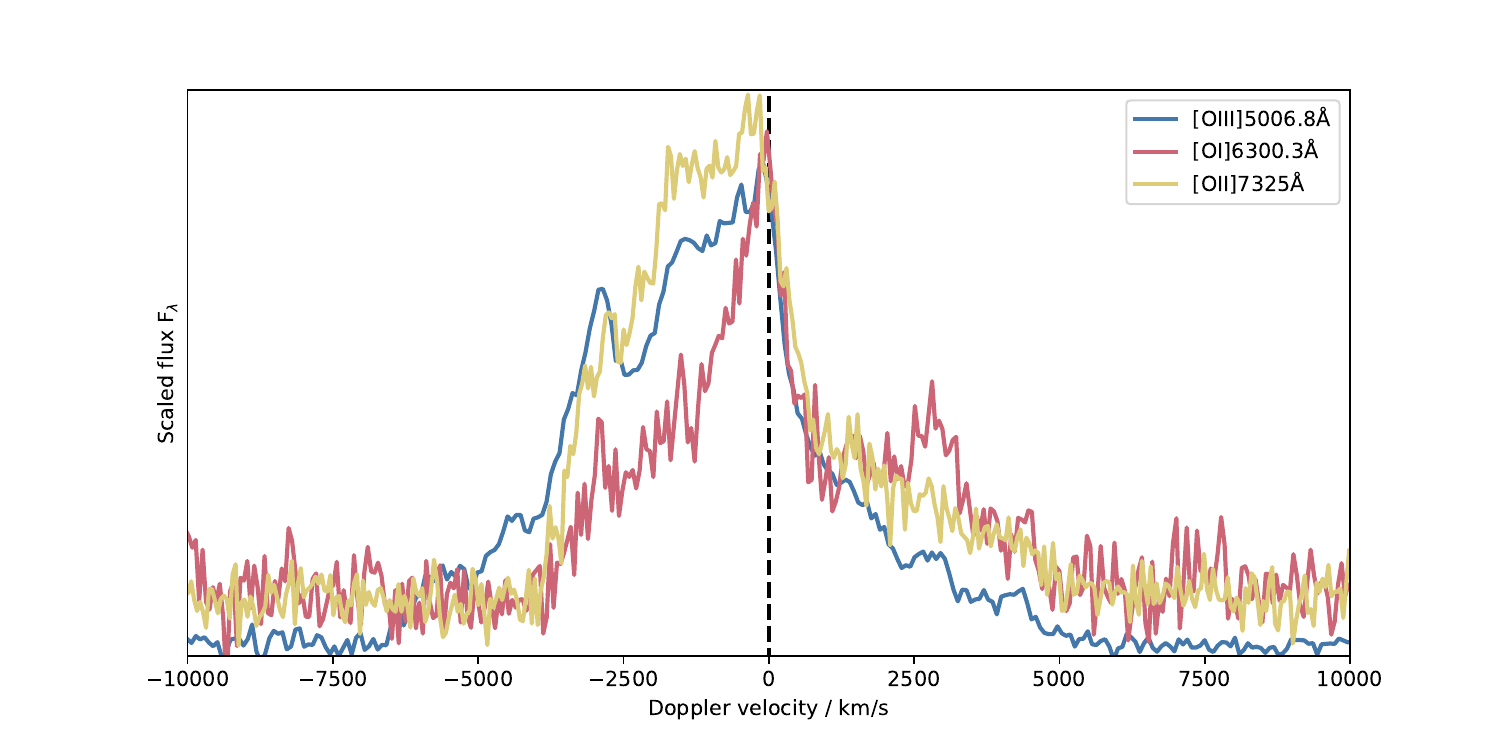}
      \caption{[O\,III]$\lambda\lambda$4959,5007 emission complex in velocity space centered around 5006.8Å (top) and same for [O\,I]$\lambda\lambda$6300,6364 centered around 6300Å (second panel). The oxygen lines of SN1979C with the addition of [O\,II]$\lambda\lambda$7320,7330 centered at 7320Å (third panel) and the same lines for SNR 4303-5 (bottom panel). The spectra are normalized arbitrarily to have similar peak intensities.}
         \label{sourcea_etc}
   \end{figure}

In Fig. \ref{sourcea_etc} we show the scaled [O\,III]$\lambda\lambda$4959,5007 and [O\,I]$\lambda\lambda$6300,6364 lines of SNR 4303-46, SNR 4449-1 and SN 1979C. SN 1979C's emission lines are clearly blueshifted when centered at the host galaxy's rest wavelengths (similar figures for the other O-rich SNRs are shown in Appendix D). This could be an indication of some O-rich outflows in SN1979C or an obstruction of the receding shell due to high dust extinction. Still, the red wings of SN 1979C's and SNR 4303-46's oxygen lines roughly trace each other, which suggests a similar origin, with SN 1979C only having excess blueshifted emitting material. Similarities to SNR 4449-1 are more clear, with the [O\,I]$\lambda\lambda$6300,6364 lines having effectively the same shape. [O\,III]$\lambda\lambda$4959,5007 lines are also similar, with SNR 4303-46 having a slight blueshifted excess and SNR 4449-1 exhibiting a red bump. Both also show a similar structure that could be explained with blueshifted and redshifted oxygen doublets as previously discussed by \citealt{Milisavljevic08}. We measured a total flux of [O\,III]$\lambda\lambda$4959,5007 for SNR 4449-1 of 4.6$\times$ 10$^{39}$ erg/s in the 1996 HST spectra. This is an order of magnitude brighter than SNR 4303-46.

SNR 4303-46 is also detected with HST, JWST and Chandra. Figures \ref{fig:sourcea_hst} and \ref{fig:sourcea_jwst} in the appendix show the cutouts from HST and JWST images. We identify a strong point source in filter HST/F555W and JWST/F2100W slightly offset from the MUSE detection centerpoint. In all cases we observe a point source. From the HST images this limits the size of the source to 6 pc, which in turn limits the age of the SNR to less than a thousand years: assuming an ejecta that has a constant speed of 4 000 km/s (based on the [O\,III] doublet) it would reach a radius of 3 pc in about 740 years. Considering the ejecta experiences only deceleration normal circumstances, this age would then be the upper limit for the age of the SNR.


Figure \ref{jwst_data} shows the similarity between SN 1979C and SNR 4303-46 JWST SEDs along with our other O-rich sources that we have been detected in the JWST observations. Adopting a distance modulus of 31.03, SNR 4303-52 reaches M$_{AB}$ = -13.14 in F2100W. While SNR 4303-46 is two magnitudes fainter, the similarity in their overall shape is apparent. The slight plateau near the F1130W filter could be an indication of silicate dust, which has an absorption feature around 14 $\mu$m.

The SNR is also detected with Chandra in 2001 and 2014, as presentend in Table \ref{tab:xray} in \nameref{xray_phot}. The two epochs have very similar luminosities and by combining the two observations, we estimate a luminosity L$_x$ = 5.4$\times$ 10$^{38}$ erg/s. The same source is also likely observed by ROSAT between July 1996 and January 1998 with a luminosity of 4.3$\pm$1.5$\times$ 10$^{38}$ erg/s; source g indentified in \citealt{tschoke2000} is consistent with the Chandra source when considering ROSAT's lower spatial resolution. This is comparable to the Chandra observations and could be an indication that the source X-ray luminosity is constant over this period. 

\subsubsection{A connection between SNR 4303-53 and SN 1926A?}
As is apparent in Fig. \ref{xray_source_specs}, our detection of SNR 4303-53 appears spatially close to the position of SN 1926A. Given that one of our hypothesis presented in this work is that our O-rich SNRs may be related - in time - with old SNe, we decided to furthere investigate a possible connection of these two events. SN 1926A was discovered and followed up by Max Wolf and Karl Reinmuth at the Heidelberg Observatory with the 72 cm Walz reflector telescope \citep{1926discovery}. The plates have been scanned and are available online\footnote{http://www.lsw.uni-heidelberg.de/projects/scanproject}. We analysed a photographic plate dated May 11th 1926. Stars were identified on the old plate and matched to those with positions from Gaia data release 3 \citep{Gaia16, Gaia23}. This provided an astrometric solution with a standard deviation of 1.5" between the reported coordinates and the star locations. While this is quite a large, the distance between SNR 4303-53 and SN1926A is 7.8", which equates to a 5-sigma separation. Following the above analysis, we do no believe that these to events are physically associated.

\subsection{SNR 4303-20}
\begin{figure}[h]
   \centering
   \includegraphics[width=\columnwidth]{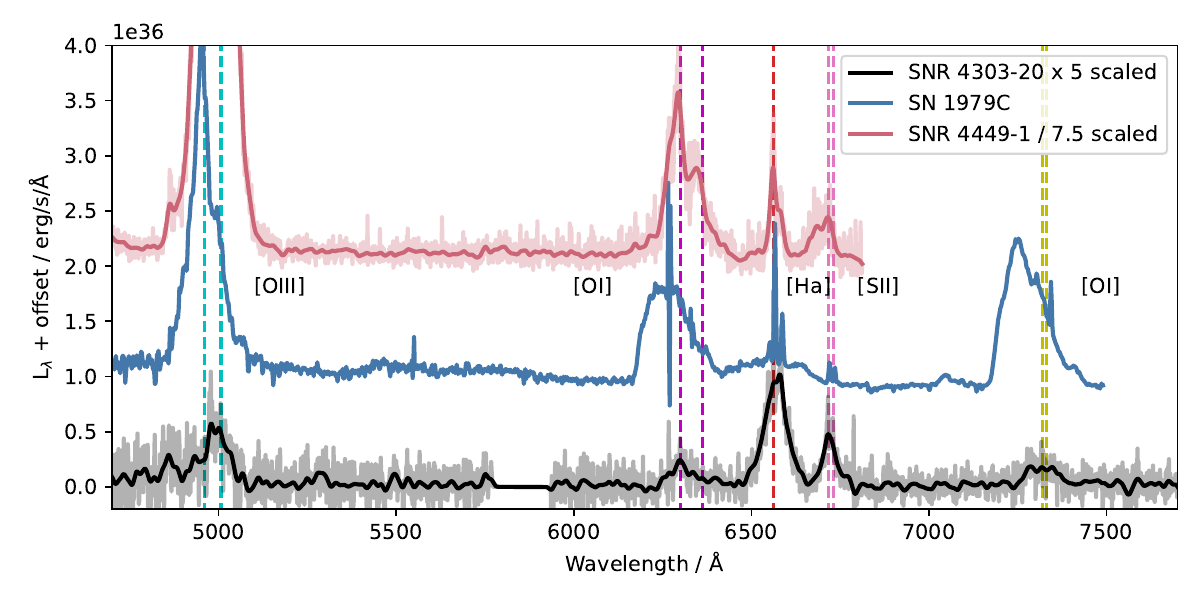}
      \caption{Comparison of SNR 4303-20, SNR 4449-1 and SN 1979C. Solid lines are smoothed spectra, shaded represents the original spectrum. Vertical lines indicate different emission lines; [O\,III] (cyan), [OI] (magenta), H$\alpha$ (red), [S\,II] (pink) and [OII] (yellow)}
         \label{snr_79_b_c}
   \end{figure}

SNR 4303-20 was detected in all linemaps used in the survey. This is due to a strong H$\alpha$+[N\,II]$\lambda$6583 complex, along with strong [S\,II] and oxygen emission (see Fig. \ref{snr_79_b_c}). These lines make the remnant to be somewhat reminiscient of a classical SNR, with the exception of the lines being broad: the H$\alpha$+[N\,II]$\lambda$6583 complex has a FWHM of about 3700 km/s, comparable to the oxygen lines observed in SNR 4303-46. This complex also seems to be blended; the [S\,II] doublet has a a total FWHM of about 1700 km/s. While the remnant's oxygen lines are not the strongest emission sources in the MUSE spectrum they are nevertheless broad as well, with [O\,III]$\lambda\lambda$4959,5007 doublet dominating over the less ionized oxygen lines. The [O\,III]$\lambda\lambda$4959,5007 luminosity was measured to be 7.02$\times$ 10$^{36}$ erg/s, over two orders of magnitude lower than SNR 4303-46.

SNR 4303-20 was also detected with JWST (Fig. \ref{jwst_data}) and Chandra (Table \ref{tab:xray}). We find no obvious candidate within 1" of the MUSE location in the HST images. The MUSE spectrum of the region is dominated by a strong blue continuum, which may occlude the weak (compared to SNR 4303-46) emission lines of the SNR.

While there is a difference between SNR 4303-20 and the literature comparisons we have presented, the sources may still have similar origins. The lower luminosity in X-ray together with the detection of optical oxygen emission could be an indication of older age, assuming that the sources are similar at early times. The emergence of the [S\,II] doublet and stronger [N\,II]$\lambda$6583 emission indicates lower densities as these lines have lower critical densities compared to the [O\,III] lines. This is further supported by the JWST SED, where we do not detect emission at the the shortest filters with MIRI, F770W and F1130W. If the mid-IR emission originates from dust, the temperature would be lower compare to SNR 4303-46 and SN 1979C.

\subsection{SNR 3627-1}
\begin{figure}[h]
   \centering
   \includegraphics[width=\columnwidth]{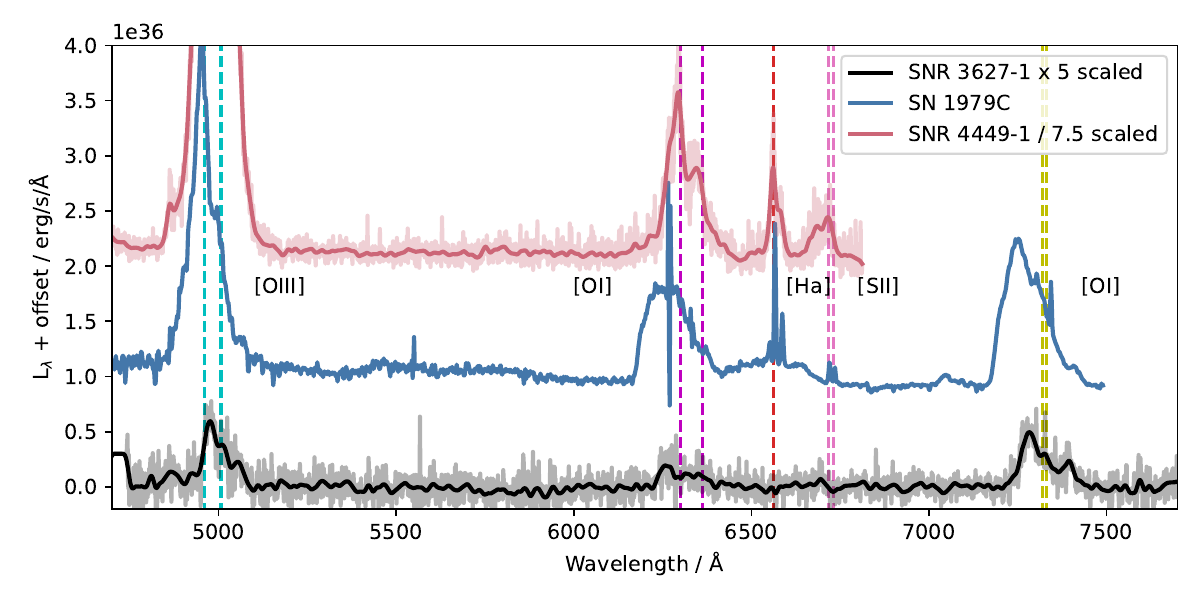}
      \caption{Comparison of SNR 3627-1, SNR 4449-1 and SN 1979C. Solid lines are smoothed spectra, shaded represents the original spectrum. Vertical lines indicate different emission lines; [O\,III] (cyan), [OI] (magenta), H$\alpha$ (red), [S\,II] (pink) and [OII] (yellow)}
         \label{snr3627-2}
   \end{figure}

SNR 3627-1 was detected only in the [O\,III] linemap and only oxygen lines are visible in the full spectrum, [O\,III] and [OII] being similar in strength. The sprectrum is faint (presented in Fig. \ref{snr3627-2}), but the linewidths are comparable to SNR 4303-20 (see Fig. \ref{snr_79_b_c}). The oxygen lines are also asymmetrical with a blueshifted peak and a longer red tail. If we assume the peaks of the [O\,III], [OII] and [OI] doublets in the spectrum to correspond to the emission lines at 5007\,\,Å, 7320\,\,Å and 6300\,\,Å respectively, peaks are blueshifted roughly by 1400 km/s. The [O\,III] luminosity is measured to be 2.89$\times$ 10$^{37}$ erg/s.

SNR 3627-1 is detected with JWST/MIRI, showing a blackbody-like SED with a maxima around 10 $\mu$m (Fig. \ref{jwst_data}). The SNR has a high S/N Chandra counterpart observed in 2008 with a luminosity of 6.5$\times$ 10$^{37}$ erg/s.

\subsection{SNR 3351-5}
\begin{figure}[h]
   \centering
   \includegraphics[width=\hsize]{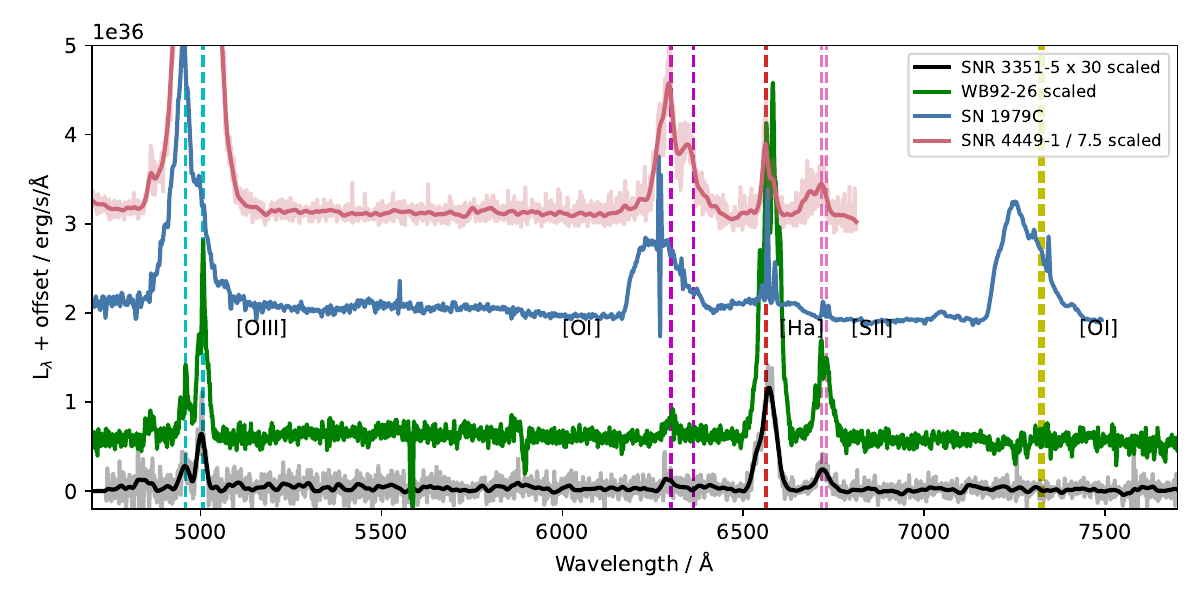}
      \caption{Comparison of SNR 3351-5, SNR 4449-1, SN 1979C and the recently discovered WB92-26 in M31 by \citep{Caldwell23}. Solid lines are smoothed spectra, shaded represents the original spectrum. Vertical lines indicate different emission lines; [O\,III] (cyan), [OI] (magenta), H$\alpha$ (red), [S\,II] (pink) and [OII] (yellow)}
         \label{snr3351}
   \end{figure}

 SNR 3351-5 was detected in all of the linemaps. The spectrum is dominated by a blended H$\alpha$+[N\,II]$\lambda$6583 emission, but [O\,III], very weak [O\,I] and [S\,II] are also detected (see Fig. \ref{snr3351}). The [O\,III] doublet is unblended and resolved with a FWHM of 1400 km/s, with a total luminosity of 7$\times$ 10$^{35}$ erg/s. This source is most reminiscient of a recent SNR detected in M31, identified as a N-rich SNR \citep{Caldwell23} and therefore we also plot this literature event in Fig. \ref{snr3351}. We do not detect the remnant in the JWST observations, but the Chandra observations detect a point source with a luminosity of 4.81$\times$ 10$^{37}$ erg/s.


\subsection{SNR 3627-17}
\begin{figure}[h]
   \centering
   \includegraphics[width=\columnwidth]{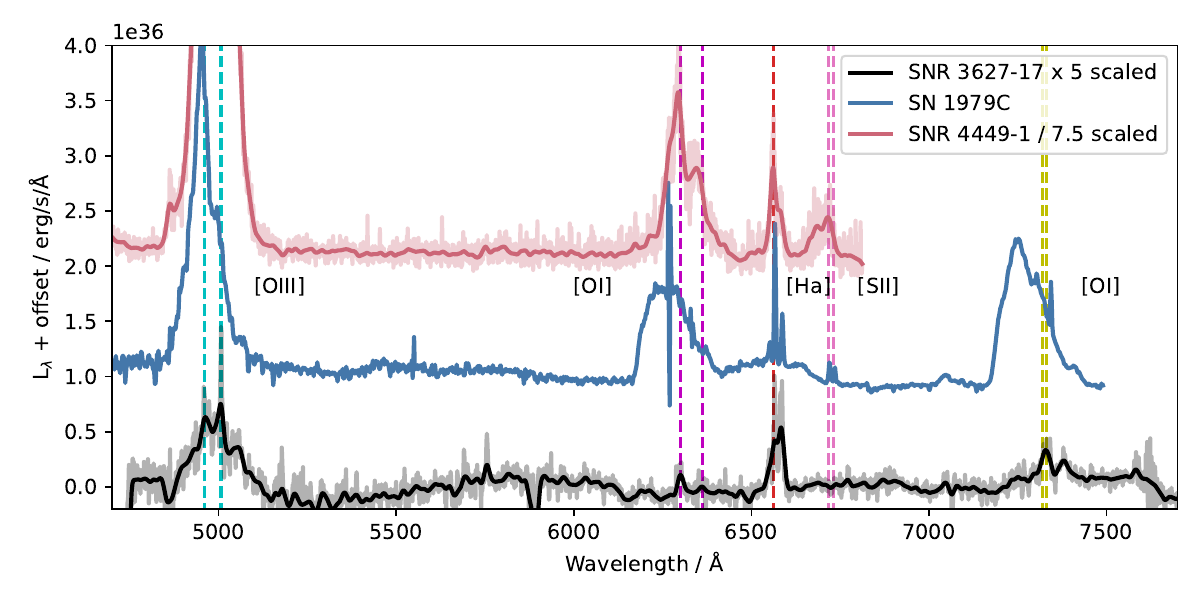}
      \caption{Comparison of SNR 3627-17, SNR 4449-1 and SN 1979C. Solid lines are smoothed spectra, shaded represents the original spectrum. Vertical lines indicate different emission lines; [O\,III] (cyan), [OI] (magenta), H$\alpha$ (red), [S\,II] (pink) and [OII] (yellow)}
         \label{snr3627-18}
   \end{figure}

SNR 3627-17 was only detected in the [O\,III] linemap. It is embedded in a cluster of blue stars. For this source, we removed the stellar continuum using STARLIGHT \citep{Fernandes05} and the SEDs of simple stellar populations at different ages with BPASS v2.2.1 \citep{Stanway18}. The resulting continuum-subtracted spectra is still noisy, but we detect [O\,III]$\lambda\lambda$4959,5007, [O\,I]$\lambda$6300 and weak H$\alpha$ and [N\,II]$\lambda$6583 (see Fig. \ref{snr3627-18}). [O\,III] has the broadest FWHM of all O-rich SNRs in our sample with a width of over 6000 km/s with a total luminosity of 9.1$\times$ 10$^{36}$ erg/s. We detect the SNR as a bright point-source in both Chandra (see Table \ref{tab:xray} with a total luminosity of 1.35$\times$ 10$^{38}$ erg/s) and JWST (Fig. \ref{jwst_data}) observations with a bright red SED; the absolute magnitude with F2100W filter is M$_{AB}$ = -12.18 mag. This remnant and SN 1979C have similar JWST SEDs; there is a small plateau between F1000W and F1130W filters. As with SNR 4303-46, this feature could be indicative of the 14 $\mu$m absorption feature inherent to silicate dust.

\subsection{SNR 4303-37}
\begin{figure}[h]
   \centering
   \includegraphics[width=\columnwidth]{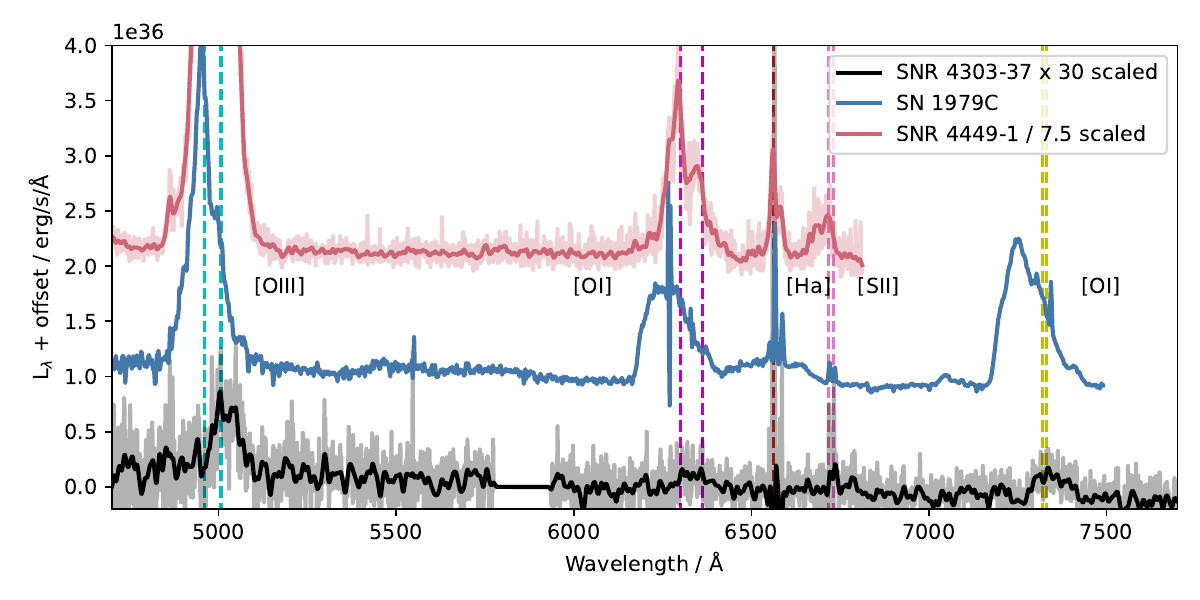}
      \caption{Comparison of SNR 4303-37, SNR 4449-1 and SN 1979C. Solid lines are smoothed spectra, shaded represents the original spectrum. Vertical lines indicate different emission lines; [O\,III] (cyan), [OI] (magenta), H$\alpha$ (red), [S\,II] (pink) and [OII] (yellow)}
         \label{snr4303-41}
   \end{figure}

SNR 4303-37 is detected as a source with a somewhat weak, but broad [O\,III] emission, with a FWHM of over 2000 km/s for the individual lines in the doublet and total luminosity of 3.7$\times$ 10$^{36}$ erg/s, with very weak [OII] and [OI] detections (see Fig. \ref{snr4303-41}). No other line is obvious, but it is reminiscient of SNR 3627-1 and also SNR 4303-20, but without the H$\alpha$+[N\,II]$\lambda$6583 and [S\,II] lines (Figs. \ref{snr_79_b_c} and \ref{snr3627-2}). The Chandra data has only a few photons, which is consistent with the diffuse host galaxy flux, giving an upper limit luminosity of 2.1$\times$ 10$^{37}$ erg/s. The remnant is clearly detected in JWST/MIRI with a blue SED and with an absolute magnitude M$_{AB}$ = -11.99 mag in filter F770W.

\subsection{SNR 1566-4}
\begin{figure}[h]
   \centering
   \includegraphics[width=\columnwidth]{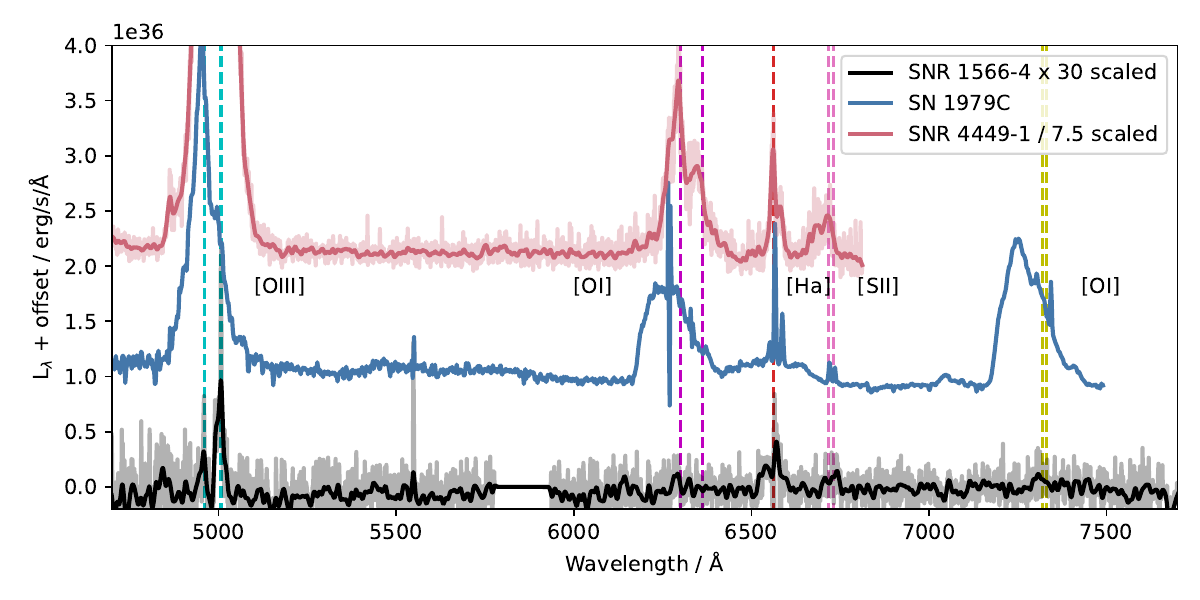}
      \caption{Comparison of SNR 1566-4, SNR 4449-1 and SN 1979C. Solid lines are smoothed spectra, shaded represents the original spectrum. Vertical lines indicate different emission lines; [O\,III] (cyan), [OI] (magenta), H$\alpha$ (red), [S\,II] (pink) and [OII] (yellow)}
         \label{snr1566-5}
   \end{figure}

SNR 1566-4 is detected in the [O\,III] and [S\,II] linemaps. Only a blueshifted [O\,III]5007Å is clearly detected with possible blended H$\alpha$+[N\,II]$\lambda$6583 and a weak [S\,II] emission (see Fig. \ref{snr1566-5}). The [O\,III]$\lambda\lambda$4959,5007 lines have FWHM of over 1200 km/s and a total luminosity of 3.7$\times$ 10$^{36}$ erg/s. It is distinctly different from the other O-rich SNRs in our sample. This could indicate an older age compared to the others, considering the similarity to SNR 0540-69.3 and its reported age of $\approx$1000 years \citep{Larsson21}. The SNR is not detected with Chandra or JWST. The X-ray fluxes listed in Table \ref{tab:xray} is based on a single X-ray photon and should be taken as an upper limit.

\section{Discussion}

In this article we have presented a new method to search for and identify extragalactic SNRs. We make use of the capabilities of a modern IFU; the fact that SNRs have broad emission lines compared to other emission sources such as photoionized nebulae. In addition, we make use of the compactness of extragalactic SNRs; with the optical quality of MUSE observations the remnants are observed as unresolved point sources. These two considerations allow us to isolate SNRs from the diffuse gas of a host galaxy and other contaminants. We applied our method to a sample of 19 nearby galaxies, the current PHANGS-MUSE catalog. We found a total of 307 SNRs. Of these, seven were found to be O-rich.

The small fraction of O-rich SNRs within the full SNR population indicates that they are observationally rare. They are also only found in 4 of 19 galaxies in our sample. The fraction of O-rich SNRs to all SNRs is consistent with the SNR population in the MW and the Magellanic Clouds. In the literature, out of the roughly 300 SNRs in the MW, only three have been designated to be O-rich. If both of these samples are considered almost complete complete, the rate of O-rich SNRs is only a 1-2\% of the SNR population. The Magellanic Clouds host three more previously classified O-rich SNRs and one more has been found in NGC 4449 for a total of seven O-rich SNRs. Our new sample thus doubles the number of known O-rich SNRs. The observed rarity could be due to two reasons:

\begin{itemize}
    \item O-rich SNRs are intrinsically rare. This could be because they come from the most massive stars or from a rare progenitor scenario. Other factors affecting rarity include unusually dense CSM and interaction with SN ejecta, or a central compact object interacting with the SN ejecta.
    \item O-rich SNRs are only a brief stage in SNR evolution. Age estimates of the known remnants span from a few decades up to a few thousand years. If this stage lasts at most only a few thousand years after the explosion, we will only rarely observe remnants at this stage.
\end{itemize}

\subsection{Observed properties of extragalactic O-rich SNRs}

Optical wavelength identification of SNRs has previously been based on empirical line ratios indicating a shocked region. Therefore, their spectra by definition show strong forbidden emission lines of singly-ionised sulphur and nitrogen. O-rich SNRs on the other hand are dominated by emission lines of oxygen with different degrees of ionisation, exemplified by SNR 4449-1. The O-lines in some cases are extremely luminous and have high FWHM compared to normal SNRs, upwards of 10$^{38}$ erg/s and FWHMs exceeding 2000 km/s.

Our sample of O-rich SNRs also have strong detections in both X-rays and the mid-IR. These measurements are extracted from archival observations and hence do not temporally align with the MUSE observations. This also allows us in some cases to give a minimum age to some of the remnants, e.g. SNR 4303-46 was serendipitiously detected by ROSITA already between 1996 and 1998 giving it a current minimum age as 28 years.

\subsection{The physical nature of O-rich SNRs}

Following the numbers of different SNRs discovered within our sample, we can conclude that O-rich SNRs are rare, but we cannot definitely conclude if this is an intrinsic property or observational effect. Our sample of galaxies in this article also host several historical SNe. We do not detect these SNe in the MUSE dataset, but instead we detect several O-rich SNRs with no known progenitor SN. While the cadence and depth of SN searches within these galaxies have been irregular and sometimes sparse in the hundred years before all-sky surveys, and a more thorough search through old photographic plates should be done in a follow-up study, we can still draw some conclusions on the non-detections. The fact that we do not detect the known SNe with MUSE shows that the SNe `normally' fade away quickly. Even if this `O-rich phase' is a common feature in SNe, it may be typically too faint to be detected. But in some cases we observe bright O-rich SNRs, making them unusual and intrinsically rare. On the other hand, the O-rich phase could be something that happens only to a certain subset of SNe, making them intrinsically rare. Long-term observations of nearby future SNe and a renewed recovery program of historical SNe could shed more light on this issue.
If we assume that there is no additional power source accelerating the SN ejecta on long timescales (e.g. at thousands of years post explosion), we can infer that the O-rich SNRs are likely young based on the O-line velocities, especially when compared to normal optical SNRs we find outside the MW. The spectral similarity with some historical extragalactic SNe, e.g. SN 1979C, hint that these remnants are indeed middle-aged SNe that have gone undetected.

The high luminosity of O-rich SNRs and their probable longevity requires us to consider the powering methods of the emission. Radioactive nickel and its daughter species are unlikely to be producing the amount of energy we observe in these remnants decades since the explosion. Ejecta-CSM interaction has previously been proposed to be powering the strong emission. Ejecta interacting with dense CSM could cause strong reverse shocks into an O-rich ejecta causing luminous oxygen emission. This scenario would require high mass losses before the explosion with CSM density profiles deviating from r$^{-2}$ in order to produce a strong reverse shock for decades or hundreds of years. The source of luminosity could also be a compact object injecting extra energy into the ejecta. This could be a magnetar shedding energy from the spin-down or a black hole accreting matter.

Recently, a peculiar type Ib SN, SN 2012au, was observed by \cite{Milisavljevic2018} to exhibit forbidden oxygen emission lines in optical light six years after the explosion. The authors note the possibility of a pulsar or a magnetar embedded in a O-rich ejecta. Similarly, \cite{Chevalier92} had previously showed that a pulsar inside an oxygen-dominated ejecta would produce shells with different ionisations of oxygen. This may explain SN 2012au, as discussed by \cite{Milisavljevic2018}, and notably also SNR 3627-1 and SNR 4303-37 in this paper. This may be an indication that SN 2012au and the O-rich SNRs are similar objects. Follow-ups on both the SN and the SNRs should reveal this e.g. if the SN O-line keep broadening and the X-ray luminosity stays stable.

\cite{Omand23} has also showed promising numerical models for a magnetar inside ejectas of different compositions. If the compact central object is indeed the main powering mechanism of these SNRs, these models seem to indicate the preference of a highly O-rich ejecta. Otherwise other metals, especially sulfur, would have equal or higher luminosities coming from forbidden emission lines. These lines are also likely not quenched due to high densities, because the critical densities for sulfur lines observable with MUSE are higher than the oxygen lines. Therefore, it is likely that the observed ejecta is O-rich and S-poor. Likewise, this may also give indications of the mass cut for the formation of the compact object. As the proto-neutron star forms, it accretes $\sim$ 2 M$_{\odot}$ of matter. This also includes at least some of the Si-shell of the collapsing star. The Si-shell is also somewhat enriched with sulfur, contrasting to the more S-poor O-shell. It may be that the Si-shell needs to be accreted more or less completely in order to account for the non-detections of sulfur and the strong oxygen emissions.

The cause of the oxygen emission should also ideally explain the strong observed X-ray and mid-IR emissions. Mid-IR implies the precense of warm dust and in the brightest O-rich SNRs in our sample we observe a feature in the SED that can be interpreted as a 14 $\mu$m silicon feature. This would still need to be reconciled with the strong X-ray emission. If the X-rays are part of a blackbody emission, this would mean extremely hot gas, which would predicate the destruction of dust. On the other hand, the observed mid-IR excess could be the tail-end of strong synchrotron emission, indicating the presence of a magnetar. If we assume that the X-rays are non-thermal in origin, i.e. arising from a compact object, the presence of dust would be more probable in outer regions, away from the compact object, but the mid-IR excess could still be non-thermal in nature. In any case, detailed and deep observations of these objects with radio and infrared (JWST spectra) is needed to better understand their nature.

Considering that these objects resemble both older SNe and specific younger SNRs, they are an important link between SNe and SNRs. The progenitor SN in the case of Cas A was identified as a type IIb, a type of core-collapse supernova that initially resembles a H-rich type II, but later evolves into a H-poor type Ib \citep{Filippenko93}. On the other hand, the SN 1979C used as a comparison in this article was an unusual type II; the SN reached M$_B\sim$ -20, declining linearly with signs of CSM. It was also a bright radio supernova. The apparent rarity of O-rich SNRs could also be due to high-mass progenitors. In fact, two O-rich SNRs in our sample are situated in a strongly star-forming regions and SNR 4449-1 is also in a starbursting galaxy. A high mass of the progenitor, implying high mass-losses, may then favour the CSM interaction scenario.

\section{Conclusions}

In this work, we have reported the discovery of 307 SNRs in the PHANGS-MUSE galaxy catalog. Of these, seven are O-rich SNRs. This doubles the number O-rich SNRs in literature. We also note the similarity of these objects to some long-lasting SNe that exploded a few decades ago. The seven events were found in a search of 19 nearby galaxies and in a total of four host galaxies. This work has thus further outlined the rarity of O-rich SNRs. The methods presented in this paper will be expanded to other nearby galaxies in order to improve our understanding of these objects and to characterise their rarity in the SN/SNR population.

SNRs, and specifically O-rich events, give constraints on the stellar evolutionary pathways that lead to these explosion remnants. In the case of significant CSM interaction, stellar evolution models would need to be able to reproduce the high mass-losses needed to produce the observed results. In compact object powered scenarios the stellar evolution and explosion models need to account for the observed rarity of O-rich SNRs, i.e. why are only some compact central objects related to some specific SNRs? Additional observations and modelling of O-rich SNRs is required to corroborate our findings using optical-wavelength spectroscopy, and further our understanding of the origin and powering mechanism of these SN remnants.

    
\begin{acknowledgements}
This work was funded by ANID, Millennium Science Initiative, ICN12\_009. HK was funded by the Research Council of Finland projects 324504, 328898, and 353019. KM acknowledges support from the Japan Society for the Promotion of Science (JSPS) KAKENHI grant (JP20H00174), and by the JSPS Open Partnership Bilateral Joint Research Project (JPJSBP120229923 for Japan-Finland, and JPJSBP120239901 for Japan-Chile). SM acknowledge support from the Research Council of Finland project 350458. Based on observations collected at the European Southern Observatory under ESO programme(s) 1100.B.0651(A-D), 0102.D-0769(A) and/or data obtained from the ESO Science Archive Facility with DOI(s) under https://doi.org/10.18727/archive/47.

This work has made use of data from the European Space Agency (ESA) mission
{\it Gaia} (\url{https://www.cosmos.esa.int/gaia}), processed by the {\it Gaia}
Data Processing and Analysis Consortium (DPAC,
\url{https://www.cosmos.esa.int/web/gaia/dpac/consortium}). Funding for the DPAC
has been provided by national institutions, in particular the institutions
participating in the {\it Gaia} Multilateral Agreement.

\end{acknowledgements}


\bibliographystyle{aa}
\bibliography{References}

\begin{thebibliography}{57}
\expandafter\ifx\csname natexlab\endcsname\relax\def\natexlab#1{#1}\fi

\bibitem[{{Anand} {et~al.}(2021){Anand}, {Lee}, {Van Dyk}, {Leroy}, {Rosolowsky}, {Schinnerer}, {Larson}, {Kourkchi}, {Kreckel}, {Scheuermann}, {Rizzi}, {Thilker}, {Tully}, {Bigiel}, {Blanc}, {Boquien}, {Chandar}, {Dale}, {Emsellem}, {Deger}, {Glover}, {Grasha}, {Groves}, {S. Klessen}, {Kruijssen}, {Querejeta}, {S{\'a}nchez-Bl{\'a}zquez}, {Schruba}, {Turner}, {Ubeda}, {Williams}, \& {Whitmore}}]{Anand21}
{Anand}, G.~S., {Lee}, J.~C., {Van Dyk}, S.~D., {et~al.} 2021, \mnras, 501, 3621

\bibitem[{{Bacon} {et~al.}(2010){Bacon}, {Accardo}, {Adjali}, {Anwand}, {Bauer}, {Biswas}, {Blaizot}, {Boudon}, {Brau-Nogue}, {Brinchmann}, {Caillier}, {Capoani}, {Carollo}, {Contini}, {Couderc}, {Daguis{\'e}}, {Deiries}, {Delabre}, {Dreizler}, {Dubois}, {Dupieux}, {Dupuy}, {Emsellem}, {Fechner}, {Fleischmann}, {Fran{\c{c}}ois}, {Gallou}, {Gharsa}, {Glindemann}, {Gojak}, {Guiderdoni}, {Hansali}, {Hahn}, {Jarno}, {Kelz}, {Koehler}, {Kosmalski}, {Laurent}, {Le Floch}, {Lilly}, {Lizon}, {Loupias}, {Manescau}, {Monstein}, {Nicklas}, {Olaya}, {Pares}, {Pasquini}, {P{\'e}contal-Rousset}, {Pell{\'o}}, {Petit}, {Popow}, {Reiss}, {Remillieux}, {Renault}, {Roth}, {Rupprecht}, {Serre}, {Schaye}, {Soucail}, {Steinmetz}, {Streicher}, {Stuik}, {Valentin}, {Vernet}, {Weilbacher}, {Wisotzki}, \& {Yerle}}]{bacon10}
{Bacon}, R., {Accardo}, M., {Adjali}, L., {et~al.} 2010, in Society of Photo-Optical Instrumentation Engineers (SPIE) Conference Series, Vol. 7735, Ground-based and Airborne Instrumentation for Astronomy III, ed. I.~S. {McLean}, S.~K. {Ramsay}, \& H.~{Takami}, 773508

\bibitem[{{Balick} \& {Heckman}(1978)}]{balick78}
{Balick}, B. \& {Heckman}, T. 1978, \apjl, 226, L7

\bibitem[{{Bauer} {et~al.}(2008){Bauer}, {Dwarkadas}, {Brandt}, {Immler}, {Smartt}, {Bartel}, \& {Bietenholz}}]{Bauer08}
{Bauer}, F.~E., {Dwarkadas}, V.~V., {Brandt}, W.~N., {et~al.} 2008, \apj, 688, 1210

\bibitem[{{Bietenholz} {et~al.}(2010){Bietenholz}, {Bartel}, {Milisavljevic}, {Fesen}, {Challis}, \& {Kirshner}}]{bietenholz10}
{Bietenholz}, M.~F., {Bartel}, N., {Milisavljevic}, D., {et~al.} 2010, \mnras, 409, 1594

\bibitem[{{Blair} {et~al.}(1983){Blair}, {Kirshner}, \& {Winkler}}]{blair83}
{Blair}, W.~P., {Kirshner}, R.~P., \& {Winkler}, P.~F., J. 1983, \apj, 272, 84

\bibitem[{{Blair} \& {Long}(1997)}]{blair97}
{Blair}, W.~P. \& {Long}, K.~S. 1997, \apjs, 108, 261

\bibitem[{{Blair} {et~al.}(2015){Blair}, {Winkler}, {Long}, {Whitmore}, {Kim}, {Soria}, {Kuntz}, {Plucinsky}, {Dopita}, \& {Stockdale}}]{blair15}
{Blair}, W.~P., {Winkler}, P.~F., {Long}, K.~S., {et~al.} 2015, \apj, 800, 118

\bibitem[{{Caldwell} \& {Raymond}(2023)}]{Caldwell23}
{Caldwell}, N. \& {Raymond}, J. 2023, arXiv e-prints, arXiv:2308.00754

\bibitem[{{Campbell}(1926)}]{1926discovery}
{Campbell}, L. 1926, Popular Astronomy, 34, 402

\bibitem[{{Cardelli} {et~al.}(1989){Cardelli}, {Clayton}, \& {Mathis}}]{Cardelli89}
{Cardelli}, J.~A., {Clayton}, G.~C., \& {Mathis}, J.~S. 1989, \apj, 345, 245

\bibitem[{{Chevalier} \& {Fransson}(1992)}]{Chevalier92}
{Chevalier}, R.~A. \& {Fransson}, C. 1992, \apj, 395, 540

\bibitem[{{Cid Fernandes} {et~al.}(2021){Cid Fernandes}, {Carvalho}, {Sanchez}, {de Amorim}, \& {Ruschel-Dutra}}]{Fernandes21}
{Cid Fernandes}, R., {Carvalho}, M.~S., {Sanchez}, S.~F., {de Amorim}, A.~L., \& {Ruschel-Dutra}, D. 2021, arXiv e-prints, arXiv:2101.12022

\bibitem[{{De Looze} {et~al.}(2017){De Looze}, {Barlow}, {Swinyard}, {Rho}, {Gomez}, {Matsuura}, \& {Wesson}}]{delooze17}
{De Looze}, I., {Barlow}, M.~J., {Swinyard}, B.~M., {et~al.} 2017, \mnras, 465, 3309

\bibitem[{{de Vaucouleurs} {et~al.}(1981){de Vaucouleurs}, {de Vaucouleurs}, {Buta}, {Ables}, \& {Hewitt}}]{deVaucouleurs1981}
{de Vaucouleurs}, G., {de Vaucouleurs}, A., {Buta}, R., {Ables}, H.~D., \& {Hewitt}, A.~V. 1981, \pasp, 93, 36

\bibitem[{Developers {et~al.}(2023)Developers, Averbukh, Bradley, Buikhuizen, Busko, Cherinka, Conroy, Earl, Fox, Geda, \& et~al.}]{jdadf}
Developers, J., Averbukh, J., Bradley, L., {et~al.} 2023, if you use this software, please cite it as below.

\bibitem[{{Emsellem} {et~al.}(2022){Emsellem}, {Schinnerer}, {Santoro}, {Belfiore}, {Pessa}, {McElroy}, {Blanc}, {Congiu}, {Groves}, {Ho}, {Kreckel}, {Razza}, {Sanchez-Blazquez}, {Egorov}, {Faesi}, {Klessen}, {Leroy}, {Meidt}, {Querejeta}, {Rosolowsky}, {Scheuermann}, {Anand}, {Barnes}, {Be{\v{s}}li{\'c}}, {Bigiel}, {Boquien}, {Cao}, {Chevance}, {Dale}, {Eibensteiner}, {Glover}, {Grasha}, {Henshaw}, {Hughes}, {Koch}, {Kruijssen}, {Lee}, {Liu}, {Pan}, {Pety}, {Saito}, {Sandstrom}, {Schruba}, {Sun}, {Thilker}, {Usero}, {Watkins}, \& {Williams}}]{Emsellem22}
{Emsellem}, E., {Schinnerer}, E., {Santoro}, F., {et~al.} 2022, \aap, 659, A191

\bibitem[{Fernandes {et~al.}(2005)Fernandes, Mateus, Sodré, Stasińska, \& Gomes}]{Fernandes05}
Fernandes, R.~C., Mateus, A., Sodré, L., Stasińska, G., \& Gomes, J.~M. 2005, Monthly Notices of the Royal Astronomical Society, 358, 363

\bibitem[{{Fern{\'a}ndez Arenas} {et~al.}(2018){Fern{\'a}ndez Arenas}, {Terlevich}, {Terlevich}, {Melnick}, {Ch{\'a}vez}, {Bresolin}, {Telles}, {Plionis}, \& {Basilakos}}]{Arenas18}
{Fern{\'a}ndez Arenas}, D., {Terlevich}, E., {Terlevich}, R., {et~al.} 2018, \mnras, 474, 1250

\bibitem[{{Fesen} {et~al.}(1999){Fesen}, {Gerardy}, {Filippenko}, {Matheson}, {Chevalier}, {Kirshner}, {Schmidt}, {Challis}, {Fransson}, {Leibundgut}, \& {van Dyk}}]{Fesen99}
{Fesen}, R.~A., {Gerardy}, C.~L., {Filippenko}, A.~V., {et~al.} 1999, \aj, 117, 725

\bibitem[{{Fesen} {et~al.}(2018){Fesen}, {Weil}, {Cisneros}, {Blair}, \& {Raymond}}]{Fesen18}
{Fesen}, R.~A., {Weil}, K.~E., {Cisneros}, I.~A., {Blair}, W.~P., \& {Raymond}, J.~C. 2018, \mnras, 481, 1786

\bibitem[{{Filippenko} {et~al.}(1993){Filippenko}, {Matheson}, \& {Ho}}]{Filippenko93}
{Filippenko}, A.~V., {Matheson}, T., \& {Ho}, L.~C. 1993, \apjl, 415, L103

\bibitem[{{Fruscione} {et~al.}(2006){Fruscione}, {McDowell}, {Allen}, {Brickhouse}, {Burke}, {Davis}, {Durham}, {Elvis}, {Galle}, {Harris}, {Huenemoerder}, {Houck}, {Ishibashi}, {Karovska}, {Nicastro}, {Noble}, {Nowak}, {Primini}, {Siemiginowska}, {Smith}, \& {Wise}}]{2006SPIE.6270E..1VF}
{Fruscione}, A., {McDowell}, J.~C., {Allen}, G.~E., {et~al.} 2006, in Society of Photo-Optical Instrumentation Engineers (SPIE) Conference Series, Vol. 6270, Society of Photo-Optical Instrumentation Engineers (SPIE) Conference Series, ed. D.~R. {Silva} \& R.~E. {Doxsey}, 62701V

\bibitem[{{Gaia Collaboration} {et~al.}(2016){Gaia Collaboration}, {Prusti}, {de Bruijne}, {Brown}, {Vallenari}, {Babusiaux}, {Bailer-Jones}, {Bastian}, {Biermann}, {Evans}, {Eyer}, {Jansen}, {Jordi}, {Klioner}, {Lammers}, {Lindegren}, {Luri}, {Mignard}, {Milligan}, {Panem}, {Poinsignon}, {Pourbaix}, {Randich}, {Sarri}, {Sartoretti}, {Siddiqui}, {Soubiran}, {Valette}, {van Leeuwen}, {Walton}, {Aerts}, {Arenou}, {Cropper}, {Drimmel}, {H{\o}g}, {Katz}, {Lattanzi}, {O'Mullane}, {Grebel}, {Holland}, {Huc}, {Passot}, {Bramante}, {Cacciari}, {Casta{\~n}eda}, {Chaoul}, {Cheek}, {De Angeli}, {Fabricius}, {Guerra}, {Hern{\'a}ndez}, {Jean-Antoine-Piccolo}, {Masana}, {Messineo}, {Mowlavi}, {Nienartowicz}, {Ord{\'o}{\~n}ez-Blanco}, {Panuzzo}, {Portell}, {Richards}, {Riello}, {Seabroke}, {Tanga}, {Th{\'e}venin}, {Torra}, {Els}, {Gracia-Abril}, {Comoretto}, {Garcia-Reinaldos}, {Lock}, {Mercier}, {Altmann}, {Andrae}, {Astraatmadja}, {Bellas-Velidis}, {Benson}, {Berthier}, {Blomme}, {Busso}, {Carry}, {Cellino}, {Clementini},
  {Cowell}, {Creevey}, {Cuypers}, {Davidson}, {De Ridder}, {de Torres}, {Delchambre}, {Dell'Oro}, {Ducourant}, {Fr{\'e}mat}, {Garc{\'\i}a-Torres}, {Gosset}, {Halbwachs}, {Hambly}, {Harrison}, {Hauser}, {Hestroffer}, {Hodgkin}, {Huckle}, {Hutton}, {Jasniewicz}, {Jordan}, {Kontizas}, {Korn}, {Lanzafame}, {Manteiga}, {Moitinho}, {Muinonen}, {Osinde}, {Pancino}, {Pauwels}, {Petit}, {Recio-Blanco}, {Robin}, {Sarro}, {Siopis}, {Smith}, {Smith}, {Sozzetti}, {Thuillot}, {van Reeven}, {Viala}, {Abbas}, {Abreu Aramburu}, {Accart}, {Aguado}, {Allan}, {Allasia}, {Altavilla}, {{\'A}lvarez}, {Alves}, {Anderson}, {Andrei}, {Anglada Varela}, {Antiche}, {Antoja}, {Ant{\'o}n}, {Arcay}, {Atzei}, {Ayache}, {Bach}, {Baker}, {Balaguer-N{\'u}{\~n}ez}, {Barache}, {Barata}, {Barbier}, {Barblan}, {Baroni}, {Barrado y Navascu{\'e}s}, {Barros}, {Barstow}, {Becciani}, {Bellazzini}, {Bellei}, {Bello Garc{\'\i}a}, {Belokurov}, {Bendjoya}, {Berihuete}, {Bianchi}, {Bienaym{\'e}}, {Billebaud}, {Blagorodnova}, {Blanco-Cuaresma}, {Boch},
  {Bombrun}, {Borrachero}, {Bouquillon}, {Bourda}, {Bouy}, {Bragaglia}, {Breddels}, {Brouillet}, {Br{\"u}semeister}, {Bucciarelli}, {Budnik}, {Burgess}, {Burgon}, {Burlacu}, {Busonero}, {Buzzi}, {Caffau}, {Cambras}, {Campbell}, {Cancelliere}, {Cantat-Gaudin}, {Carlucci}, {Carrasco}, {Castellani}, {Charlot}, {Charnas}, {Charvet}, {Chassat}, {Chiavassa}, {Clotet}, {Cocozza}, {Collins}, {Collins}, {Costigan}, {Crifo}, {Cross}, {Crosta}, {Crowley}, {Dafonte}, {Damerdji}, {Dapergolas}, {David}, {David}, {De Cat}, {de Felice}, {de Laverny}, {De Luise}, {De March}, {de Martino}, {de Souza}, {Debosscher}, {del Pozo}, {Delbo}, {Delgado}, {Delgado}, {di Marco}, {Di Matteo}, {Diakite}, {Distefano}, {Dolding}, {Dos Anjos}, {Drazinos}, {Dur{\'a}n}, {Dzigan}, {Ecale}, {Edvardsson}, {Enke}, {Erdmann}, {Escolar}, {Espina}, {Evans}, {Eynard Bontemps}, {Fabre}, {Fabrizio}, {Faigler}, {Falc{\~a}o}, {Farr{\`a}s Casas}, {Faye}, {Federici}, {Fedorets}, {Fern{\'a}ndez-Hern{\'a}ndez}, {Fernique}, {Fienga}, {Figueras}, {Filippi},
  {Findeisen}, {Fonti}, {Fouesneau}, {Fraile}, {Fraser}, {Fuchs}, {Furnell}, {Gai}, {Galleti}, {Galluccio}, {Garabato}, {Garc{\'\i}a-Sedano}, {Gar{\'e}}, {Garofalo}, {Garralda}, {Gavras}, {Gerssen}, {Geyer}, {Gilmore}, {Girona}, {Giuffrida}, {Gomes}, {Gonz{\'a}lez-Marcos}, {Gonz{\'a}lez-N{\'u}{\~n}ez}, {Gonz{\'a}lez-Vidal}, {Granvik}, {Guerrier}, {Guillout}, {Guiraud}, {G{\'u}rpide}, {Guti{\'e}rrez-S{\'a}nchez}, {Guy}, {Haigron}, {Hatzidimitriou}, {Haywood}, {Heiter}, {Helmi}, {Hobbs}, {Hofmann}, {Holl}, {Holland}, {Hunt}, {Hypki}, {Icardi}, {Irwin}, {Jevardat de Fombelle}, {Jofr{\'e}}, {Jonker}, {Jorissen}, {Julbe}, {Karampelas}, {Kochoska}, {Kohley}, {Kolenberg}, {Kontizas}, {Koposov}, {Kordopatis}, {Koubsky}, {Kowalczyk}, {Krone-Martins}, {Kudryashova}, {Kull}, {Bachchan}, {Lacoste-Seris}, {Lanza}, {Lavigne}, {Le Poncin-Lafitte}, {Lebreton}, {Lebzelter}, {Leccia}, {Leclerc}, {Lecoeur-Taibi}, {Lemaitre}, {Lenhardt}, {Leroux}, {Liao}, {Licata}, {Lindstr{\o}m}, {Lister}, {Livanou}, {Lobel}, {L{\"o}ffler},
  {L{\'o}pez}, {Lopez-Lozano}, {Lorenz}, {Loureiro}, {MacDonald}, {Magalh{\~a}es Fernandes}, {Managau}, {Mann}, {Mantelet}, {Marchal}, {Marchant}, {Marconi}, {Marie}, {Marinoni}, {Marrese}, {Marschalk{\'o}}, {Marshall}, {Mart{\'\i}n-Fleitas}, {Martino}, {Mary}, {Matijevi{\v{c}}}, {Mazeh}, {McMillan}, {Messina}, {Mestre}, {Michalik}, {Millar}, {Miranda}, {Molina}, {Molinaro}, {Molinaro}, {Moln{\'a}r}, {Moniez}, {Montegriffo}, {Monteiro}, {Mor}, {Mora}, {Morbidelli}, {Morel}, {Morgenthaler}, {Morley}, {Morris}, {Mulone}, {Muraveva}, {Musella}, {Narbonne}, {Nelemans}, {Nicastro}, {Noval}, {Ord{\'e}novic}, {Ordieres-Mer{\'e}}, {Osborne}, {Pagani}, {Pagano}, {Pailler}, {Palacin}, {Palaversa}, {Parsons}, {Paulsen}, {Pecoraro}, {Pedrosa}, {Pentik{\"a}inen}, {Pereira}, {Pichon}, {Piersimoni}, {Pineau}, {Plachy}, {Plum}, {Poujoulet}, {Pr{\v{s}}a}, {Pulone}, {Ragaini}, {Rago}, {Rambaux}, {Ramos-Lerate}, {Ranalli}, {Rauw}, {Read}, {Regibo}, {Renk}, {Reyl{\'e}}, {Ribeiro}, {Rimoldini}, {Ripepi}, {Riva}, {Rixon},
  {Roelens}, {Romero-G{\'o}mez}, {Rowell}, {Royer}, {Rudolph}, {Ruiz-Dern}, {Sadowski}, {Sagrist{\`a} Sell{\'e}s}, {Sahlmann}, {Salgado}, {Salguero}, {Sarasso}, {Savietto}, {Schnorhk}, {Schultheis}, {Sciacca}, {Segol}, {Segovia}, {Segransan}, {Serpell}, {Shih}, {Smareglia}, {Smart}, {Smith}, {Solano}, {Solitro}, {Sordo}, {Soria Nieto}, {Souchay}, {Spagna}, {Spoto}, {Stampa}, {Steele}, {Steidelm{\"u}ller}, {Stephenson}, {Stoev}, {Suess}, {S{\"u}veges}, {Surdej}, {Szabados}, {Szegedi-Elek}, {Tapiador}, {Taris}, {Tauran}, {Taylor}, {Teixeira}, {Terrett}, {Tingley}, {Trager}, {Turon}, {Ulla}, {Utrilla}, {Valentini}, {van Elteren}, {Van Hemelryck}, {van Leeuwen}, {Varadi}, {Vecchiato}, {Veljanoski}, {Via}, {Vicente}, {Vogt}, {Voss}, {Votruba}, {Voutsinas}, {Walmsley}, {Weiler}, {Weingrill}, {Werner}, {Wevers}, {Whitehead}, {Wyrzykowski}, {Yoldas}, {{\v{Z}}erjal}, {Zucker}, {Zurbach}, {Zwitter}, {Alecu}, {Allen}, {Allende Prieto}, {Amorim}, {Anglada-Escud{\'e}}, {Arsenijevic}, {Azaz}, {Balm}, {Beck}, {Bernstein},
  {Bigot}, {Bijaoui}, {Blasco}, {Bonfigli}, {Bono}, {Boudreault}, {Bressan}, {Brown}, {Brunet}, {Bunclark}, {Buonanno}, {Butkevich}, {Carret}, {Carrion}, {Chemin}, {Ch{\'e}reau}, {Corcione}, {Darmigny}, {de Boer}, {de Teodoro}, {de Zeeuw}, {Delle Luche}, {Domingues}, {Dubath}, {Fodor}, {Fr{\'e}zouls}, {Fries}, {Fustes}, {Fyfe}, {Gallardo}, {Gallegos}, {Gardiol}, {Gebran}, {Gomboc}, {G{\'o}mez}, {Grux}, {Gueguen}, {Heyrovsky}, {Hoar}, {Iannicola}, {Isasi Parache}, {Janotto}, {Joliet}, {Jonckheere}, {Keil}, {Kim}, {Klagyivik}, {Klar}, {Knude}, {Kochukhov}, {Kolka}, {Kos}, {Kutka}, {Lainey}, {LeBouquin}, {Liu}, {Loreggia}, {Makarov}, {Marseille}, {Martayan}, {Martinez-Rubi}, {Massart}, {Meynadier}, {Mignot}, {Munari}, {Nguyen}, {Nordlander}, {Ocvirk}, {O'Flaherty}, {Olias Sanz}, {Ortiz}, {Osorio}, {Oszkiewicz}, {Ouzounis}, {Palmer}, {Park}, {Pasquato}, {Peltzer}, {Peralta}, {P{\'e}turaud}, {Pieniluoma}, {Pigozzi}, {Poels}, {Prat}, {Prod'homme}, {Raison}, {Rebordao}, {Risquez}, {Rocca-Volmerange}, {Rosen},
  {Ruiz-Fuertes}, {Russo}, {Sembay}, {Serraller Vizcaino}, {Short}, {Siebert}, {Silva}, {Sinachopoulos}, {Slezak}, {Soffel}, {Sosnowska}, {Strai{\v{z}}ys}, {ter Linden}, {Terrell}, {Theil}, {Tiede}, {Troisi}, {Tsalmantza}, {Tur}, {Vaccari}, {Vachier}, {Valles}, {Van Hamme}, {Veltz}, {Virtanen}, {Wallut}, {Wichmann}, {Wilkinson}, {Ziaeepour}, \& {Zschocke}}]{Gaia16}
{Gaia Collaboration}, {Prusti}, T., {de Bruijne}, J.~H.~J., {et~al.} 2016, \aap, 595, A1

\bibitem[{{Gaia Collaboration} {et~al.}(2023){Gaia Collaboration}, {Vallenari}, {Brown}, {Prusti}, {de Bruijne}, {Arenou}, {Babusiaux}, {Biermann}, {Creevey}, {Ducourant}, {Evans}, {Eyer}, {Guerra}, {Hutton}, {Jordi}, {Klioner}, {Lammers}, {Lindegren}, {Luri}, {Mignard}, {Panem}, {Pourbaix}, {Randich}, {Sartoretti}, {Soubiran}, {Tanga}, {Walton}, {Bailer-Jones}, {Bastian}, {Drimmel}, {Jansen}, {Katz}, {Lattanzi}, {van Leeuwen}, {Bakker}, {Cacciari}, {Casta{\~n}eda}, {De Angeli}, {Fabricius}, {Fouesneau}, {Fr{\'e}mat}, {Galluccio}, {Guerrier}, {Heiter}, {Masana}, {Messineo}, {Mowlavi}, {Nicolas}, {Nienartowicz}, {Pailler}, {Panuzzo}, {Riclet}, {Roux}, {Seabroke}, {Sordo}, {Th{\'e}venin}, {Gracia-Abril}, {Portell}, {Teyssier}, {Altmann}, {Andrae}, {Audard}, {Bellas-Velidis}, {Benson}, {Berthier}, {Blomme}, {Burgess}, {Busonero}, {Busso}, {C{\'a}novas}, {Carry}, {Cellino}, {Cheek}, {Clementini}, {Damerdji}, {Davidson}, {de Teodoro}, {Nu{\~n}ez Campos}, {Delchambre}, {Dell'Oro}, {Esquej},
  {Fern{\'a}ndez-Hern{\'a}ndez}, {Fraile}, {Garabato}, {Garc{\'\i}a-Lario}, {Gosset}, {Haigron}, {Halbwachs}, {Hambly}, {Harrison}, {Hern{\'a}ndez}, {Hestroffer}, {Hodgkin}, {Holl}, {Jan{\ss}en}, {Jevardat de Fombelle}, {Jordan}, {Krone-Martins}, {Lanzafame}, {L{\"o}ffler}, {Marchal}, {Marrese}, {Moitinho}, {Muinonen}, {Osborne}, {Pancino}, {Pauwels}, {Recio-Blanco}, {Reyl{\'e}}, {Riello}, {Rimoldini}, {Roegiers}, {Rybizki}, {Sarro}, {Siopis}, {Smith}, {Sozzetti}, {Utrilla}, {van Leeuwen}, {Abbas}, {{\'A}brah{\'a}m}, {Abreu Aramburu}, {Aerts}, {Aguado}, {Ajaj}, {Aldea-Montero}, {Altavilla}, {{\'A}lvarez}, {Alves}, {Anders}, {Anderson}, {Anglada Varela}, {Antoja}, {Baines}, {Baker}, {Balaguer-N{\'u}{\~n}ez}, {Balbinot}, {Balog}, {Barache}, {Barbato}, {Barros}, {Barstow}, {Bartolom{\'e}}, {Bassilana}, {Bauchet}, {Becciani}, {Bellazzini}, {Berihuete}, {Bernet}, {Bertone}, {Bianchi}, {Binnenfeld}, {Blanco-Cuaresma}, {Blazere}, {Boch}, {Bombrun}, {Bossini}, {Bouquillon}, {Bragaglia}, {Bramante}, {Breedt},
  {Bressan}, {Brouillet}, {Brugaletta}, {Bucciarelli}, {Burlacu}, {Butkevich}, {Buzzi}, {Caffau}, {Cancelliere}, {Cantat-Gaudin}, {Carballo}, {Carlucci}, {Carnerero}, {Carrasco}, {Casamiquela}, {Castellani}, {Castro-Ginard}, {Chaoul}, {Charlot}, {Chemin}, {Chiaramida}, {Chiavassa}, {Chornay}, {Comoretto}, {Contursi}, {Cooper}, {Cornez}, {Cowell}, {Crifo}, {Cropper}, {Crosta}, {Crowley}, {Dafonte}, {Dapergolas}, {David}, {David}, {de Laverny}, {De Luise}, {De March}, {De Ridder}, {de Souza}, {de Torres}, {del Peloso}, {del Pozo}, {Delbo}, {Delgado}, {Delisle}, {Demouchy}, {Dharmawardena}, {Di Matteo}, {Diakite}, {Diener}, {Distefano}, {Dolding}, {Edvardsson}, {Enke}, {Fabre}, {Fabrizio}, {Faigler}, {Fedorets}, {Fernique}, {Fienga}, {Figueras}, {Fournier}, {Fouron}, {Fragkoudi}, {Gai}, {Garcia-Gutierrez}, {Garcia-Reinaldos}, {Garc{\'\i}a-Torres}, {Garofalo}, {Gavel}, {Gavras}, {Gerlach}, {Geyer}, {Giacobbe}, {Gilmore}, {Girona}, {Giuffrida}, {Gomel}, {Gomez}, {Gonz{\'a}lez-N{\'u}{\~n}ez},
  {Gonz{\'a}lez-Santamar{\'\i}a}, {Gonz{\'a}lez-Vidal}, {Granvik}, {Guillout}, {Guiraud}, {Guti{\'e}rrez-S{\'a}nchez}, {Guy}, {Hatzidimitriou}, {Hauser}, {Haywood}, {Helmer}, {Helmi}, {Sarmiento}, {Hidalgo}, {Hilger}, {H{\l}adczuk}, {Hobbs}, {Holland}, {Huckle}, {Jardine}, {Jasniewicz}, {Jean-Antoine Piccolo}, {Jim{\'e}nez-Arranz}, {Jorissen}, {Juaristi Campillo}, {Julbe}, {Karbevska}, {Kervella}, {Khanna}, {Kontizas}, {Kordopatis}, {Korn}, {K{\'o}sp{\'a}l}, {Kostrzewa-Rutkowska}, {Kruszy{\'n}ska}, {Kun}, {Laizeau}, {Lambert}, {Lanza}, {Lasne}, {Le Campion}, {Lebreton}, {Lebzelter}, {Leccia}, {Leclerc}, {Lecoeur-Taibi}, {Liao}, {Licata}, {Lindstr{\o}m}, {Lister}, {Livanou}, {Lobel}, {Lorca}, {Loup}, {Madrero Pardo}, {Magdaleno Romeo}, {Managau}, {Mann}, {Manteiga}, {Marchant}, {Marconi}, {Marcos}, {Marcos Santos}, {Mar{\'\i}n Pina}, {Marinoni}, {Marocco}, {Marshall}, {Martin Polo}, {Mart{\'\i}n-Fleitas}, {Marton}, {Mary}, {Masip}, {Massari}, {Mastrobuono-Battisti}, {Mazeh}, {McMillan}, {Messina}, {Michalik},
  {Millar}, {Mints}, {Molina}, {Molinaro}, {Moln{\'a}r}, {Monari}, {Mongui{\'o}}, {Montegriffo}, {Montero}, {Mor}, {Mora}, {Morbidelli}, {Morel}, {Morris}, {Muraveva}, {Murphy}, {Musella}, {Nagy}, {Noval}, {Oca{\~n}a}, {Ogden}, {Ordenovic}, {Osinde}, {Pagani}, {Pagano}, {Palaversa}, {Palicio}, {Pallas-Quintela}, {Panahi}, {Payne-Wardenaar}, {Pe{\~n}alosa Esteller}, {Penttil{\"a}}, {Pichon}, {Piersimoni}, {Pineau}, {Plachy}, {Plum}, {Poggio}, {Pr{\v{s}}a}, {Pulone}, {Racero}, {Ragaini}, {Rainer}, {Raiteri}, {Rambaux}, {Ramos}, {Ramos-Lerate}, {Re Fiorentin}, {Regibo}, {Richards}, {Rios Diaz}, {Ripepi}, {Riva}, {Rix}, {Rixon}, {Robichon}, {Robin}, {Robin}, {Roelens}, {Rogues}, {Rohrbasser}, {Romero-G{\'o}mez}, {Rowell}, {Royer}, {Ruz Mieres}, {Rybicki}, {Sadowski}, {S{\'a}ez N{\'u}{\~n}ez}, {Sagrist{\`a} Sell{\'e}s}, {Sahlmann}, {Salguero}, {Samaras}, {Sanchez Gimenez}, {Sanna}, {Santove{\~n}a}, {Sarasso}, {Schultheis}, {Sciacca}, {Segol}, {Segovia}, {S{\'e}gransan}, {Semeux}, {Shahaf}, {Siddiqui}, {Siebert},
  {Siltala}, {Silvelo}, {Slezak}, {Slezak}, {Smart}, {Snaith}, {Solano}, {Solitro}, {Souami}, {Souchay}, {Spagna}, {Spina}, {Spoto}, {Steele}, {Steidelm{\"u}ller}, {Stephenson}, {S{\"u}veges}, {Surdej}, {Szabados}, {Szegedi-Elek}, {Taris}, {Taylor}, {Teixeira}, {Tolomei}, {Tonello}, {Torra}, {Torra}, {Torralba Elipe}, {Trabucchi}, {Tsounis}, {Turon}, {Ulla}, {Unger}, {Vaillant}, {van Dillen}, {van Reeven}, {Vanel}, {Vecchiato}, {Viala}, {Vicente}, {Voutsinas}, {Weiler}, {Wevers}, {Wyrzykowski}, {Yoldas}, {Yvard}, {Zhao}, {Zorec}, {Zucker}, \& {Zwitter}}]{Gaia23}
{Gaia Collaboration}, {Vallenari}, A., {Brown}, A.~G.~A., {et~al.} 2023, \aap, 674, A1

\bibitem[{{G{\'o}mez-Gonz{\'a}lez} {et~al.}(2021){G{\'o}mez-Gonz{\'a}lez}, {Mayya}, {Toal{\'a}}, {Arthur}, {Zaragoza-Cardiel}, \& {Guerrero}}]{gomezgonzales21}
{G{\'o}mez-Gonz{\'a}lez}, V.~M.~A., {Mayya}, Y.~D., {Toal{\'a}}, J.~A., {et~al.} 2021, \mnras, 500, 2076

\bibitem[{{Krause} {et~al.}(2008){Krause}, {Birkmann}, {Usuda}, {Hattori}, {Goto}, {Rieke}, \& {Misselt}}]{Krause2008}
{Krause}, O., {Birkmann}, S.~M., {Usuda}, T., {et~al.} 2008, Science, 320, 1195

\bibitem[{{Kuncarayakti} {et~al.}(2016){Kuncarayakti}, {Maeda}, {Anderson}, {Hamuy}, {Nomoto}, {Galbany}, \& {Doi}}]{kuncarayakti16}
{Kuncarayakti}, H., {Maeda}, K., {Anderson}, J.~P., {et~al.} 2016, \mnras, 458, 2063

\bibitem[{{Lacey} {et~al.}(2007){Lacey}, {Goss}, \& {Mizouni}}]{Lacey07}
{Lacey}, C.~K., {Goss}, W.~M., \& {Mizouni}, L.~K. 2007, \aj, 133, 2156

\bibitem[{{Larsson} {et~al.}(2021){Larsson}, {Sollerman}, {Lyman}, {Spyromilio}, {Tenhu}, {Fransson}, \& {Lundqvist}}]{Larsson21}
{Larsson}, J., {Sollerman}, J., {Lyman}, J.~D., {et~al.} 2021, \apj, 922, 265

\bibitem[{{Lee} {et~al.}(2023){Lee}, {Sandstrom}, {Leroy}, {Thilker}, {Schinnerer}, {Rosolowsky}, {Larson}, {Egorov}, {Williams}, {Schmidt}, {Emsellem}, {Anand}, {Barnes}, {Belfiore}, {Be{\v{s}}li{\'c}}, {Bigiel}, {Blanc}, {Bolatto}, {Boquien}, {den Brok}, {Cao}, {Chandar}, {Chastenet}, {Chevance}, {Chiang}, {Congiu}, {Dale}, {Deger}, {Eibensteiner}, {Faesi}, {Glover}, {Grasha}, {Groves}, {Hassani}, {Henny}, {Henshaw}, {Hoyer}, {Hughes}, {Jeffreson}, {Jim{\'e}nez-Donaire}, {Kim}, {Kim}, {Klessen}, {Koch}, {Kreckel}, {Kruijssen}, {Li}, {Liu}, {Lopez}, {Maschmann}, {Chen}, {Meidt}, {Murphy}, {Neumann}, {Neumayer}, {Pan}, {Pessa}, {Pety}, {Querejeta}, {Pinna}, {Rodr{\'\i}guez}, {Saito}, {S{\'a}nchez-Bl{\'a}zquez}, {Santoro}, {Sardone}, {Smith}, {Sormani}, {Scheuermann}, {Stuber}, {Sutter}, {Sun}, {Teng}, {Tre{\ss}}, {Usero}, {Watkins}, {Whitmore}, \& {Razza}}]{Lee23}
{Lee}, J.~C., {Sandstrom}, K.~M., {Leroy}, A.~K., {et~al.} 2023, \apjl, 944, L17

\bibitem[{{Lee} {et~al.}(2022){Lee}, {Whitmore}, {Thilker}, {Deger}, {Larson}, {Ubeda}, {Anand}, {Boquien}, {Chandar}, {Dale}, {Emsellem}, {Leroy}, {Rosolowsky}, {Schinnerer}, {Schmidt}, {Lilly}, {Turner}, {Van Dyk}, {White}, {Barnes}, {Belfiore}, {Bigiel}, {Blanc}, {Cao}, {Chevance}, {Congiu}, {Egorov}, {Glover}, {Grasha}, {Groves}, {Henshaw}, {Hughes}, {Klessen}, {Koch}, {Kreckel}, {Kruijssen}, {Liu}, {Lopez}, {Mayker}, {Meidt}, {Murphy}, {Pan}, {Pety}, {Querejeta}, {Razza}, {Saito}, {S{\'a}nchez-Bl{\'a}zquez}, {Santoro}, {Sardone}, {Scheuermann}, {Schruba}, {Sun}, {Usero}, {Watkins}, \& {Williams}}]{Lee22}
{Lee}, J.~C., {Whitmore}, B.~C., {Thilker}, D.~A., {et~al.} 2022, \apjs, 258, 10

\bibitem[{{Leroy} {et~al.}(2021){Leroy}, {Schinnerer}, {Hughes}, {Rosolowsky}, {Pety}, {Schruba}, {Usero}, {Blanc}, {Chevance}, {Emsellem}, {Faesi}, {Herrera}, {Liu}, {Meidt}, {Querejeta}, {Saito}, {Sandstrom}, {Sun}, {Williams}, {Anand}, {Barnes}, {Behrens}, {Belfiore}, {Benincasa}, {Be{\v{s}}li{\'c}}, {Bigiel}, {Bolatto}, {den Brok}, {Cao}, {Chandar}, {Chastenet}, {Chiang}, {Congiu}, {Dale}, {Deger}, {Eibensteiner}, {Egorov}, {Garc{\'\i}a-Rodr{\'\i}guez}, {Glover}, {Grasha}, {Henshaw}, {Ho}, {Kepley}, {Kim}, {Klessen}, {Kreckel}, {Koch}, {Kruijssen}, {Larson}, {Lee}, {Lopez}, {Machado}, {Mayker}, {McElroy}, {Murphy}, {Ostriker}, {Pan}, {Pessa}, {Puschnig}, {Razza}, {S{\'a}nchez-Bl{\'a}zquez}, {Santoro}, {Sardone}, {Scheuermann}, {Sliwa}, {Sormani}, {Stuber}, {Thilker}, {Turner}, {Utomo}, {Watkins}, \& {Whitmore}}]{Leroy21}
{Leroy}, A.~K., {Schinnerer}, E., {Hughes}, A., {et~al.} 2021, \apjs, 257, 43

\bibitem[{{Long} {et~al.}(2012{\natexlab{a}}){Long}, {Blair}, {Godfrey}, {Kuntz}, {Plucinsky}, {Soria}, {Stockdale}, {Whitmore}, \& {Winkler}}]{long12}
{Long}, K.~S., {Blair}, W.~P., {Godfrey}, L.~E.~H., {et~al.} 2012{\natexlab{a}}, \apj, 756, 18

\bibitem[{{Long} {et~al.}(2012{\natexlab{b}}){Long}, {Blair}, {Godfrey}, {Kuntz}, {Plucinsky}, {Soria}, {Stockdale}, {Whitmore}, \& {Winkler}}]{Knox12}
{Long}, K.~S., {Blair}, W.~P., {Godfrey}, L.~E.~H., {et~al.} 2012{\natexlab{b}}, \apj, 756, 18

\bibitem[{{Makarenko} {et~al.}(2023){Makarenko}, {Walch}, {Clarke}, {Seifried}, {Naab}, {N{\"u}rnberger}, \& {Rathjen}}]{Makarenko23}
{Makarenko}, E.~I., {Walch}, S., {Clarke}, S.~D., {et~al.} 2023, \mnras, 523, 1421

\bibitem[{{Mart{\'\i}nez-Rodr{\'\i}guez} {et~al.}(2024){Mart{\'\i}nez-Rodr{\'\i}guez}, {Galbany}, {Badenes}, {Anderson}, {Dom{\'\i}nguez}, {Kuncarayakti}, {Lyman}, {S{\'a}nchez}, {V{\'\i}lchez}, {Smith}, \& {Milisavljevic}}]{martinez-rodriguez24}
{Mart{\'\i}nez-Rodr{\'\i}guez}, H., {Galbany}, L., {Badenes}, C., {et~al.} 2024, \apj, 963, 125

\bibitem[{{Matonick} \& {Fesen}(1997)}]{Matonick97}
{Matonick}, D.~M. \& {Fesen}, R.~A. 1997, \apjs, 112, 49

\bibitem[{{Meaburn} {et~al.}(2010){Meaburn}, {Redman}, {Boumis}, \& {Harvey}}]{meaburn10}
{Meaburn}, J., {Redman}, M.~P., {Boumis}, P., \& {Harvey}, E. 2010, \mnras, 408, 1249

\bibitem[{{Micelotta} {et~al.}(2018){Micelotta}, {Matsuura}, \& {Sarangi}}]{micelotta18}
{Micelotta}, E.~R., {Matsuura}, M., \& {Sarangi}, A. 2018, \ssr, 214, 53

\bibitem[{{Milisavljevic} \& {Fesen}(2008)}]{Milisavljevic08}
{Milisavljevic}, D. \& {Fesen}, R.~A. 2008, \apj, 677, 306

\bibitem[{{Milisavljevic} {et~al.}(2012){Milisavljevic}, {Fesen}, {Chevalier}, {Kirshner}, {Challis}, \& {Turatto}}]{Milisavljevic12}
{Milisavljevic}, D., {Fesen}, R.~A., {Chevalier}, R.~A., {et~al.} 2012, \apj, 751, 25

\bibitem[{Milisavljevic {et~al.}(2018)Milisavljevic, Patnaude, Chevalier, Raymond, Fesen, Margutti, Conner, \& Banovetz}]{Milisavljevic2018}
Milisavljevic, D., Patnaude, D.~J., Chevalier, R.~A., {et~al.} 2018, The Astrophysical Journal Letters, 864, L36

\bibitem[{Niculescu-Duvaz {et~al.}(2022)Niculescu-Duvaz, Barlow, Bevan, Wesson, Milisavljevic, De Looze, Clayton, Krafton, Matsuura, \& Brady}]{niculescuduvaz22}
Niculescu-Duvaz, M., Barlow, M.~J., Bevan, A., {et~al.} 2022, Monthly Notices of the Royal Astronomical Society, 515, 4302

\bibitem[{{Omand} \& {Jerkstrand}(2023)}]{Omand23}
{Omand}, C.~M.~B. \& {Jerkstrand}, A. 2023, \aap, 673, A107

\bibitem[{{Patnaude} \& {Fesen}(2003)}]{Patnaude2003}
{Patnaude}, D.~J. \& {Fesen}, R.~A. 2003, \apj, 587, 221

\bibitem[{{Patnaude} {et~al.}(2011){Patnaude}, {Loeb}, \& {Jones}}]{Patnaude11}
{Patnaude}, D.~J., {Loeb}, A., \& {Jones}, C. 2011, \na, 16, 187

\bibitem[{Schenck {et~al.}(2014)Schenck, Park, Burrows, Hughes, Lee, \& Mori}]{Schenck2014}
Schenck, A., Park, S., Burrows, D.~N., {et~al.} 2014, The Astrophysical Journal, 791, 50

\bibitem[{{Sch{\"o}nberner} {et~al.}(2018){Sch{\"o}nberner}, {Balick}, \& {Jacob}}]{Schonberner18}
{Sch{\"o}nberner}, D., {Balick}, B., \& {Jacob}, R. 2018, \aap, 609, A126

\bibitem[{{Seaquist} \& {Bignell}(1978)}]{seaquist78}
{Seaquist}, E.~R. \& {Bignell}, R.~C. 1978, \apjl, 226, L5

\bibitem[{{Smith} {et~al.}(1993){Smith}, {Kirshner}, {Blair}, {Long}, \& {Winkler}}]{Smith1993}
{Smith}, R.~C., {Kirshner}, R.~P., {Blair}, W.~P., {Long}, K.~S., \& {Winkler}, P.~F. 1993, \apj, 407, 564

\bibitem[{{Stanway} \& {Eldridge}(2018)}]{Stanway18}
{Stanway}, E.~R. \& {Eldridge}, J.~J. 2018, \mnras, 479, 75

\bibitem[{{Tinyanont} {et~al.}(2016){Tinyanont}, {Kasliwal}, {Fox}, {Lau}, {Smith}, {Williams}, {Jencson}, {Perley}, {Dykhoff}, {Gehrz}, {Johansson}, {Van Dyk}, {Masci}, {Cody}, \& {Prince}}]{Tinyanont16}
{Tinyanont}, S., {Kasliwal}, M.~M., {Fox}, O.~D., {et~al.} 2016, \apj, 833, 231

\bibitem[{{Tschöke} {et~al.}(2000){Tschöke}, {Hensler}, \& {Junkes}}]{tschoke2000}
{Tschöke}, D., {Hensler}, G., \& {Junkes}, N. 2000, \aap, 360, 447

\bibitem[{Vogt \& Dopita(2010)}]{Vogt2010}
Vogt, F. \& Dopita, M.~A. 2010, Astrophysics and Space Science, 331, 521

\bibitem[{{Weiler} {et~al.}(1981){Weiler}, {van der Hulst}, {Sramek}, \& {Panagia}}]{weiler81}
{Weiler}, K.~W., {van der Hulst}, J.~M., {Sramek}, R.~A., \& {Panagia}, N. 1981, \apjl, 243, L151

\bibitem[{{Wesson} {et~al.}(2023){Wesson}, {Bevan}, {Barlow}, {De Looze}, {Matsuura}, {Clayton}, \& {Andrews}}]{Wesson2023}
{Wesson}, R., {Bevan}, A.~M., {Barlow}, M.~J., {et~al.} 2023, \mnras [\eprint[arXiv]{2308.09028}]

\end{thebibliography}

\onecolumn

\section*{Appendix A}\label{xray_phot}
This appendix presents the Chandra X-ray observations and JWST photometry.

\begin{table}[h]
    \centering
    \begin{tabular}{l|c|c|l|c}
    Source Name	&	Absorbed model flux	&	Flux 90\% conf.		&	L$_x$ 0.5-7.0 keV	&	Adopted distance \\
    \hline
    		&	erg/cm$^2$/s		&	erg/cm$^2$/s		&	$\cdot$ 10$^{37}$ erg/s	&	Mpc \\
    \hline
    4303-46	&	1.74e-14		&	(1.47e-14, 2.01e-14)	&	60.1 (50.0,69.4)	&	16.99$\pm$3.04 \\
    4303-20	&	1.21e-15		&	(5.82e-16, 2.12e-15)	&	4.18 (2.0,7.3)	&	16.99$\pm$3.04 \\
    3627-1	&	7.14e-15		&	(5.43e-15, 8.86e-15)	&	10.9 (8.3,13.6)		&	11.32$\pm$0.48 \\
    3351-5	&	4.02e-15		&	(2.82e-15, 5.51e-15)	&	4.78 (3.3,6.5)	&	9.96$\pm$0.33 \\
    3627-17	&	1.22e-14		&	(9.97e-15, 1.44e-14)	&	18.7 (15.3,22.1)	&	11.32$\pm$0.48 \\
    4303-41	&	6.73e-16        &   (1.83e-16, 1.44e-15)    &   2.32 (0.6,5.0)  &   16.99$\pm$3.04\\
    1566-4  &   3.82e-15		&	(1.22e-16, 1.57e-14)	&	14.3 (0.4,58.8)&	17.69$\pm$2.00 \\
    \end{tabular}
    \caption{Measurements of Chandra X-ray fluxes and luminosities based on the absorbed model flux. Luminosity range within the 90\% interval is presented in the parentheses. We assume a power law model with a power law index of 2. The last two source have very few photons, 4303-41 has three and 1566-5 only has one. Qualitatively these are likely incidental and the fluxes are just an absolute upper limit. Distances to the host galaxies are adopted from \cite{Anand21}.}
    \label{tab:xray}
\end{table}

\begin{table}[h]
    \centering
    \begin{tabular}{l|c|c|c|c}
                &   F770W   &   F1000W  &   F1130W  &   F2100W  \\
    \hline
    Source Name &   m$_{AB}$&   m$_{AB}$&   m$_{AB}$&   m$_{AB}$\\
    \hline
    4303-46 &   >21.0   &   20.1    &   19.83   &   17.89   \\
    4303-20 &   >24.0   &   >22.2   &   20.62   &   19.99   \\
    3627-1  &    19.79  &   19.54   &   19.57   &   20.55  \\
    3627-17 &    17.82  &   16.41   &   16.25   &   14.727  \\
    4303-37 &    20.04  &   21.69   &   20.35        &    20.43       \\
    SN1979C &   19.54   &   17.90   &   17.87   &   16.46   \\
    \end{tabular}
    \caption{Archival JWST observations of sources. All observations are part of the PHANGS-JWST program. The measurements were done with Jdaviz, using an aperture with a radius of four pixels. Background emission was estimated with an annular ring with and inner radius of four pixels and width of two pixels. For the upper limits we use the fluxes obtained from a similar aperture with no background subtraction.}
    \label{tab:jwst}
\end{table}

\section*{Appendix B} \label{snr-list}
In this appendix we present all of SNRs discovered in our survey of PHANGS-MUSE galaxies. The selected lines are used in Fig. \ref{dpuser_plot} as a diagnostic tool to identify them as SNRs.
\begin{longtable}{rllrrr}
\caption{NGC1365}\\
\toprule
 SNR No. &           RA &          Dec & F([SII]) & F([NII]) & F(H$\alpha$) \\
&&&erg/s/cm$^2$&erg/s/cm$^2$&erg/s/cm$^2$\\
\midrule
\endfirsthead
\caption[]{NGC1365} \\
\toprule
 SNR No. &           RA &          Dec & F([SII]) & F([NII]) & F(H$\alpha$) \\
&&&erg/s/cm$^2$&erg/s/cm$^2$&erg/s/cm$^2$\\
\midrule
\endhead
\midrule
\multicolumn{6}{r}{{Continued on next page}} \\
\midrule
\endfoot

\bottomrule
\endlastfoot
1 & 3h33m30.45s & -36d09m45.28s & 9.8e-17 & 6.69e-17 & 8.19e-17\\
2 & 3h33m26.69s & -36d09m43.28s & 1.49e-16 & 5.78e-17 & 1.13e-16\\
3 & 3h33m28.84s & -36d09m39.27s & 7.65e-17 & 1.08e-16 & 2.71e-17\\
4 & 3h33m30.33s & -36d08m59.93s & 3.77e-16 & 2.17e-16 & 1.74e-16\\
5 & 3h33m32.82s & -36d08m53.38s & 1.7e-17 & 2.8e-17 & 2.04e-17\\
6 & 3h33m32.62s & -36d08m46.17s & 2.97e-17 & 4.8e-17 & 2.64e-17\\
7 & 3h33m29.47s & -36d08m26.73s & 9.62e-17 & 7.43e-17 & 6.54e-17\\
8 & 3h33m33.21s & -36d07m20.95s & 2.8e-17 & 3.97e-17 & 4.47e-17\\
\end{longtable}
\begin{longtable}{rllrrr}
\caption{NGC4535}\\
\toprule
 SNR No. &           RA &          Dec & F([SII]) & F([NII]) & F(H$\alpha$) \\
&&&erg/s/cm$^2$&erg/s/cm$^2$&erg/s/cm$^2$\\
\midrule
\endfirsthead
\caption[]{NGC4535} \\
\toprule
 SNR No. &           RA &          Dec & F([SII]) & F([NII]) & F(H$\alpha$) \\
&&&erg/s/cm$^2$&erg/s/cm$^2$&erg/s/cm$^2$\\
\midrule
\endhead
\midrule
\multicolumn{6}{r}{{Continued on next page}} \\
\midrule
\endfoot

\bottomrule
\endlastfoot
1 & 12h34m21.53s & 8d10m23.66s & 5.24e-16 & 3.47e-16 & 3.37e-16\\
2 & 12h34m21.74s & 8d10m32.37s & 1.66e-16 & 1.28e-16 & 1.47e-16\\
3 & 12h34m16.79s & 8d10m38.34s & 1.07e-16 & 1.28e-16 & 1.19e-16\\
4 & 12h34m17.05s & 8d10m47.71s & 6.45e-17 & 1.07e-16 & 8.13e-17\\
5 & 12h34m18.65s & 8d10m57.26s & 2.75e-16 & 2.75e-16 & 2.8e-16\\
6 & 12h34m20.59s & 8d10m58.32s & 1.16e-16 & 1.59e-16 & 1.2e-16\\
7 & 12h34m20.79s & 8d11m01.48s & 3.39e-16 & 4.98e-16 & 3.15e-16\\
8 & 12h34m19.78s & 8d11m06.11s & 1.19e-16 & 1.29e-16 & 1.09e-16\\
9 & 12h34m19.56s & 8d11m11.69s & 2.56e-16 & 6.56e-16 & 3.28e-16\\
10 & 12h34m19.30s & 8d11m13.97s & 2.33e-16 & 2.6e-16 & 1.91e-16\\
11 & 12h34m18.61s & 8d11m20.32s & 9.78e-17 & 2.42e-16 & 1.74e-16\\
12 & 12h34m17.47s & 8d11m21.99s & 6.68e-17 & 7.99e-17 & 5.84e-17\\
13 & 12h34m23.82s & 8d11m25.93s & 3.07e-16 & 3.04e-16 & 2.55e-16\\
14 & 12h34m23.64s & 8d11m36.69s & 6.24e-16 & 9.53e-16 & 7.47e-16\\
15 & 12h34m18.02s & 8d11m41.44s & 5.39e-17 & 1.31e-16 & 1.24e-16\\
16 & 12h34m20.10s & 8d11m50.13s & 5.83e-16 & 9.55e-16 & 5.84e-16\\
17 & 12h34m20.59s & 8d12m07.16s & 1.38e-16 & 2.95e-16 & 1.72e-16\\
18 & 12h34m22.09s & 8d12m27.58s & 5.15e-16 & 5.33e-16 & 3.55e-16\\
19 & 12h34m20.49s & 8d12m38.11s & 1.43e-17 & 3.58e-17 & 1.1e-17\\
20 & 12h34m20.61s & 8d12m44.98s & 2.51e-16 & 3.26e-16 & 2.83e-16\\
21 & 12h34m22.85s & 8d12m54.77s & 9.46e-17 & 4.83e-16 & 1.33e-16\\
\end{longtable}
\begin{longtable}{rllrrr}
\caption{IC5332}\\
\toprule
 SNR No. &           RA &          Dec & F([SII]) & F([NII]) & F(H$\alpha$) \\
&&&erg/s/cm$^2$&erg/s/cm$^2$&erg/s/cm$^2$\\
\midrule
\endfirsthead
\caption[]{IC5332} \\
\toprule
 SNR No. &           RA &          Dec & F([SII]) & F([NII]) & F(H$\alpha$) \\
&&&erg/s/cm$^2$&erg/s/cm$^2$&erg/s/cm$^2$\\
\midrule
\endhead
\midrule
\multicolumn{6}{r}{{Continued on next page}} \\
\midrule
\endfoot

\bottomrule
\endlastfoot
1 & 23h34m25.92s & -36d06m42.49s & 6.19e-16 & 4.26e-16 & 8.08e-16\\
2 & 23h34m28.24s & -36d06m16.90s & 8.56e-17 & 5.94e-17 & 1.19e-16\\
3 & 23h34m28.02s & -36d06m16.00s & 1.91e-16 & 1.68e-16 & 2.69e-16\\
4 & 23h34m29.55s & -36d06m12.32s & 4.47e-17 & 3.44e-17 & 6.07e-17\\
5 & 23h34m31.09s & -36d05m54.18s & 5.25e-17 & 2.58e-17 & 6.18e-17\\
6 & 23h34m23.71s & -36d05m50.65s & 1.43e-16 & 5.08e-17 & 1.38e-16\\
7 & 23h34m24.42s & -36d05m21.57s & 1.75e-16 & 6.29e-17 & 2.28e-16\\
\end{longtable}
\begin{longtable}{rllrrr}
\caption{NGC4321}\\
\toprule
 SNR No. &           RA &          Dec & F([SII]) & F([NII]) & F(H$\alpha$) \\
&&&erg/s/cm$^2$&erg/s/cm$^2$&erg/s/cm$^2$\\
\midrule
\endfirsthead
\caption[]{NGC4321} \\
\toprule
 SNR No. &           RA &          Dec & F([SII]) & F([NII]) & F(H$\alpha$) \\
&&&erg/s/cm$^2$&erg/s/cm$^2$&erg/s/cm$^2$\\
\midrule
\endhead
\midrule
\multicolumn{6}{r}{{Continued on next page}} \\
\midrule
\endfoot

\bottomrule
\endlastfoot
1 & 12h22m56.17s & 15d48m11.38s & 1.18e-16 & 2.11e-16 & 2.4e-16\\
2 & 12h22m53.20s & 15d48m32.77s & 1.25e-15 & 8.62e-16 & 6.74e-16\\
3 & 12h22m52.66s & 15d48m36.12s & 1.42e-16 & 2.58e-16 & 1.78e-16\\
4 & 12h22m58.87s & 15d48m57.31s & 1.59e-16 & 4.18e-16 & 3.27e-16\\
5 & 12h22m51.46s & 15d49m01.23s & 6.11e-17 & 1.86e-16 & 9.81e-17\\
6 & 12h22m58.34s & 15d49m04.59s & 3.43e-17 & 1.45e-16 & 7.93e-17\\
7 & 12h22m54.68s & 15d49m08.46s & 4.39e-16 & 5.13e-16 & 2.96e-16\\
8 & 12h22m54.83s & 15d49m17.59s & 9.91e-16 & 1.1e-15 & 3.58e-16\\
9 & 12h23m00.40s & 15d49m18.30s & 2.97e-16 & 2.32e-16 & 1.76e-16\\
10 & 12h22m51.80s & 15d49m25.33s & 5.57e-17 & 1.16e-16 & 6.85e-17\\
11 & 12h22m54.46s & 15d49m36.36s & 8.53e-16 & 6.01e-16 & 3.5e-16\\
12 & 12h22m54.01s & 15d49m42.97s & 2.52e-17 & 7.1e-17 & 4.1e-17\\
13 & 12h22m52.46s & 15d49m52.72s & 9.11e-17 & 1.37e-16 & 8.03e-17\\
14 & 12h22m47.46s & 15d49m58.93s & 2.4e-16 & 3.31e-16 & 3.02e-16\\
15 & 12h22m54.86s & 15d50m00.30s & 9.07e-17 & 2.54e-16 & 1.09e-16\\
16 & 12h22m51.76s & 15d50m04.46s & 6.03e-17 & 6.5e-17 & 3.69e-17\\
17 & 12h22m54.37s & 15d50m08.43s & 1.94e-17 & 4.33e-17 & 2.91e-17\\
18 & 12h22m48.98s & 15d50m10.79s & 4.78e-16 & 3.15e-16 & 2.4e-16\\
19 & 12h22m51.92s & 15d50m15.64s & 3.44e-17 & 6.73e-17 & 4.8e-17\\
20 & 12h22m54.49s & 15d50m15.86s & 2.59e-16 & 5.8e-16 & 3.23e-16\\
21 & 12h22m50.07s & 15d50m19.55s & 1.3e-16 & 3.63e-16 & 2.9e-16\\
22 & 12h22m54.31s & 15d50m24.45s & 7.04e-17 & 9.82e-17 & 9.9e-17\\
23 & 12h23m00.61s & 15d50m29.06s & 4.76e-17 & 5.84e-17 & 4.18e-17\\
24 & 12h22m56.22s & 15d50m29.54s & 2.49e-17 & 4.38e-17 & 2.14e-17\\
25 & 12h22m54.09s & 15d50m50.96s & 1.46e-16 & 1.33e-16 & 1.63e-16\\
26 & 12h22m54.14s & 15d50m52.43s & 4.95e-17 & 6.96e-17 & 6.19e-17\\
\end{longtable}
\begin{longtable}{rllrrr}
\caption{NGC1672}\\
\toprule
 SNR No. &           RA &          Dec & F([SII]) & F([NII]) & F(H$\alpha$) \\
&&&erg/s/cm$^2$&erg/s/cm$^2$&erg/s/cm$^2$\\
\midrule
\endfirsthead
\caption[]{NGC1672} \\
\toprule
 SNR No. &           RA &          Dec & F([SII]) & F([NII]) & F(H$\alpha$) \\
&&&erg/s/cm$^2$&erg/s/cm$^2$&erg/s/cm$^2$\\
\midrule
\endhead
\midrule
\multicolumn{6}{r}{{Continued on next page}} \\
\midrule
\endfoot

\bottomrule
\endlastfoot
1 & 4h45m42.40s & -59d15m04.88s & 2.18e-17 & 1.83e-17 & 2.09e-17\\
2 & 4h45m29.97s & -59d14m59.71s & 2.1e-16 & 2.38e-16 & 1.83e-16\\
3 & 4h45m53.34s & -59d14m53.18s & 1.4e-16 & 1.16e-16 & 1.29e-16\\
4 & 4h45m35.36s & -59d14m51.77s & 1.14e-16 & 8.61e-17 & 7.47e-17\\
5 & 4h45m33.46s & -59d14m51.32s & 2.71e-16 & 1.69e-16 & 2.06e-16\\
6 & 4h45m35.98s & -59d14m46.70s & 5.31e-17 & 9.42e-17 & 1.28e-16\\
7 & 4h45m37.15s & -59d14m38.41s & 1.01e-16 & 1.24e-16 & 1.24e-16\\
8 & 4h45m43.99s & -59d14m38.01s & 4.01e-17 & 9.24e-17 & 6.84e-17\\
9 & 4h45m53.68s & -59d14m29.99s & 8.19e-16 & 5.3e-16 & 3.55e-16\\
10 & 4h45m54.94s & -59d14m08.15s & 7.16e-17 & 3.97e-17 & 9.46e-17\\
11 & 4h45m48.26s & -59d13m49.72s & 4.6e-17 & 5.18e-17 & 5.17e-17\\
\end{longtable}
\begin{longtable}{rllrrr}
\caption{NGC5068}\\
\toprule
 SNR No. &           RA &          Dec & F([SII]) & F([NII]) & F(H$\alpha$) \\
&&&erg/s/cm$^2$&erg/s/cm$^2$&erg/s/cm$^2$\\
\midrule
\endfirsthead
\caption[]{NGC5068} \\
\toprule
 SNR No. &           RA &          Dec & F([SII]) & F([NII]) & F(H$\alpha$) \\
&&&erg/s/cm$^2$&erg/s/cm$^2$&erg/s/cm$^2$\\
\midrule
\endhead
\midrule
\multicolumn{6}{r}{{Continued on next page}} \\
\midrule
\endfoot

\bottomrule
\endlastfoot
1 & 13h18m53.05s & -21d03m55.70s & 1.83e-16 & 9.18e-17 & 1.56e-16\\
2 & 13h18m47.41s & -21d03m53.00s & 9.53e-16 & 3.02e-16 & 1.16e-15\\
3 & 13h18m53.73s & -21d03m35.88s & 7.35e-17 & 6.2e-17 & 1.74e-16\\
4 & 13h18m56.66s & -21d03m35.69s & 2.12e-16 & 7.14e-17 & 1.86e-16\\
5 & 13h18m53.06s & -21d03m28.40s & 9.14e-17 & 3.99e-17 & 1.12e-16\\
6 & 13h18m52.67s & -21d03m20.03s & 9.28e-17 & 2.48e-17 & 1.2e-16\\
7 & 13h18m49.73s & -21d02m26.91s & 3.55e-16 & 1.22e-16 & 2.62e-16\\
8 & 13h18m51.09s & -21d01m51.40s & 1.37e-16 & 7.83e-17 & 1.34e-16\\
9 & 13h18m53.49s & -21d01m46.20s & 1.25e-16 & 5.83e-17 & 1.57e-16\\
10 & 13h18m44.82s & -21d01m10.86s & 2.86e-16 & 6.18e-17 & 2.88e-16\\
11 & 13h18m46.10s & -21d00m54.04s & 6.54e-17 & 2.77e-17 & 7.64e-17\\
12 & 13h18m50.72s & -21d00m48.49s & 1.4e-16 & 3.78e-17 & 6.56e-17\\
13 & 13h18m48.32s & -21d00m25.02s & 4.46e-16 & 1.05e-16 & 5.07e-16\\
\end{longtable}
\begin{longtable}{rllrrr}
\caption{NGC2835}\\
\toprule
 SNR No. &           RA &          Dec & F([SII]) & F([NII]) & F(H$\alpha$) \\
&&&erg/s/cm$^2$&erg/s/cm$^2$&erg/s/cm$^2$\\
\midrule
\endfirsthead
\caption[]{NGC2835} \\
\toprule
 SNR No. &           RA &          Dec & F([SII]) & F([NII]) & F(H$\alpha$) \\
&&&erg/s/cm$^2$&erg/s/cm$^2$&erg/s/cm$^2$\\
\midrule
\endhead
\midrule
\multicolumn{6}{r}{{Continued on next page}} \\
\midrule
\endfoot

\bottomrule
\endlastfoot
1 & 9h17m51.80s & -22d22m03.10s & 1.53e-16 & 1.14e-16 & 1.55e-16\\
2 & 9h17m48.41s & -22d21m46.45s & 1.48e-16 & 3.42e-17 & 1.08e-16\\
3 & 9h17m54.22s & -22d21m34.13s & 5.06e-17 & 3.56e-17 & 3.13e-17\\
4 & 9h17m51.45s & -22d21m29.79s & 3.43e-17 & 4.46e-17 & 5.5e-17\\
5 & 9h17m57.81s & -22d20m59.41s & 6.42e-17 & 1.89e-17 & 6.1e-17\\
6 & 9h17m51.48s & -22d20m49.45s & 5.78e-17 & 4.91e-17 & 7.15e-17\\
7 & 9h17m51.38s & -22d20m43.43s & 4.9e-17 & 3.75e-17 & 5.82e-17\\
8 & 9h17m52.32s & -22d20m28.40s & 1.6e-16 & 7.02e-17 & 1.48e-16\\
9 & 9h17m48.14s & -22d20m17.13s & 1.38e-16 & 2.68e-17 & 9.1e-17\\
\end{longtable}
\begin{longtable}{rllrrr}
\caption{NGC1385}\\
\toprule
 SNR No. &           RA &          Dec & F([SII]) & F([NII]) & F(H$\alpha$) \\
&&&erg/s/cm$^2$&erg/s/cm$^2$&erg/s/cm$^2$\\
\midrule
\endfirsthead
\caption[]{NGC1385} \\
\toprule
 SNR No. &           RA &          Dec & F([SII]) & F([NII]) & F(H$\alpha$) \\
&&&erg/s/cm$^2$&erg/s/cm$^2$&erg/s/cm$^2$\\
\midrule
\endhead
\midrule
\multicolumn{6}{r}{{Continued on next page}} \\
\midrule
\endfoot

\bottomrule
\endlastfoot
1 & 3h37m30.77s & -24d30m14.78s & 1.02e-16 & 4.92e-17 & 9.9e-17\\
2 & 3h37m29.54s & -24d30m05.68s & 6.35e-17 & 6.87e-17 & 7.84e-17\\
3 & 3h37m30.65s & -24d30m05.31s & 1.64e-16 & 8.58e-17 & 1.42e-16\\
4 & 3h37m28.03s & -24d29m58.38s & 4.14e-15 & 2.02e-15 & 1.64e-15\\
5 & 3h37m29.31s & -24d29m56.33s & 1.04e-16 & 1.38e-16 & 1.48e-16\\
6 & 3h37m30.53s & -24d29m35.09s & 1.96e-17 & 1.47e-17 & 3.14e-17\\
7 & 3h37m29.84s & -24d29m26.91s & 2.37e-17 & 2.39e-17 & 5.18e-17\\
8 & 3h37m30.13s & -24d29m11.41s & 1.8e-16 & 2e-16 & 1.87e-16\\
\end{longtable}
\begin{longtable}{rllrrr}
\caption{NGC1566}\\
\toprule
 SNR No. &           RA &          Dec & F([SII]) & F([NII]) & F(H$\alpha$) \\
&&&erg/s/cm$^2$&erg/s/cm$^2$&erg/s/cm$^2$\\
\midrule
\endfirsthead
\caption[]{NGC1566} \\
\toprule
 SNR No. &           RA &          Dec & F([SII]) & F([NII]) & F(H$\alpha$) \\
&&&erg/s/cm$^2$&erg/s/cm$^2$&erg/s/cm$^2$\\
\midrule
\endhead
\midrule
\multicolumn{6}{r}{{Continued on next page}} \\
\midrule
\endfoot

\bottomrule
\endlastfoot
1 & 4h20m02.67s & -54d57m34.19s & 2.23e-16 & 2.52e-16 & 3.17e-16\\
2 & 4h20m01.44s & -54d57m24.65s & 1.58e-15 & 9.45e-16 & 8.17e-16\\
3 & 4h20m03.43s & -54d57m19.83s & 3.02e-16 & 2.58e-16 & 2.63e-16\\
4 & 4h20m04.91s & -54d57m05.90s & 0.0 & 0.0 & 0.0\\
5 & 4h19m56.33s & -54d57m00.05s & 7.33e-17 & 8.39e-17 & 6.82e-17\\
6 & 4h20m05.29s & -54d56m55.59s & 2.8e-16 & 3.21e-16 & 2.76e-16\\
7 & 4h20m05.96s & -54d56m55.32s & 2.47e-16 & 4.18e-16 & 2.61e-16\\
8 & 4h20m04.59s & -54d56m51.88s & 3.67e-16 & 6.07e-16 & 4.87e-16\\
9 & 4h20m01.96s & -54d56m48.61s & 2.64e-16 & 3.51e-16 & 2.11e-16\\
10 & 4h20m04.45s & -54d56m46.74s & 5.11e-16 & 8.33e-16 & 6.65e-16\\
11 & 4h20m02.61s & -54d56m41.79s & 8.68e-17 & 1.41e-16 & 1.2e-16\\
12 & 4h20m00.53s & -54d56m38.53s & 8.73e-17 & 9.62e-17 & 6.33e-17\\
13 & 4h20m03.55s & -54d56m37.32s & 3.93e-17 & 1.77e-16 & 1.34e-16\\
14 & 4h20m05.19s & -54d56m37.17s & 2.7e-16 & 3.76e-16 & 2.86e-16\\
15 & 4h20m03.63s & -54d56m35.38s & 3.45e-17 & 4.49e-17 & 3.69e-17\\
16 & 4h20m05.94s & -54d56m27.04s & 2.91e-16 & 2.04e-16 & 2.16e-16\\
17 & 4h20m04.56s & -54d56m26.93s & 1.25e-16 & 2.7e-16 & 1.88e-16\\
18 & 4h20m05.27s & -54d56m25.12s & 4.3e-16 & 3.68e-16 & 5.05e-16\\
19 & 4h20m05.12s & -54d56m18.22s & 1.96e-15 & 1.2e-15 & 8.02e-16\\
20 & 4h20m04.01s & -54d56m12.78s & 2.67e-16 & 6.2e-16 & 4.26e-16\\
21 & 4h20m03.00s & -54d56m08.70s & 2.43e-16 & 3.33e-16 & 3.23e-16\\
22 & 4h20m03.98s & -54d56m00.15s & 1.91e-16 & 4.92e-16 & 4.25e-16\\
23 & 4h19m59.14s & -54d55m50.08s & 3.97e-17 & 1.1e-16 & 1.13e-16\\
24 & 4h20m02.43s & -54d55m47.73s & 1.15e-16 & 2.17e-16 & 1.24e-16\\
25 & 4h20m02.16s & -54d55m47.64s & 2.93e-16 & 5.86e-16 & 4.22e-16\\
26 & 4h19m55.82s & -54d55m43.25s & 1.92e-16 & 1.76e-16 & 2.52e-16\\
27 & 4h19m56.44s & -54d55m38.14s & 4.84e-17 & 8.83e-17 & 7.18e-17\\
28 & 4h19m59.23s & -54d55m33.97s & 3.94e-17 & 6.43e-17 & 5.89e-17\\
29 & 4h20m03.30s & -54d55m27.44s & 2.31e-17 & 3.07e-17 & 3.04e-17\\
30 & 4h20m06.49s & -54d55m25.61s & 6.86e-17 & 1.1e-16 & 8.69e-17\\
31 & 4h20m06.26s & -54d55m24.62s & 1.72e-16 & 2.16e-16 & 2.3e-16\\
32 & 4h20m04.19s & -54d55m06.45s & 1.03e-15 & 5e-16 & 3.81e-16\\
33 & 4h19m59.12s & -54d54m52.12s & 5.74e-17 & 2.18e-16 & 1.86e-16\\
\end{longtable}
\begin{longtable}{rllrrr}
\caption{NGC3351}\\
\toprule
 SNR No. &           RA &          Dec & F([SII]) & F([NII]) & F(H$\alpha$) \\
&&&erg/s/cm$^2$&erg/s/cm$^2$&erg/s/cm$^2$\\
\midrule
\endfirsthead
\caption[]{NGC3351} \\
\toprule
 SNR No. &           RA &          Dec & F([SII]) & F([NII]) & F(H$\alpha$) \\
&&&erg/s/cm$^2$&erg/s/cm$^2$&erg/s/cm$^2$\\
\midrule
\endhead
\midrule
\multicolumn{6}{r}{{Continued on next page}} \\
\midrule
\endfoot

\bottomrule
\endlastfoot
1 & 10h43m56.16s & 11d40m56.48s & 6.74e-17 & 1.18e-16 & 6.67e-17\\
2 & 10h43m56.82s & 11d41m22.07s & 2.96e-16 & 5.71e-16 & 2.88e-16\\
3 & 10h43m52.46s & 11d41m48.39s & 1.56e-16 & 2.94e-16 & 1.71e-16\\
4 & 10h43m57.40s & 11d42m13.47s & 3.92e-16 & 1.23e-15 & 4e-16\\
5 & 10h43m56.69s & 11d42m50.70s & 0.0 & 0.0 & 0.0\\
6 & 10h44m02.62s & 11d42m54.54s & 1.66e-16 & 3.17e-16 & 2.17e-16\\
7 & 10h43m58.00s & 11d43m06.45s & 5.61e-17 & 1.11e-16 & 6.66e-17\\
8 & 10h43m56.61s & 11d43m22.95s & 2.59e-16 & 5.45e-16 & 3.63e-16\\
9 & 10h43m57.37s & 11d43m24.59s & 9.23e-16 & 4.41e-16 & 3.02e-16\\
\end{longtable}
\begin{longtable}{rllrrr}
\caption{NGC4254}\\
\toprule
 SNR No. &           RA &          Dec & F([SII]) & F([NII]) & F(H$\alpha$) \\
&&&erg/s/cm$^2$&erg/s/cm$^2$&erg/s/cm$^2$\\
\midrule
\endfirsthead
\caption[]{NGC4254} \\
\toprule
 SNR No. &           RA &          Dec & F([SII]) & F([NII]) & F(H$\alpha$) \\
&&&erg/s/cm$^2$&erg/s/cm$^2$&erg/s/cm$^2$\\
\midrule
\endhead
\midrule
\multicolumn{6}{r}{{Continued on next page}} \\
\midrule
\endfoot

\bottomrule
\endlastfoot
1 & 12h18m48.56s & 14d23m52.50s & 1.86e-16 & 8.57e-17 & 1.21e-16\\
2 & 12h18m46.48s & 14d24m01.28s & 1.7e-16 & 1.49e-16 & 1.13e-16\\
3 & 12h18m48.68s & 14d24m13.57s & 5.86e-17 & 8.76e-17 & 8.45e-17\\
4 & 12h18m49.60s & 14d24m20.41s & 4.81e-17 & 8.54e-17 & 7.11e-17\\
5 & 12h18m46.10s & 14d24m23.54s & 1.48e-16 & 1.5e-16 & 1.1e-16\\
6 & 12h18m51.30s & 14d24m29.87s & 1.25e-16 & 3.58e-16 & 2.63e-16\\
7 & 12h18m48.29s & 14d24m37.05s & 9.86e-16 & 7.55e-16 & 4.71e-16\\
8 & 12h18m49.13s & 14d24m38.86s & 5.92e-16 & 6.59e-16 & 2.65e-16\\
9 & 12h18m55.28s & 14d24m43.19s & 1.44e-16 & 1.64e-16 & 7.65e-17\\
10 & 12h18m48.33s & 14d24m43.37s & 9.12e-17 & 8.77e-17 & 5.07e-17\\
11 & 12h18m50.82s & 14d24m45.40s & 4.14e-16 & 3.43e-16 & 2.67e-16\\
12 & 12h18m48.57s & 14d24m46.68s & 1.44e-16 & 2.98e-16 & 1.29e-16\\
13 & 12h18m47.26s & 14d24m46.89s & 1.03e-16 & 1.82e-16 & 1.43e-16\\
14 & 12h18m49.06s & 14d24m47.76s & 7.93e-17 & 2.08e-16 & 8.77e-17\\
15 & 12h18m48.46s & 14d24m50.81s & 6.73e-16 & 1.05e-15 & 4.78e-16\\
16 & 12h18m47.45s & 14d24m52.64s & 1.55e-16 & 5.05e-16 & 4e-16\\
17 & 12h18m48.12s & 14d24m52.75s & 1.42e-15 & 1.72e-15 & 4.91e-16\\
18 & 12h18m52.91s & 14d24m52.76s & 6.26e-17 & 9.31e-17 & 8.34e-17\\
19 & 12h18m48.43s & 14d24m55.03s & 1.92e-16 & 2.49e-16 & 1.32e-16\\
20 & 12h18m49.10s & 14d24m56.58s & 1.05e-15 & 1.44e-15 & 6.35e-16\\
21 & 12h18m49.20s & 14d25m00.34s & 7.49e-17 & 2.87e-16 & 1.33e-16\\
22 & 12h18m48.26s & 14d25m00.34s & 6.28e-16 & 1.18e-15 & 4.58e-16\\
23 & 12h18m48.23s & 14d25m05.06s & 8.76e-17 & 2.05e-16 & 7.64e-17\\
24 & 12h18m52.84s & 14d25m07.63s & 3.13e-16 & 3.81e-16 & 2.95e-16\\
25 & 12h18m51.51s & 14d25m08.43s & 2.22e-15 & 1.76e-15 & 1.14e-15\\
26 & 12h18m50.63s & 14d25m09.46s & 1.43e-15 & 3.03e-15 & 1.02e-15\\
27 & 12h18m55.90s & 14d25m09.82s & 1.77e-16 & 1.57e-16 & 1.69e-16\\
28 & 12h18m47.55s & 14d25m10.68s & 1.83e-16 & 2.07e-16 & 1.47e-16\\
29 & 12h18m53.05s & 14d25m15.51s & 1.52e-16 & 1.49e-16 & 7.48e-17\\
30 & 12h18m57.16s & 14d25m17.37s & 7.91e-17 & 1.03e-16 & 1.18e-16\\
31 & 12h18m48.22s & 14d25m17.91s & 7.86e-17 & 1.98e-16 & 1.31e-16\\
32 & 12h18m50.33s & 14d25m18.52s & 2.83e-16 & 5.55e-16 & 2.47e-16\\
33 & 12h18m52.35s & 14d25m21.41s & 3.51e-16 & 3.76e-16 & 3.39e-16\\
34 & 12h18m50.36s & 14d25m24.48s & 9.28e-17 & 9.64e-17 & 6.75e-17\\
35 & 12h18m49.37s & 14d25m27.69s & 6.12e-16 & 7.56e-16 & 3.85e-16\\
36 & 12h18m46.90s & 14d25m31.54s & 4.24e-17 & 5.15e-17 & 5.02e-17\\
37 & 12h18m49.11s & 14d25m34.51s & 4.4e-16 & 6.31e-16 & 3.86e-16\\
38 & 12h18m51.75s & 14d25m48.27s & 7.76e-17 & 1.44e-16 & 1.44e-16\\
39 & 12h18m56.92s & 14d25m53.24s & 7.02e-17 & 6.95e-17 & 1.89e-16\\
40 & 12h18m56.08s & 14d26m14.90s & 7.92e-17 & 9.36e-17 & 1.46e-16\\
41 & 12h18m48.37s & 14d26m17.51s & 6.08e-17 & 5.33e-17 & 6.11e-17\\
42 & 12h18m51.80s & 14d26m20.27s & 4.17e-17 & 5.69e-17 & 3.34e-17\\
43 & 12h18m52.77s & 14d26m28.27s & 4.04e-17 & 4.39e-17 & 2.86e-17\\
44 & 12h18m53.69s & 14d26m36.76s & 1.2e-16 & 9.56e-17 & 7.67e-17\\
45 & 12h18m47.12s & 14d26m37.36s & 8.49e-17 & 7.19e-17 & 8.03e-17\\
46 & 12h18m54.83s & 14d26m42.58s & 7.68e-17 & 7.52e-17 & 1.04e-16\\
47 & 12h18m53.61s & 14d26m57.57s & 2.06e-16 & 1.12e-16 & 1.38e-16\\
48 & 12h18m52.44s & 14d27m00.92s & 4.12e-16 & 2.95e-16 & 3.76e-16\\
\end{longtable}
\begin{longtable}{rllrrr}
\caption{NGC3627}\\
\toprule
 SNR No. &           RA &          Dec & F([SII]) & F([NII]) & F(H$\alpha$) \\
&&&erg/s/cm$^2$&erg/s/cm$^2$&erg/s/cm$^2$\\
\midrule
\endfirsthead
\caption[]{NGC3627} \\
\toprule
 SNR No. &           RA &          Dec & F([SII]) & F([NII]) & F(H$\alpha$) \\
&&&erg/s/cm$^2$&erg/s/cm$^2$&erg/s/cm$^2$\\
\midrule
\endhead
\midrule
\multicolumn{6}{r}{{Continued on next page}} \\
\midrule
\endfoot

\bottomrule
\endlastfoot
1 & 11h20m13.00s & 12d57m37.09s & 0.0 & 0.0 & 0.0\\
2 & 11h20m15.38s & 12d57m37.16s & 1.57e-16 & 1.09e-16 & 1.3e-16\\
3 & 11h20m15.30s & 12d57m53.59s & 2e-16 & 3.01e-16 & 2.7e-16\\
4 & 11h20m11.97s & 12d58m27.22s & 3.35e-16 & 6e-16 & 4.73e-16\\
5 & 11h20m14.52s & 12d58m56.49s & 1.76e-16 & 3.17e-16 & 2.33e-16\\
6 & 11h20m12.22s & 12d59m00.12s & 1.07e-16 & 2.55e-16 & 1.48e-16\\
7 & 11h20m17.94s & 12d59m04.84s & 8.83e-16 & 4.57e-16 & 3.11e-16\\
8 & 11h20m17.45s & 12d59m06.76s & 1.37e-16 & 3.11e-16 & 2.38e-16\\
9 & 11h20m18.82s & 12d59m24.75s & 1.72e-16 & 2.75e-16 & 2.66e-16\\
10 & 11h20m18.81s & 12d59m32.45s & 1.73e-15 & 1.3e-15 & 9.41e-16\\
11 & 11h20m13.53s & 12d59m44.92s & 2.5e-15 & 5.72e-15 & 3.62e-15\\
12 & 11h20m18.20s & 12d59m45.07s & 1.48e-15 & 1.23e-15 & 8.44e-16\\
13 & 11h20m19.14s & 12d59m47.28s & 3.97e-16 & 4.42e-16 & 7.13e-16\\
14 & 11h20m14.27s & 13d00m13.64s & 8.13e-17 & 1.38e-16 & 9.19e-17\\
15 & 11h20m17.99s & 13d00m14.07s & 2.59e-17 & 5.28e-17 & 2.64e-17\\
16 & 11h20m14.20s & 13d00m14.75s & 1.9e-16 & 5.54e-16 & 4.97e-16\\
17 & 11h20m13.40s & 13d00m26.64s & 3.74e-16 & 3.4e-16 & 6.7e-17\\
18 & 11h20m13.27s & 13d00m36.23s & 2.03e-15 & 2.17e-15 & 9.92e-16\\
\end{longtable}
\begin{longtable}{rllrrr}
\caption{NGC4303}\\
\toprule
 SNR No. &           RA &          Dec & F([SII]) & F([NII]) & F(H$\alpha$) \\
&&&erg/s/cm$^2$&erg/s/cm$^2$&erg/s/cm$^2$\\
\midrule
\endfirsthead
\caption[]{NGC4303} \\
\toprule
 SNR No. &           RA &          Dec & F([SII]) & F([NII]) & F(H$\alpha$) \\
&&&erg/s/cm$^2$&erg/s/cm$^2$&erg/s/cm$^2$\\
\midrule
\endhead
\midrule
\multicolumn{6}{r}{{Continued on next page}} \\
\midrule
\endfoot

\bottomrule
\endlastfoot
1 & 12h22m00.78s & 4d27m05.81s & 4.3e-17 & 4.85e-17 & 5.28e-17\\
2 & 12h21m52.21s & 4d27m09.89s & 2.22e-16 & 2.31e-16 & 2.5e-16\\
3 & 12h21m53.76s & 4d27m13.85s & 3.33e-16 & 4.41e-16 & 4.13e-16\\
4 & 12h21m55.59s & 4d27m15.38s & 1.12e-16 & 1.23e-16 & 1.55e-16\\
5 & 12h21m55.70s & 4d27m19.57s & 4.08e-16 & 2.66e-16 & 2.72e-16\\
6 & 12h21m56.68s & 4d27m44.37s & 1.33e-16 & 9.96e-17 & 5.21e-17\\
7 & 12h21m56.90s & 4d27m46.69s & 5.78e-16 & 5.04e-16 & 4.66e-16\\
8 & 12h21m53.53s & 4d27m46.86s & 2.38e-16 & 2.73e-16 & 1.46e-16\\
9 & 12h21m58.86s & 4d27m47.16s & 4.1e-17 & 8.38e-17 & 7.98e-17\\
10 & 12h21m52.58s & 4d27m49.16s & 4.61e-16 & 5.73e-16 & 5e-16\\
11 & 12h21m56.27s & 4d27m50.90s & 1.44e-15 & 1.21e-15 & 8.87e-16\\
12 & 12h21m53.67s & 4d27m51.48s & 7.37e-17 & 1.25e-16 & 8.99e-17\\
13 & 12h21m51.53s & 4d28m04.25s & 8.35e-17 & 1.34e-16 & 1.13e-16\\
14 & 12h21m58.19s & 4d28m05.96s & 6.73e-17 & 8.75e-17 & 7.29e-17\\
15 & 12h21m50.88s & 4d28m08.12s & 1.02e-16 & 1.85e-16 & 1.16e-16\\
16 & 12h21m51.47s & 4d28m09.64s & 1.66e-16 & 2.44e-16 & 2.07e-16\\
17 & 12h21m58.58s & 4d28m15.33s & 1.43e-16 & 1.14e-16 & 1.32e-16\\
18 & 12h21m51.00s & 4d28m17.55s & 5.43e-17 & 1.16e-16 & 1.29e-16\\
19 & 12h21m57.81s & 4d28m18.97s & 2.95e-16 & 4.54e-16 & 2.97e-16\\
20 & 12h21m50.40s & 4d28m21.08s & 6.73e-17 & 8.17e-17 & 2.83e-17\\
21 & 12h21m50.58s & 4d28m22.34s & 1.07e-16 & 1.29e-16 & 8.06e-17\\
22 & 12h21m59.47s & 4d28m27.90s & 2.74e-16 & 3.2e-16 & 2.29e-16\\
23 & 12h21m52.84s & 4d28m29.15s & 7.33e-17 & 1.65e-16 & 1.16e-16\\
24 & 12h21m51.53s & 4d28m31.23s & 7.23e-17 & 8.86e-17 & 7.49e-17\\
25 & 12h22m00.56s & 4d28m31.84s & 1.72e-16 & 1.79e-16 & 1.34e-16\\
26 & 12h21m52.43s & 4d28m35.87s & 2.76e-16 & 2.3e-16 & 1.66e-16\\
27 & 12h21m54.47s & 4d28m38.15s & 2.68e-16 & 2.32e-16 & 1.73e-16\\
28 & 12h21m49.98s & 4d28m53.70s & 8.58e-16 & 1.05e-15 & 8.23e-16\\
29 & 12h21m54.81s & 4d28m54.96s & 1.39e-16 & 3.41e-16 & 1.69e-16\\
30 & 12h21m56.18s & 4d28m55.47s & 1.05e-16 & 1.7e-16 & 1.24e-16\\
31 & 12h21m59.95s & 4d28m55.71s & 5.98e-17 & 1.43e-16 & 1.49e-16\\
32 & 12h21m55.26s & 4d29m01.06s & 4.13e-16 & 5e-16 & 3.24e-16\\
33 & 12h21m55.80s & 4d29m02.03s & 3.87e-16 & 5.46e-16 & 2.54e-16\\
34 & 12h21m54.98s & 4d29m03.58s & 1.48e-16 & 3.11e-16 & 2.56e-16\\
35 & 12h21m55.16s & 4d29m03.59s & 4.5e-16 & 5.22e-16 & 2.6e-16\\
36 & 12h21m50.95s & 4d29m07.19s & 1.04e-15 & 6.27e-16 & 1.11e-15\\
37 & 12h21m54.65s & 4d29m08.50s & 0.0 & 0.0 & 0.0\\
38 & 12h21m55.15s & 4d29m10.11s & 7.86e-18 & 4.14e-17 & 3.76e-17\\
39 & 12h21m55.15s & 4d29m10.11s & 6.24e-17 & 3.5e-17 & 3.76e-17\\
40 & 12h21m50.17s & 4d29m12.55s & 7.21e-17 & 8.68e-17 & 6.61e-17\\
41 & 12h21m56.16s & 4d29m12.62s & 3.3e-16 & 6.54e-16 & 5.56e-16\\
42 & 12h21m50.38s & 4d29m16.63s & 5.77e-17 & 5.81e-17 & 6.66e-17\\
43 & 12h21m56.88s & 4d29m20.78s & 1.13e-16 & 2.92e-16 & 2.6e-16\\
44 & 12h21m49.02s & 4d29m31.74s & 7.1e-17 & 1.16e-16 & 1.52e-16\\
45 & 12h21m52.78s & 4d29m31.74s & 1.44e-17 & 8.53e-18 & 2.08e-17\\
46 & 12h21m54.48s & 4d29m34.29s & 1.43e-17 & 1.5e-17 & 5.96e-18\\
47 & 12h21m49.17s & 4d29m34.95s & 2.61e-17 & 2.89e-17 & 2.45e-17\\
48 & 12h21m49.36s & 4d29m37.44s & 4.07e-16 & 1.7e-16 & 2.55e-16\\
49 & 12h21m58.73s & 4d29m40.18s & 1.35e-16 & 9.38e-17 & 1.36e-16\\
\end{longtable}
\begin{longtable}{rllrrr}
\caption{NGC1433}\\
\toprule
 SNR No. &           RA &          Dec & F([SII]) & F([NII]) & F(H$\alpha$) \\
&&&erg/s/cm$^2$&erg/s/cm$^2$&erg/s/cm$^2$\\
\midrule
\endfirsthead
\caption[]{NGC1433} \\
\toprule
 SNR No. &           RA &          Dec & F([SII]) & F([NII]) & F(H$\alpha$) \\
&&&erg/s/cm$^2$&erg/s/cm$^2$&erg/s/cm$^2$\\
\midrule
\endhead
\midrule
\multicolumn{6}{r}{{Continued on next page}} \\
\midrule
\endfoot

\bottomrule
\endlastfoot
1 & 3h42m05.69s & -47d14m07.77s & 7.87e-17 & 1.15e-16 & 1.03e-16\\
2 & 3h42m09.10s & -47d14m04.90s & 1.21e-16 & 9.98e-17 & 9.38e-17\\
3 & 3h41m51.60s & -47d13m17.06s & 5.81e-17 & 1.05e-16 & 1.09e-16\\
4 & 3h42m06.94s & -47d12m49.11s & 3.71e-17 & 5.01e-17 & 4.84e-17\\
5 & 3h41m59.62s & -47d12m38.95s & 1.3e-17 & 1.56e-17 & 1.58e-17\\
6 & 3h41m53.47s & -47d12m15.53s & 5.31e-17 & 5.04e-17 & 5.82e-17\\
\end{longtable}
\begin{longtable}{rllrrr}
\caption{NGC1087}\\
\toprule
 SNR No. &           RA &          Dec & F([SII]) & F([NII]) & F(H$\alpha$) \\
&&&erg/s/cm$^2$&erg/s/cm$^2$&erg/s/cm$^2$\\
\midrule
\endfirsthead
\caption[]{NGC1087} \\
\toprule
 SNR No. &           RA &          Dec & F([SII]) & F([NII]) & F(H$\alpha$) \\
&&&erg/s/cm$^2$&erg/s/cm$^2$&erg/s/cm$^2$\\
\midrule
\endhead
\midrule
\multicolumn{6}{r}{{Continued on next page}} \\
\midrule
\endfoot

\bottomrule
\endlastfoot
1 & 2h46m27.07s & -0d30m42.36s & 1.51e-16 & 6.14e-17 & 1.16e-16\\
2 & 2h46m26.44s & -0d30m03.62s & 6.19e-17 & 2.01e-17 & 7.43e-17\\
3 & 2h46m24.56s & -0d29m55.82s & 4.56e-17 & 7.67e-17 & 5.67e-17\\
4 & 2h46m25.04s & -0d29m30.19s & 6.95e-17 & 4.54e-17 & 7.88e-17\\
5 & 2h46m25.34s & -0d29m02.38s & 7.76e-17 & 2.73e-17 & 4.43e-17\\
\end{longtable}
\begin{longtable}{rllrrr}
\caption{NGC0628}\\
\toprule
 SNR No. &           RA &          Dec & F([SII]) & F([NII]) & F(H$\alpha$) \\
&&&erg/s/cm$^2$&erg/s/cm$^2$&erg/s/cm$^2$\\
\midrule
\endfirsthead
\caption[]{NGC0628} \\
\toprule
 SNR No. &           RA &          Dec & F([SII]) & F([NII]) & F(H$\alpha$) \\
&&&erg/s/cm$^2$&erg/s/cm$^2$&erg/s/cm$^2$\\
\midrule
\endhead
\midrule
\multicolumn{6}{r}{{Continued on next page}} \\
\midrule
\endfoot

\bottomrule
\endlastfoot
1 & 1h36m42.74s & 15d45m05.80s & 1.07e-16 & 9.62e-17 & 1.38e-16\\
2 & 1h36m42.52s & 15d45m34.14s & 6.16e-17 & 7.76e-17 & 7.1e-17\\
3 & 1h36m40.53s & 15d46m09.44s & 2.44e-17 & 3.62e-17 & 3.03e-17\\
4 & 1h36m39.69s & 15d46m18.33s & 1.19e-16 & 1.6e-16 & 1.51e-16\\
5 & 1h36m40.08s & 15d46m25.80s & 4.7e-16 & 7.49e-16 & 5.94e-16\\
6 & 1h36m44.35s & 15d46m30.23s & 7.13e-17 & 1.33e-16 & 6.98e-17\\
7 & 1h36m47.31s & 15d46m36.17s & 9.01e-16 & 5.78e-16 & 7.53e-16\\
8 & 1h36m35.29s & 15d46m39.44s & 6.77e-17 & 6.4e-17 & 6.32e-17\\
9 & 1h36m43.76s & 15d46m42.65s & 3.37e-17 & 5.31e-17 & 4.92e-17\\
10 & 1h36m38.46s & 15d47m01.04s & 2.08e-17 & 3.35e-17 & 5.07e-17\\
11 & 1h36m40.85s & 15d47m02.02s & 5.27e-16 & 1.39e-15 & 9.82e-16\\
12 & 1h36m47.78s & 15d47m05.16s & 5.4e-17 & 1.56e-16 & 6.31e-17\\
13 & 1h36m41.99s & 15d47m12.24s & 2.75e-16 & 3.68e-16 & 2.95e-16\\
14 & 1h36m33.38s & 15d47m13.80s & 2.35e-17 & 2.28e-17 & 2.07e-17\\
15 & 1h36m32.83s & 15d47m19.86s & 1.58e-17 & 2.74e-17 & 3.72e-17\\
16 & 1h36m43.62s & 15d47m20.43s & 1.91e-16 & 3.02e-16 & 2.42e-16\\
17 & 1h36m43.70s & 15d47m36.62s & 1.07e-16 & 1.62e-16 & 1.44e-16\\
18 & 1h36m43.87s & 15d47m57.74s & 7.82e-17 & 2.03e-16 & 1.65e-16\\
19 & 1h36m36.58s & 15d48m05.30s & 1.8e-16 & 1.05e-16 & 1.28e-16\\
20 & 1h36m37.64s & 15d48m09.47s & 3.51e-17 & 3.15e-17 & 2.41e-17\\
21 & 1h36m37.71s & 15d48m19.45s & 4.05e-16 & 2.66e-16 & 3.46e-16\\
22 & 1h36m41.14s & 15d48m46.90s & 3.15e-16 & 1.82e-16 & 3.15e-16\\
\end{longtable}
\begin{longtable}{rllrrr}
\caption{NGC1300}\\
\toprule
 SNR No. &           RA &          Dec & F([SII]) & F([NII]) & F(H$\alpha$) \\
&&&erg/s/cm$^2$&erg/s/cm$^2$&erg/s/cm$^2$\\
\midrule
\endfirsthead
\caption[]{NGC1300} \\
\toprule
 SNR No. &           RA &          Dec & F([SII]) & F([NII]) & F(H$\alpha$) \\
&&&erg/s/cm$^2$&erg/s/cm$^2$&erg/s/cm$^2$\\
\midrule
\endhead
\midrule
\multicolumn{6}{r}{{Continued on next page}} \\
\midrule
\endfoot

\bottomrule
\endlastfoot
1 & 3h19m37.80s & -19d25m48.73s & 1e-16 & 9.91e-17 & 1.35e-16\\
2 & 3h19m35.86s & -19d24m35.83s & 1.7e-16 & 1.98e-16 & 2.13e-16\\
3 & 3h19m46.75s & -19d24m04.20s & 5.33e-17 & 6.89e-17 & 9.03e-17\\
4 & 3h19m42.67s & -19d23m56.45s & 2.14e-16 & 2.18e-16 & 1.51e-16\\
\end{longtable}
\begin{longtable}{rllrrr}
\caption{NGC1512}\\
\toprule
 SNR No. &           RA &          Dec & F([SII]) & F([NII]) & F(H$\alpha$) \\
&&&erg/s/cm$^2$&erg/s/cm$^2$&erg/s/cm$^2$\\
\midrule
\endfirsthead
\caption[]{NGC1512} \\
\toprule
 SNR No. &           RA &          Dec & F([SII]) & F([NII]) & F(H$\alpha$) \\
&&&erg/s/cm$^2$&erg/s/cm$^2$&erg/s/cm$^2$\\
\midrule
\endhead
\midrule
\multicolumn{6}{r}{{Continued on next page}} \\
\midrule
\endfoot

\bottomrule
\endlastfoot
1 & 4h03m59.55s & -43d21m42.06s & 5.49e-17 & 3.74e-17 & 4.24e-17\\
2 & 4h03m57.76s & -43d21m17.57s & 4.08e-16 & 2.62e-16 & 2.09e-16\\
3 & 4h03m53.89s & -43d20m49.27s & 2.48e-16 & 5.41e-16 & 2.51e-16\\
4 & 4h03m46.93s & -43d20m31.97s & 3.13e-17 & 3.25e-17 & 4.04e-17\\
5 & 4h03m58.64s & -43d20m10.88s & 2.59e-17 & 4.44e-17 & 4.03e-17\\
\end{longtable}
\begin{longtable}{rllrrr}
\caption{NGC7496}\\
\toprule
 SNR No. &           RA &          Dec & F([SII]) & F([NII]) & F(H$\alpha$) \\
&&&erg/s/cm$^2$&erg/s/cm$^2$&erg/s/cm$^2$\\
\midrule
\endfirsthead
\caption[]{NGC7496} \\
\toprule
 SNR No. &           RA &          Dec & F([SII]) & F([NII]) & F(H$\alpha$) \\
&&&erg/s/cm$^2$&erg/s/cm$^2$&erg/s/cm$^2$\\
\midrule
\endhead
\midrule
\multicolumn{6}{r}{{Continued on next page}} \\
\midrule
\endfoot

\bottomrule
\endlastfoot
1 & 23h09m48.53s & -43d25m57.04s & 1.46e-16 & 2.12e-16 & 2.15e-16\\
2 & 23h09m48.32s & -43d25m41.51s & 1.7e-17 & 1.7e-17 & 1.77e-17\\
3 & 23h09m48.06s & -43d25m11.27s & 1.69e-16 & 1.54e-16 & 1.02e-16\\
4 & 23h09m45.95s & -43d24m56.15s & 1.45e-15 & 1.21e-15 & 2.22e-15\\
\end{longtable}

\clearpage
\section*{Appendix C}
This appendix lists all of the data used from HST, JWST and Chandra. The tables list all filters used, observation dates, exposure times, proposal ID of the programs, their PI's and the instruments used.

\begin{table}[hp]
    \caption{The observations from HST, JWST, and Chandra used in this article. The tables list the filter  and instrument used, observation date, total exposure time, Proposal ID and the PI of the proposal they originate from}
    \centering
    \begin{tabular}{l|c|c|c|c|c}
    Filter	&	Date	&	Exposure (s)		&	Proposal ID	&	PI & Instrument \\
    \hline
    NGC 4303 & & & & & \\
    \hline
    f814W&	2020-03-29T15:57:02&	803&	15654&	Lee, Janice&    HST/WFC3 \\
	f438W&	2020-03-29T16:03:55&	1050&	15654&	Lee, Janice&    HST/WFC3 \\
	f336W&	2020-03-29T16:12:22&	1110&	15654&	Lee, Janice&    HST/WFC3 \\
	f275W&	2020-03-29T16:21:19&	2190&	15654&	Lee, Janice&    HST/WFC3 \\
	f555W&	2020-03-29T16:36:08&	670&	15654&	Lee, Janice&    HST/WFC3 \\
	f555W&	2021-07-28T11:44:28&	710&	16179&	Filippenko, Alex V.&    HST/WFC3 \\
	f814W&	2021-07-28T11:35:35&	780&	16179&	Filippenko, Alex V.&    HST/WFC3 \\
    f1000W&	2023-01-31T13:37:17&	610.5&	2107&	Lee, Janice&	JWST/MIRI \\
	f1130W&	2023-01-31T13:46:08&	1554.02&	2107&	Lee, Janice	&	JWST/MIRI \\
	f2100W&	2023-01-31T13:58:53&	1609.52&	2107&	Lee, Janice	&	JWST/MIRI \\
	f770W&	2023-01-31T13:28:02&	444&	2107&	Lee, Janice	&	JWST/MIRI \\
	f200W&	2023-01-31T18:13:01&	1202.52&	2107&	Lee, Janice	&	JWST/NIRCam \\
	f300M&	2023-01-31T18:38:36&	386.524&	2107&	Lee, Janice	&	JWST/NIRCam \\
	f335M&	2023-01-31T18:54:32&	386.524&	2107&	Lee, Janice	&	JWST/NIRCam \\
	f360M&	2023-01-31T18:13:01&	429.472&	2107&	Lee, Janice	&	JWST/NIRCam \\
    &2001-08-07T22:05:00&   28.02k&		2700561&	Santos Lleo&	Chandra/ACIS-S \\
	&2011-11-25T22:04:00&  2.92k&	13620028&	Kaastra&	Chandra/ACIS-S \\
	&2014-11-16T15:07:00&  19.8k&  15508486&	Margutti&	Chandra/ACIS-S \\
	&2016-01-20T23:02:00&  3.37k&  17700036&	Brandt&	Chandra/ACIS-S \\
	&2021-02-17T11:48:00&  7.76k&	22700333&	Baldassare&	Chandra/ACIS-S \\
    \end{tabular}
    \label{tab:observations}
\end{table}

\begin{table}[p]
    \caption*{\textit{Table \ref{tab:observations} Continued}}
    \centering
    \begin{tabular}{l|c|c|c|c|c}
    Filter	&	Date	&	Exposure (s)		&	Proposal ID	&	PI & Instrument \\
    \hline
    NGC 3627 & & & & & \\
    \hline
    f814W&	2013-11-28T22:33:11&	1119&	13477&	Kochanek, Chris S.&    HST/WFC3 \\
	f555W&	2013-11-28T22:07:51&	1119&	13477&	Kochanek, Chris S.&    HST/WFC3 \\
	f275W&	2014-02-08T14:13:21&	2361&	13364&	Calzetti, D&    HST/WFC3 \\
	f336W&	2014-02-08T15:43:35&	1110&	13364&	Calzetti, D&    HST/WFC3 \\
	f438W&	2014-02-08T16:09:13&	956&	13364&	Calzetti, D&    HST/WFC3 \\
	f555W&	2014-02-08T17:19:12&	1134&	13364&	Calzetti, D&    HST/WFC3 \\
	f814W&	2014-02-08T17:44:47&	980&	13364&	Calzetti, D&    HST/WFC3 \\
	f814W&	2019-11-11T15:14:34&	830&	15654&	Lee, Janice&    HST/WFC3 \\
	f438W&	2019-11-11T15:21:36&	1050&	15654&	Lee, Janice&    HST/WFC3 \\
	f336W&	2019-11-11T15:30:03&	1110&	15654&	Lee, Janice&    HST/WFC3 \\
	f275W&	2019-11-11T16:43:22&	2920&	15654&	Lee, Janice&    HST/WFC3 \\
	f555W&	2019-11-11T16:58:11&	910&	15654&	Lee, Janice&    HST/WFC3 \\
    f770W&	2023-01-13T11:42:47&	266.4&	2107&	Lee, Janice	&	JWST/MIRI \\
	f1000W&	2023-01-13T11:52:07&	366.3&	2107&	Lee, Janice	&	JWST/MIRI \\
	f1130W&	2023-01-13T12:01:00&	932.412&  2107&	Lee, Janice	&	JWST/MIRI \\
	f2100W&	2023-01-13T12:13:32&	965.712&	2107&	Lee, Janice	&	JWST/MIRI \\
	f200W&	2023-01-13T14:25:07&	2405.04&	2107&	Lee, Janice	&	JWST/NIRCam \\
	f300M&	2023-01-13T14:50:32&	773.048&	2107&	Lee, Janice	&	JWST/NIRCam \\
	f335M&	2023-01-13T15:06:28&	773.048&	2107&	Lee, Janice	&	JWST/NIRCam \\
	f360M&	2023-01-13T14:25:08&	858.944&	2107&	Lee, Janice	&	JWST/NIRCam \\
    &1999-11-03T09:07:00&   1.75k&	1700068&	Garmire&	Chandra/ACIS-S \\
	&2008-03-31T14:35:00&  49.54k&	9620513&	Jenkins&	Chandra/ACIS-S \\
    \end{tabular}
\end{table}

\begin{table}[p]
    \caption*{\textit{Table \ref{tab:observations} Continued}}
    \centering
    \begin{tabular}{l|c|c|c|c|c}
    Filter	&	Date	&	Exposure (s)		&	Proposal ID	&	PI & Instrument \\
    \hline
    NGC 3351 & & & & & \\
    \hline
    f275W&	2014-04-23T03:47:31&	2361&	13364&	Calzetti, Daniela&    HST/WFC3 \\
	f336W&	2014-04-23T05:42:15&	1062&	13364&	Calzetti, Daniela&    HST/WFC3 \\
	f438W&	2014-04-23T07:04:47&	908&	13364&	Calzetti, Daniela&    HST/WFC3 \\
	f555W&	2014-04-23T08:31:50&	1062&	13364&	Calzetti, Daniela&    HST/WFC3 \\
	f814W&	2014-04-23T08:56:13&	908&	13364&	Calzetti, Daniela&    HST/WFC3 \\
	f547M&	2015-03-03T16:36:12&	550&	13773&	Chandar, Rupali&    HST/WFC3 \\
	f657N&	2015-03-03T16:03:19&	2056&	13773&	Chandar, Rupali&    HST/WFC3 \\
	f275W&	2019-05-22T04:02:31&	2190&	15654&	Lee, Janice&    HST/WFC3 \\
	f336W&	2019-05-22T02:43:20&	1110&	15654&	Lee, Janice&    HST/WFC3 \\
	f438W&	2019-05-22T02:34:53&	1050&	15654&	Lee, Janice&    HST/WFC3 \\
	f555W&	2019-05-22T04:17:20&	670&	15654&	Lee, Janice&    HST/WFC3 \\
	f814W&	2019-05-22T02:27:51&	830&	15654&	Lee, Janice&    HST/WFC3 \\
	f658N&	2023-05-23T14:02:45&	2344&	17126&	Chandar, Rupali&    HST/WFC3 \\
    f770W&	2023-05-18T14:07:16&	532.8&	2107&	Lee, Janice	&	JWST/MIRI \\
	f1000W&	2023-05-18T14:16:36&	732.6&	2107&	Lee, Janice	&	JWST/MIRI \\
	f1130W&	2023-05-18T14:25:29&	1864.824&	2107&	Lee, Janice	&	JWST/MIRI \\
	f2100W&	2023-05-18T14:38:01&	1931.424&	2107&	Lee, Janice	&	JWST/MIRI \\
	f200W&	2023-05-18T17:46:36&	2405.04&	2107&	Lee, Janice	&	JWST/NIRCam \\
	f360M&	2023-05-18T17:46:36&	858.944&	2107&	Lee, Janice	&	JWST/NIRCam \\
	f300M&	2023-05-18T18:12:12&	773.048&	2107&	Lee, Janice	&	JWST/NIRCam \\
	f335M&	2023-05-18T18:27:57&	773.048&	2107&	Lee, Janice	&	JWST/NIRCam \\
    &2005-02-08T04:40:00&	39.45k&	6620375&	Swartz&	Chandra/ACIS-S	 \\
	&2005-05-08T13:26:00&	39.45k&	6620375&	Swartz&	Chandra/ACIS-S	 \\
	&2005-07-01T05:09:00&	39.55k&	6620375&	Swartz&	Chandra/ACIS-S	 \\
	&2012-04-11T05:21:00&	9.84k&	13500310&	Pooley&	Chandra/ACIS-S	 \\
	&2018-03-26T16:31:00&	9.96k&	19508623&	Patnaude&	Chandra/ACIS-S	 \\
    \end{tabular}
\end{table}

\begin{table}[p]
    \caption*{\textit{Table \ref{tab:observations} Continued}}
    \centering
    \begin{tabular}{l|c|c|c|c|c}
    Filter	&	Date	&	Exposure (s)		&	Proposal ID	&	PI & Instrument \\
    \hline
    NGC 1566 & & & & & \\
    \hline
    f275W&	2013-09-03T16:02:52&	2382&	13364&	Calzetti, Daniela&    HST/WFC3 \\
	f336W&	2013-09-03T17:32:36&	1119&	13364&	Calzetti, Daniela&    HST/WFC3 \\
	f438W&	2013-09-03T17:58:23&	965&	13364&	Calzetti, Daniela&    HST/WFC3 \\
	f555W&	2013-09-03T19:08:14&	1143&	13364&	Calzetti, Daniela&    HST/WFC3 \\
	f814W&	2013-09-03T19:33:58&	989&	13364&	Calzetti, Daniela&    HST/WFC3 \\
	f547M&	2015-07-06T22:34:03&	2420&	13816&	Bentz, Misty C.&    HST/WFC3 \\
	f555W&	2017-08-02T10:04:02&	710&	14668&	Filippenko, Alex V.&    HST/WFC3 \\
	f814W&	2017-08-02T09:55:12&	780&	14668&	Filippenko, Alex V.&    HST/WFC3 \\
	f658N&	2023-08-10T22:26:31&	2528&	17126&	Chandar, Rupali&    HST/WFC3 \\
    f770W&	2022-11-22T09:40:28&	532.8&	2107&	Lee, Janice	&	JWST/MIRI \\
	f1000W&	2022-11-22T09:49:49&	732.6&	2107&	Lee, Janice	&	JWST/MIRI \\
	f1130W&	2022-11-22T09:58:36&	1243.216&	2107&	Lee, Janice	&	JWST/MIRI \\
	f2100W&	2022-11-22T10:11:06&	1931.424&	2107&	Lee, Janice	&	JWST/MIRI \\
	f200W&	2022-11-22T13:28:51&	2405.04&	2107&	Lee, Janice	&	JWST/NIRCam \\
	f360M&	2022-11-22T13:28:51&	858.944&	2107&	Lee, Janice	&	JWST/NIRCam \\
	f300M&	2022-11-22T13:54:26&	773.048&	2107&	Lee, Janice	&	JWST/NIRCam \\
	f335M&	2022-11-22T14:10:22&	773.048&	2107&	Lee, Janice	&	JWST/NIRCam \\
    &2018-12-03T12:36:00&	3.04k&	20700424&	Baldassare&	Chandra/ACIS-S \\
    \end{tabular}
\end{table}

\clearpage
\section*{Appendix D}
The appendix presents the velocity space comparisons of oxygen emission lines of six O-rich SNRs. SNR 4303-46 is presented in the main text, Fig. \ref{sourcea_etc}
\twocolumn
   \begin{figure}[h]
   \centering
   \includegraphics[width=\hsize]{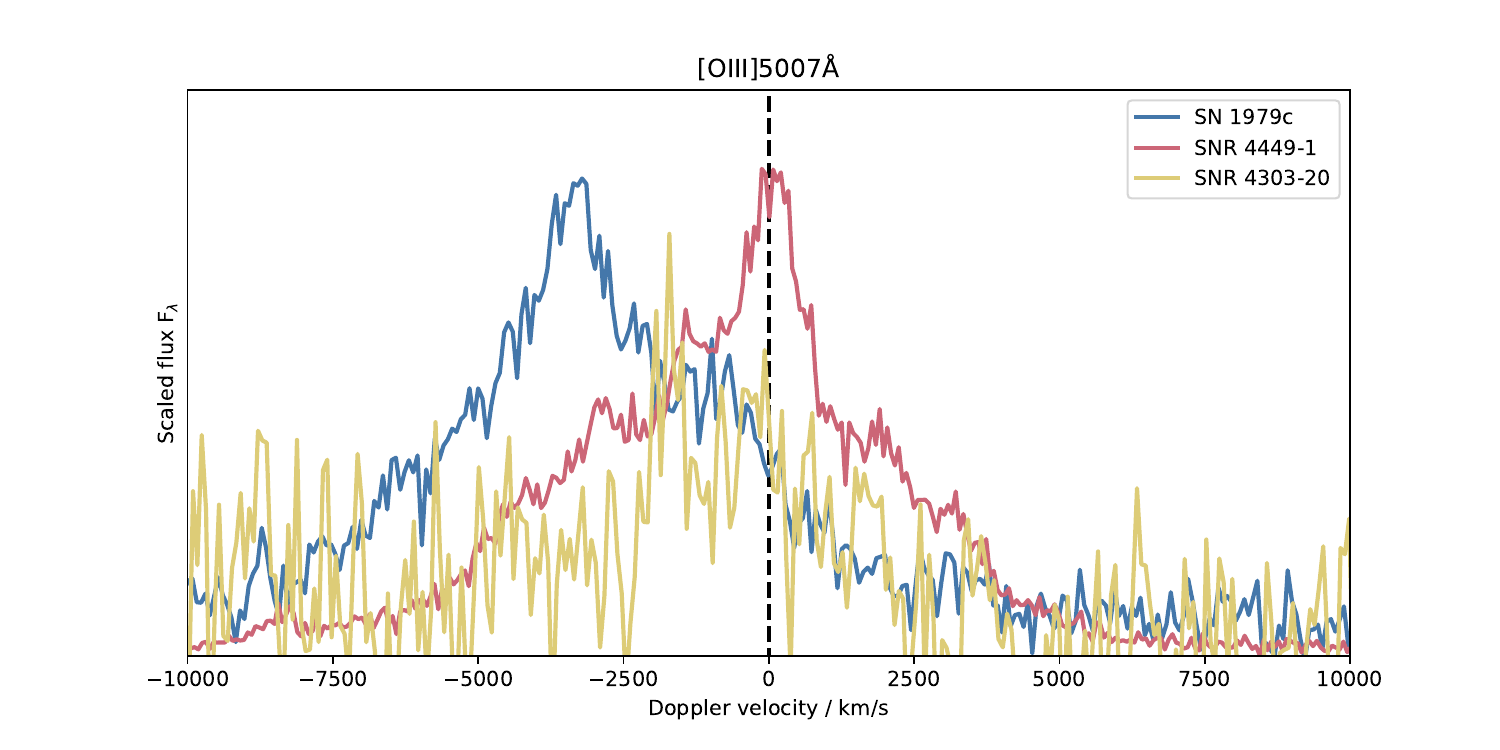}
   \includegraphics[width=\hsize]{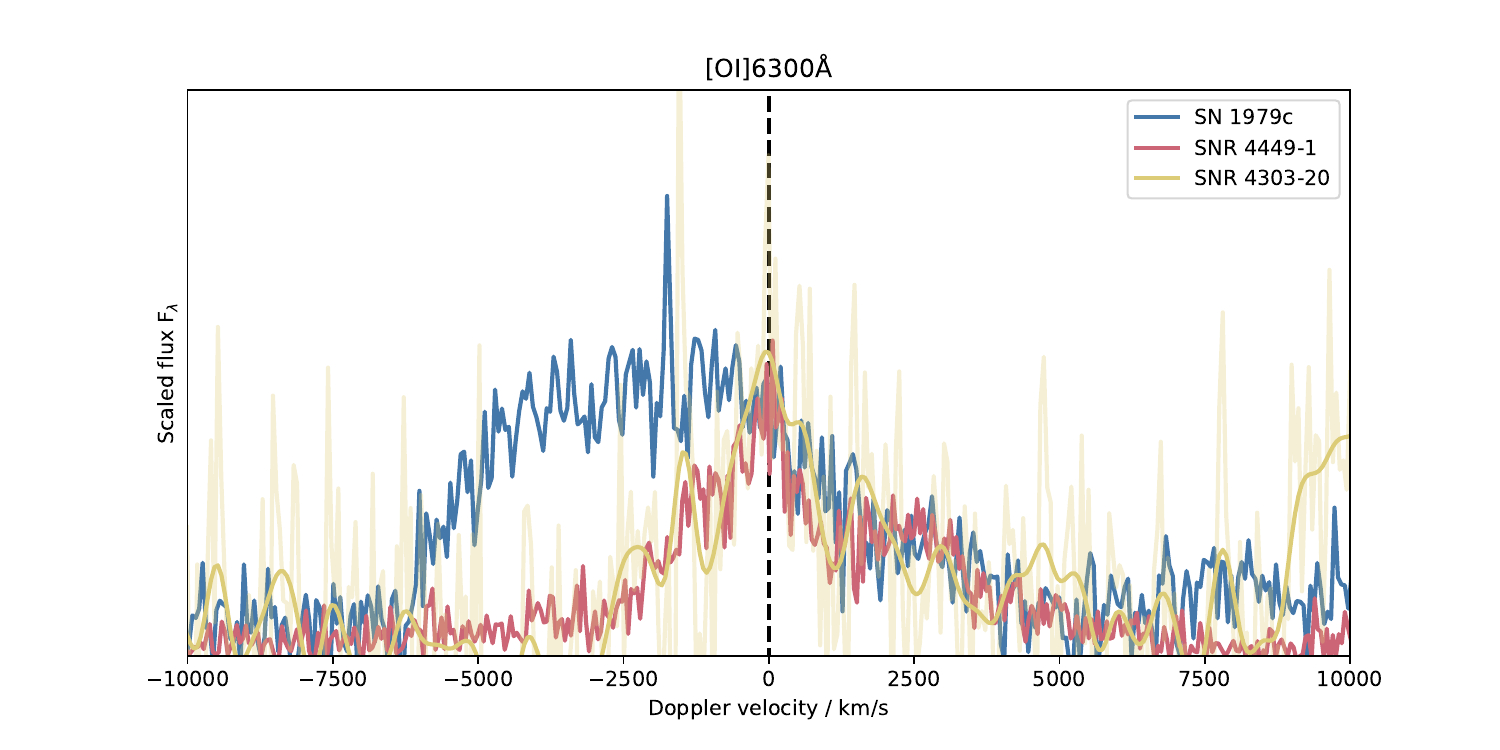}
   \includegraphics[width=\hsize]{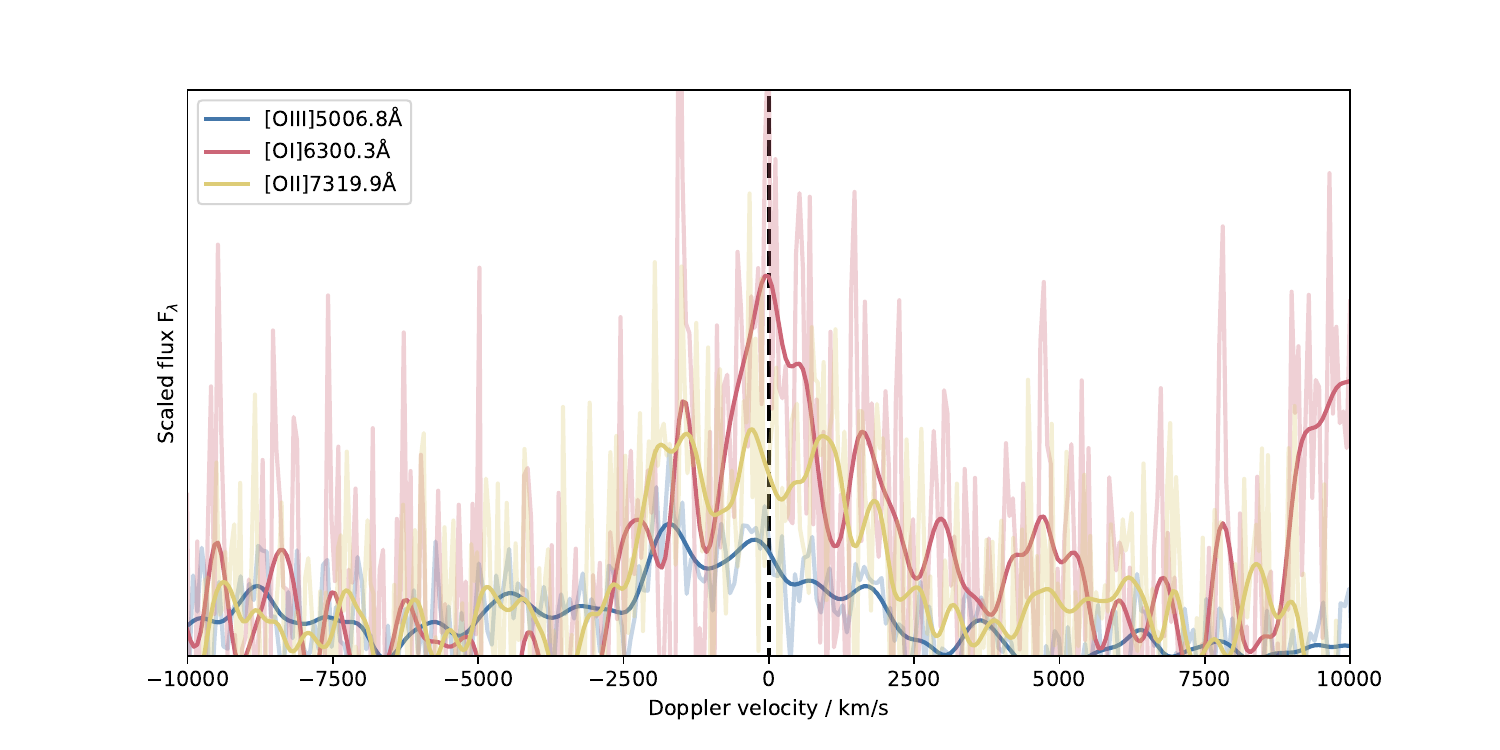}
      \caption{[O\,III]$\lambda\lambda$4959,5007 emission complex in velocity space centered around 5006.8Å (top) and same for [O\,I]$\lambda\lambda$6300,6364 centered around 6300Å (second panel). The visible light oxygen lines of SNR 4303-20 in velocity space centered on 5006.8\,\,Å, 6300.3\,\,Å and 7319.9\,\,Å (third panel). The spectra are normalized arbitrarily to have similar peak intensities.}
         \label{sourceb_etc}
   \end{figure}

   \begin{figure}[h]
   \centering
   \includegraphics[width=\hsize]{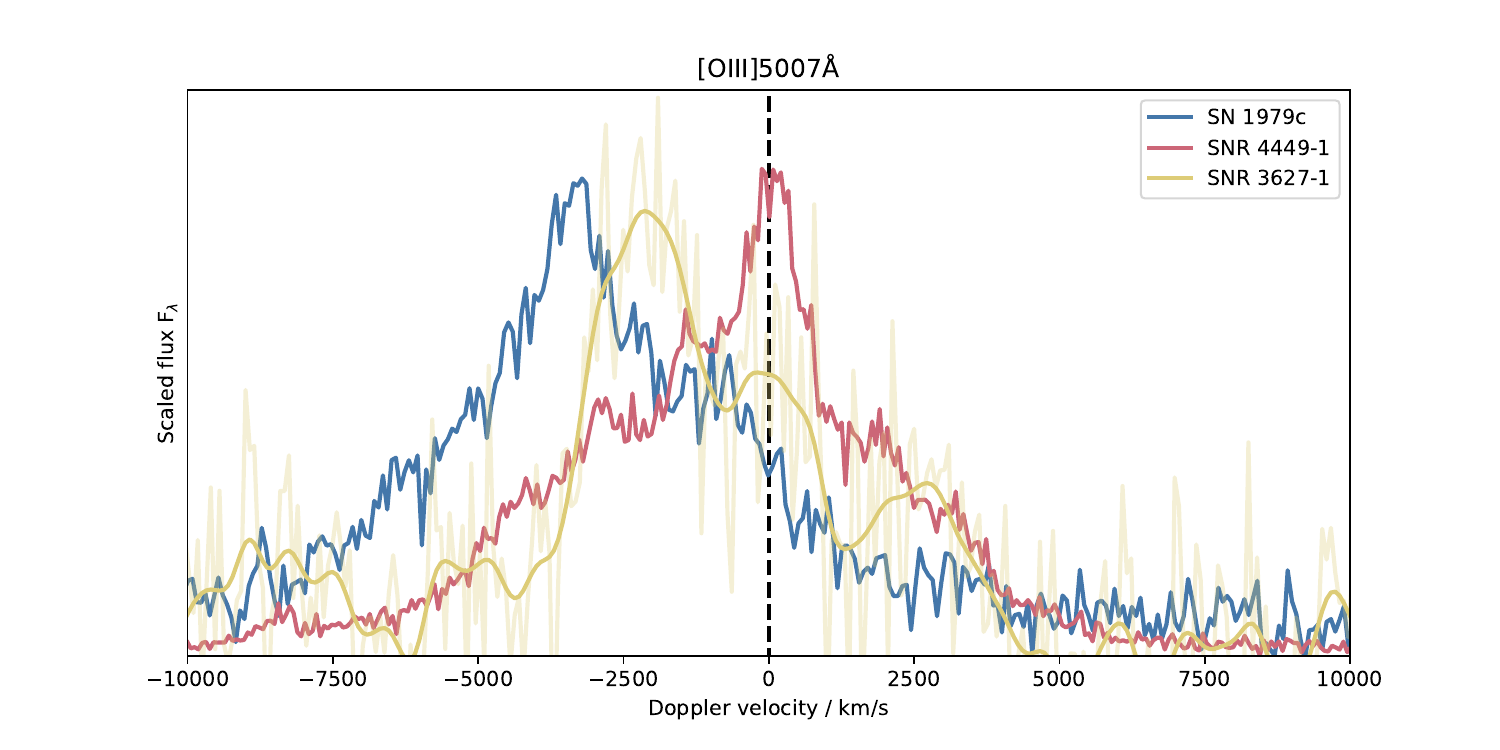}
   \includegraphics[width=\hsize]{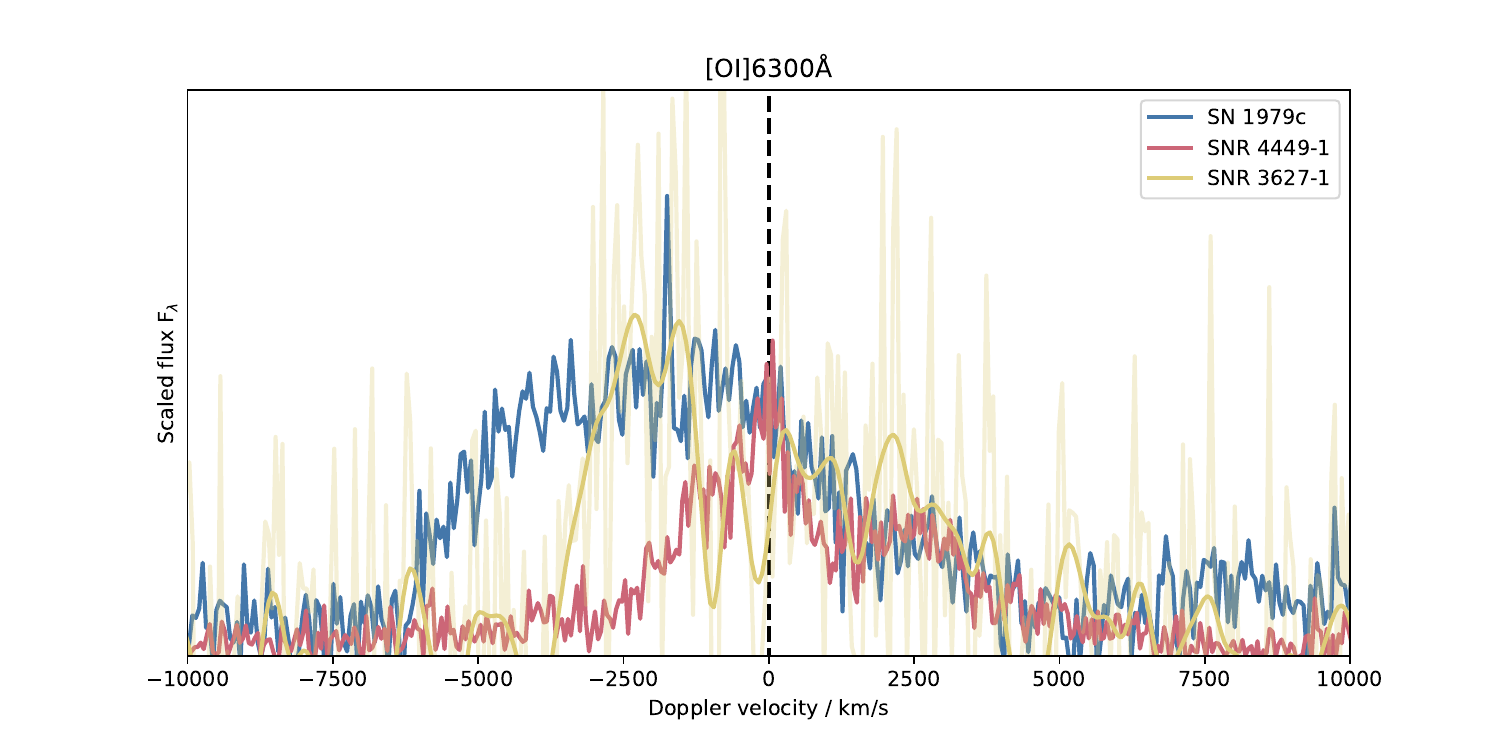}
   \includegraphics[width=\hsize]{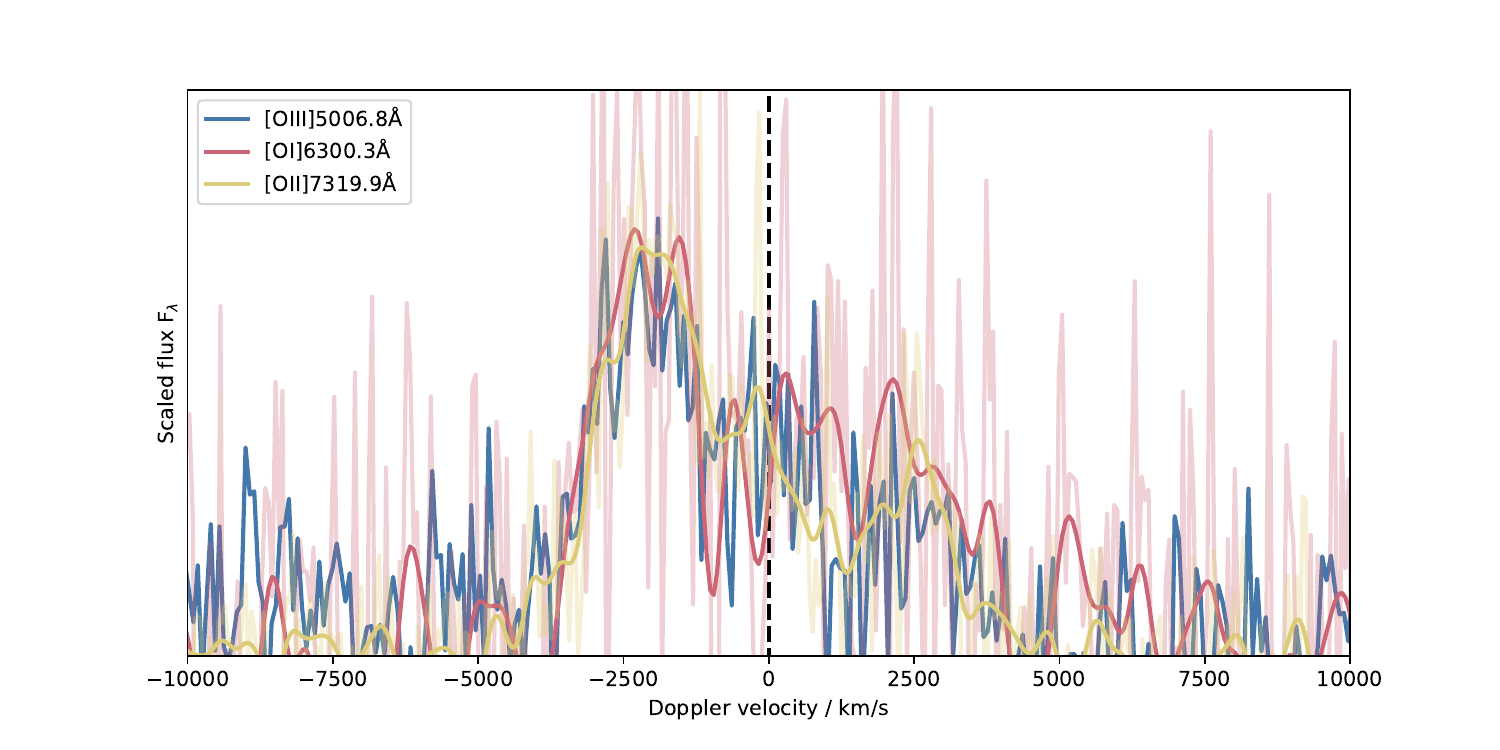}
      \caption{[O\,III]$\lambda\lambda$4959,5007 emission complex in velocity space centered around 5006.8Å (top) and same for [O\,I]$\lambda\lambda$6300,6364 centered around 6300Å (second panel). The visible light oxygen lines of SNR 3627-1 in velocity space centered on 5006.8\,\,Å, 6300.3\,\,Å and 7319.9\,\,Å (third panel). The spectra are normalized arbitrarily to have similar peak intensities.}
         \label{sourcec_etc}
   \end{figure}

    \begin{figure}[h]
   \centering
   \includegraphics[width=\hsize]{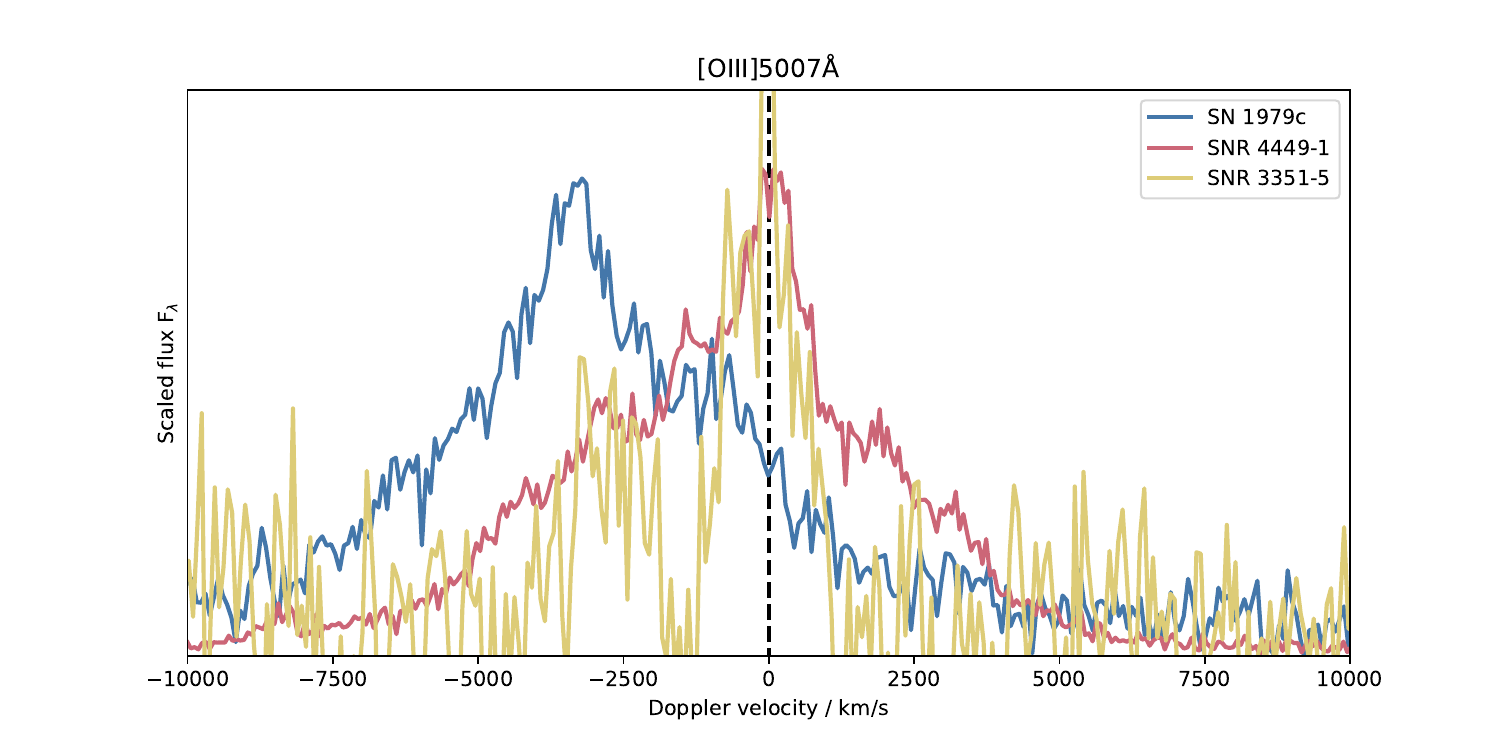}
   \includegraphics[width=\hsize]{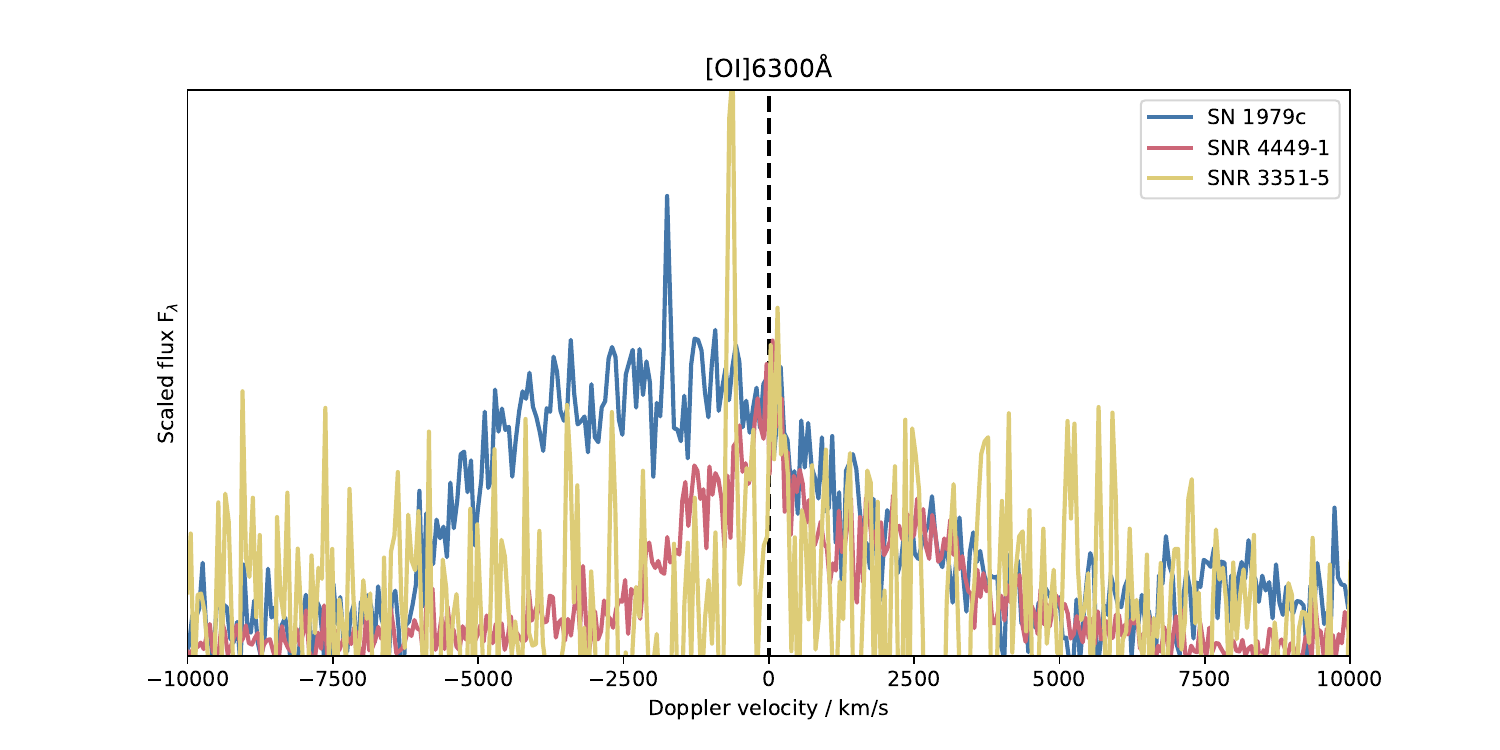}
   \includegraphics[width=\hsize]{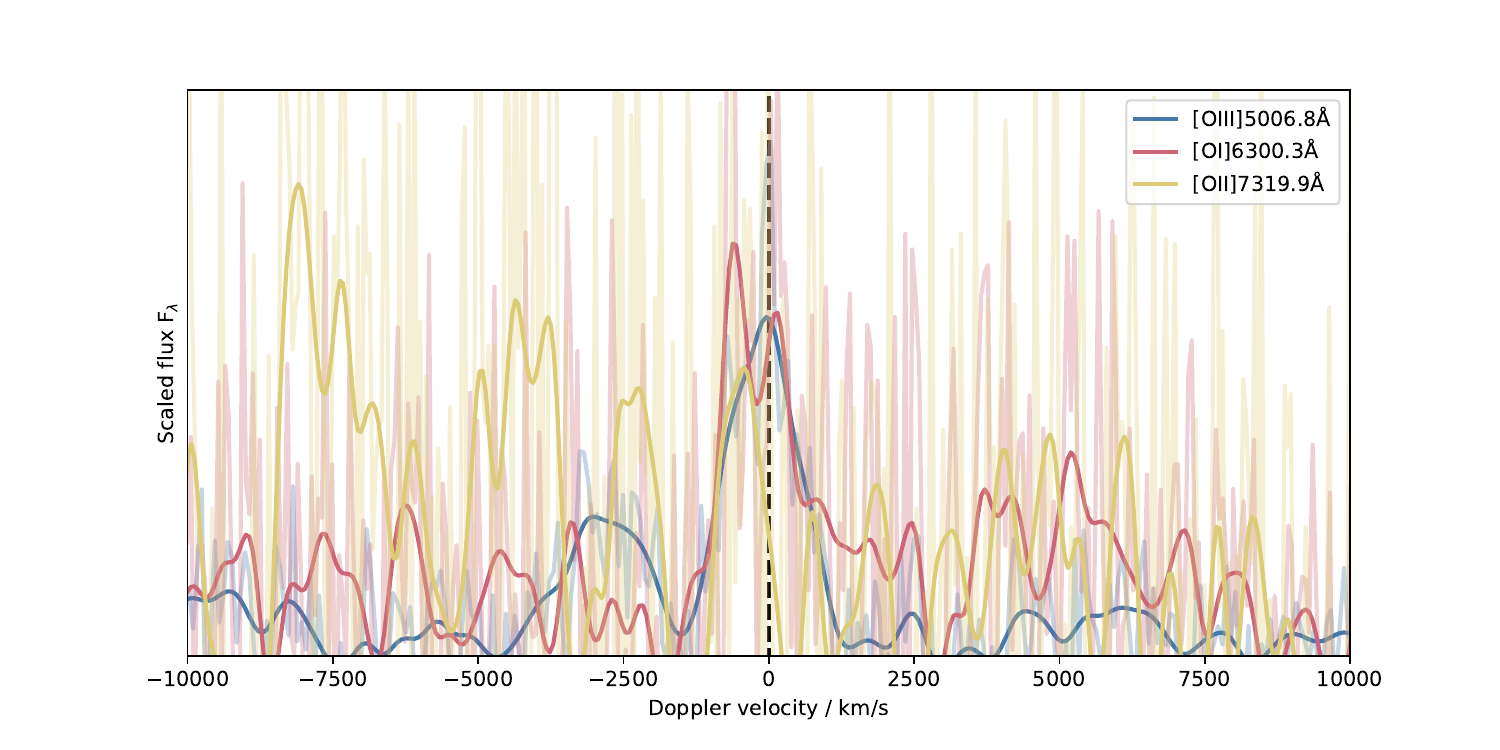}
      \caption{[O\,III]$\lambda\lambda$4959,5007 emission complex in velocity space centered around 5006.8Å (top) and same for [O\,I]$\lambda\lambda$6300,6364 centered around 6300Å (second panel). The visible light oxygen lines of SNR 3351-5 in velocity space centered on 5006.8\,\,Å, 6300.3\,\,Å and 7319.9\,\,Å (third panel). The spectra are normalized arbitrarily to have similar peak intensities.}
         \label{sourced_etc}
   \end{figure}

   \begin{figure}[h]
   \centering
   \includegraphics[width=\hsize]{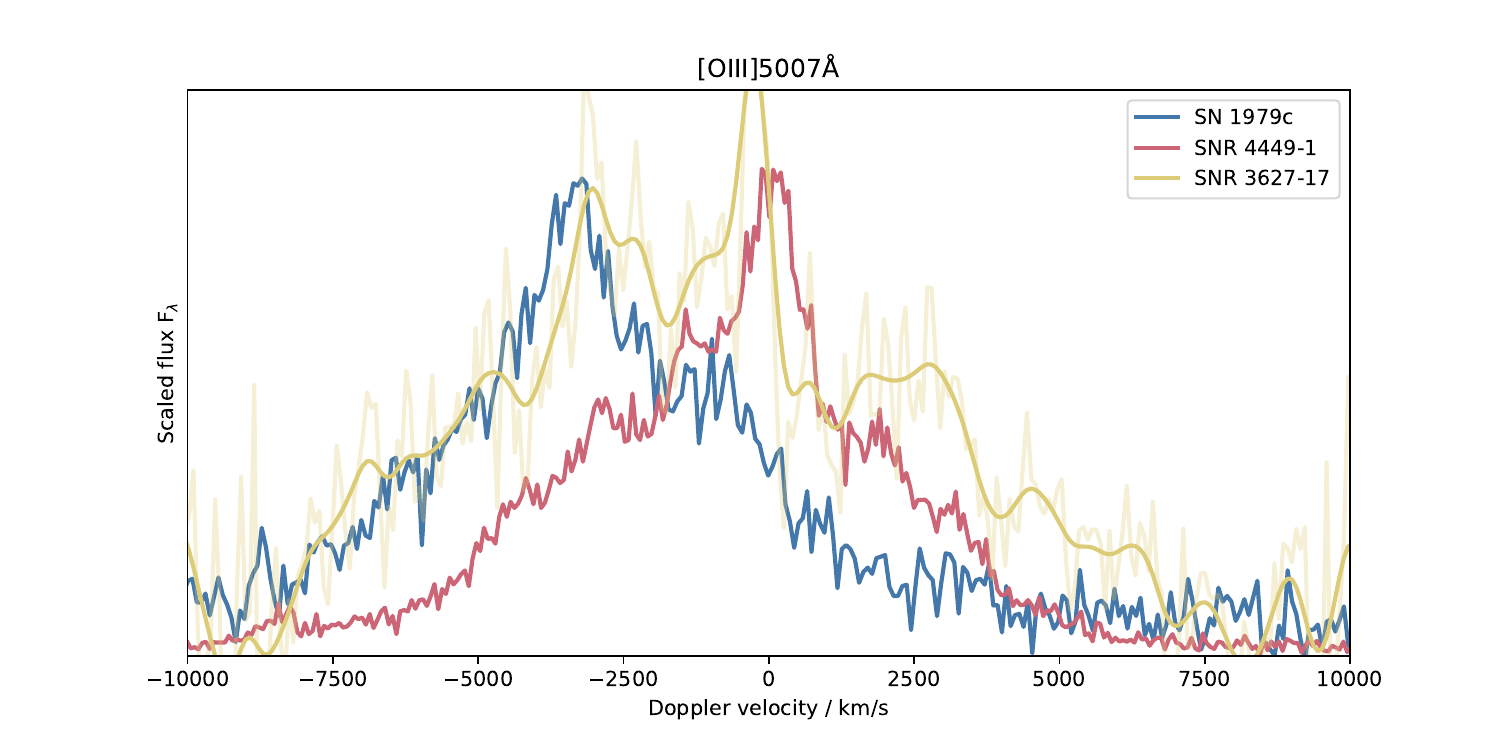}
   \includegraphics[width=\hsize]{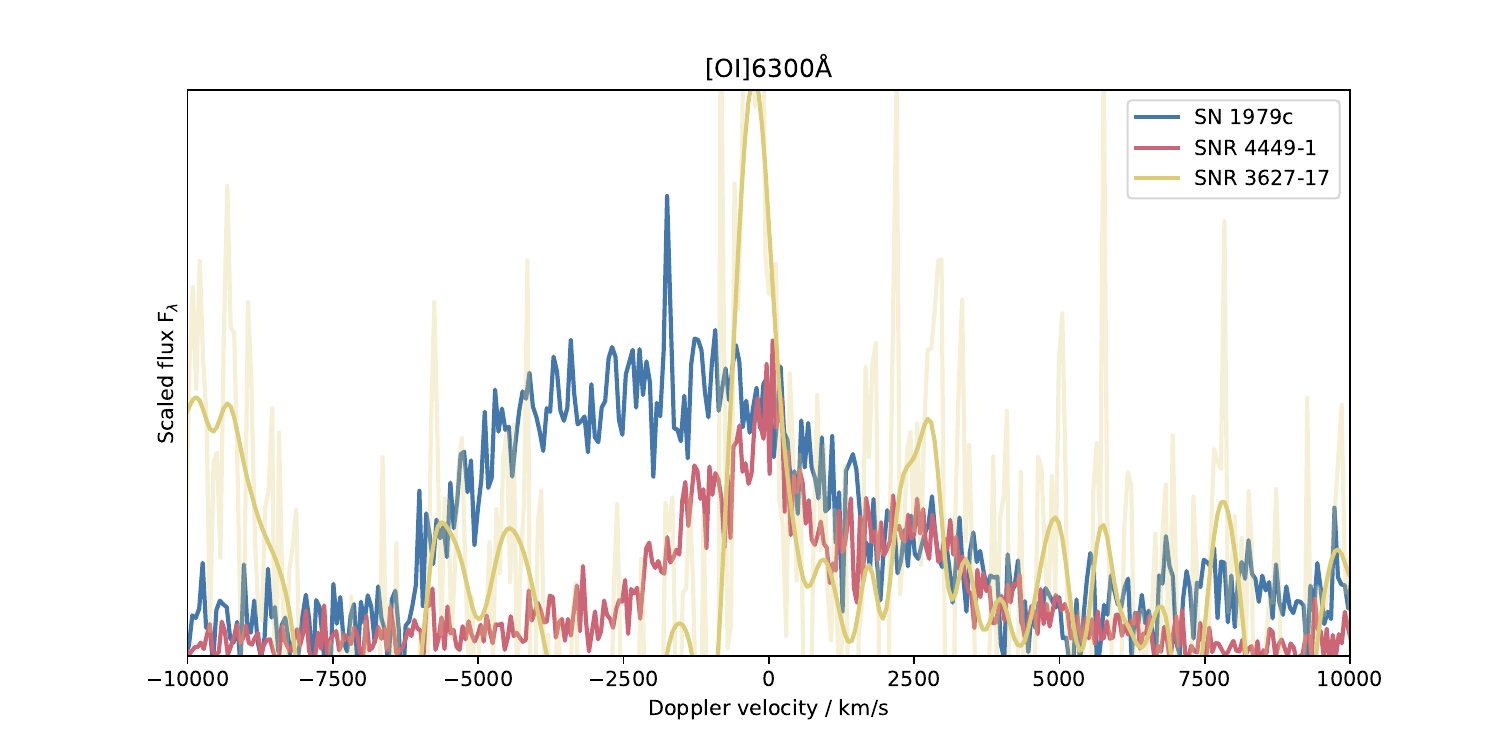}
   \includegraphics[width=\hsize]{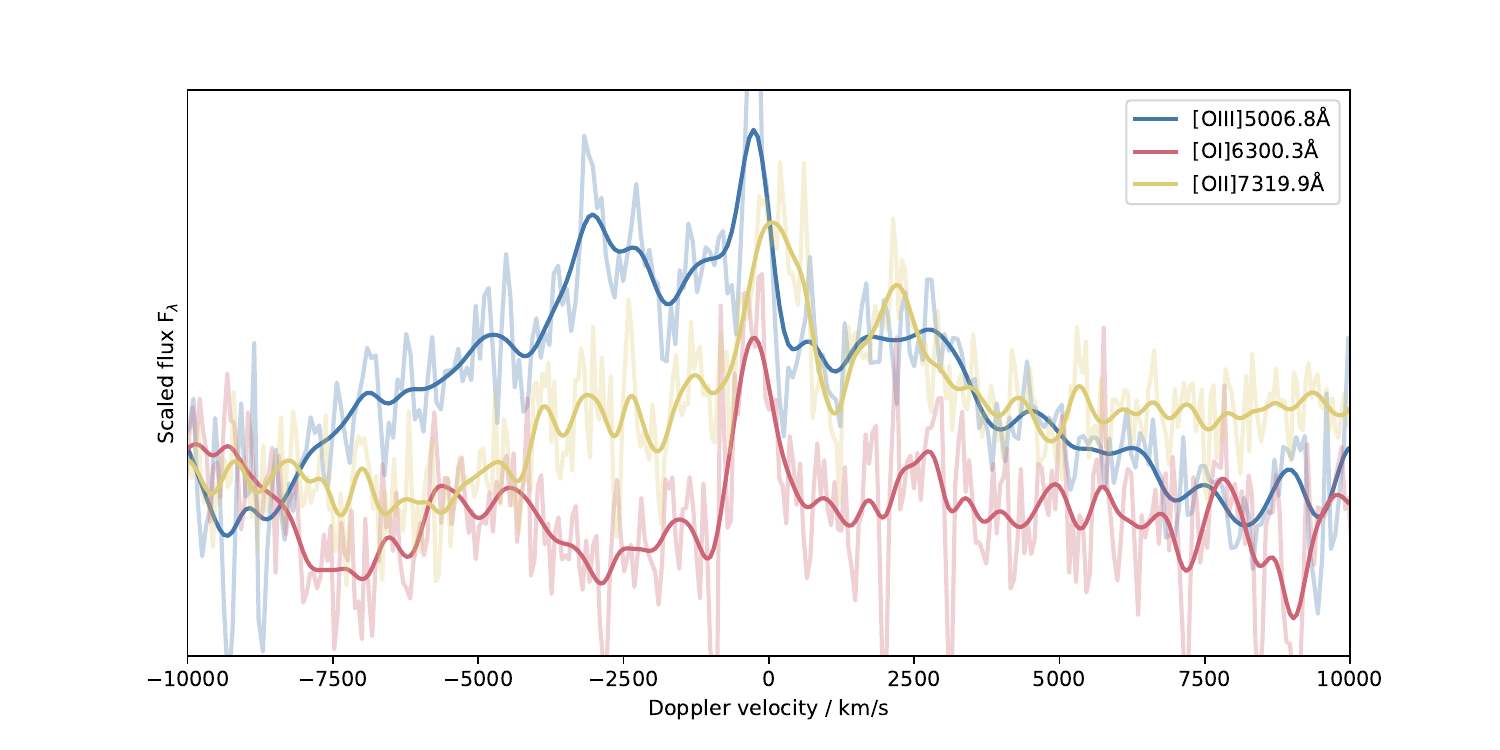}
      \caption{[O\,III]$\lambda\lambda$4959,5007 emission complex in velocity space centered around 5006.8Å (top) and same for [O\,I]$\lambda\lambda$6300,6364 centered around 6300Å (second panel). The visible light oxygen lines of SNR 3627-17 in velocity space centered on 5006.8\,\,Å, 6300.3\,\,Å and 7319.9\,\,Å (third panel). The spectra are normalized arbitrarily to have similar peak intensities.}
         \label{sourcee_etc}
   \end{figure}

   \begin{figure}[h]
   \centering
   \includegraphics[width=\hsize]{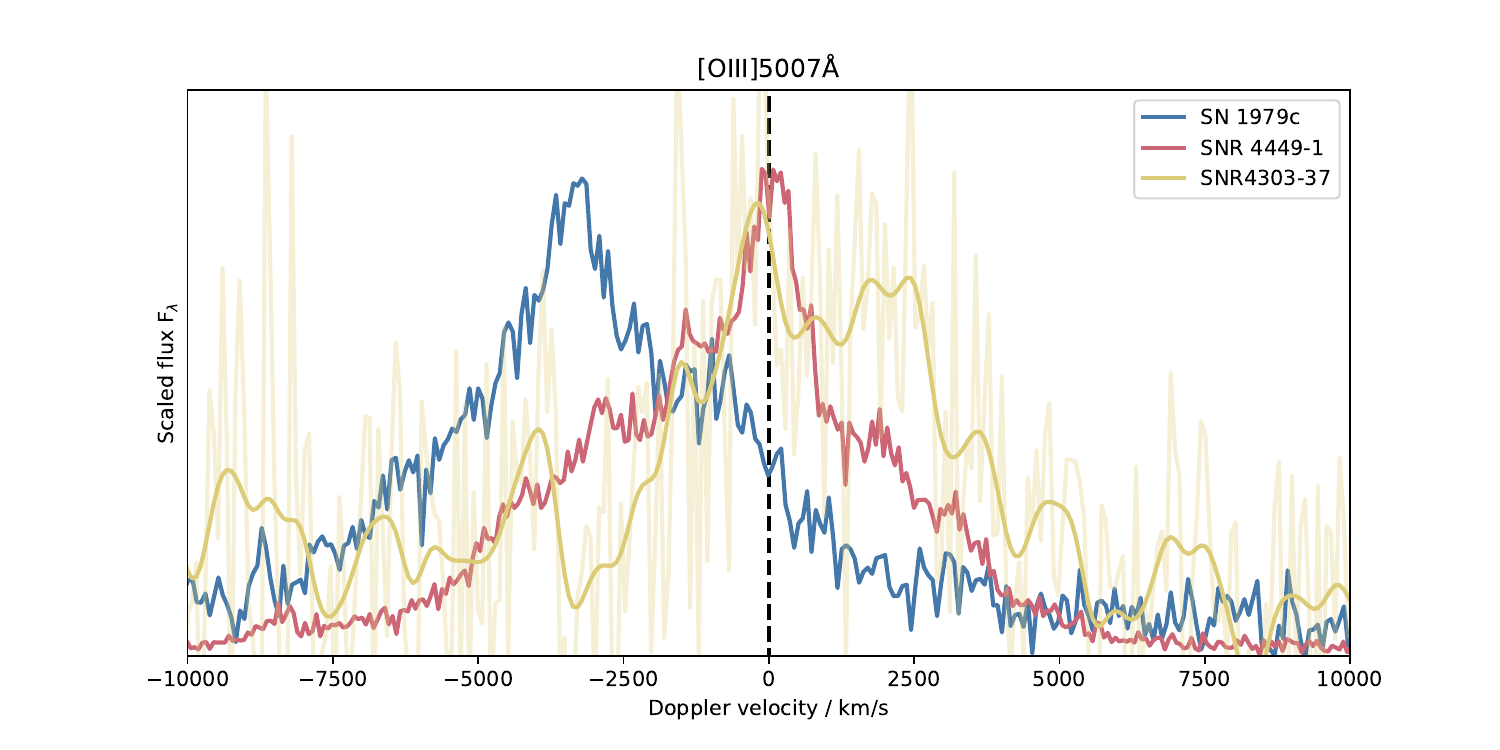}
   \includegraphics[width=\hsize]{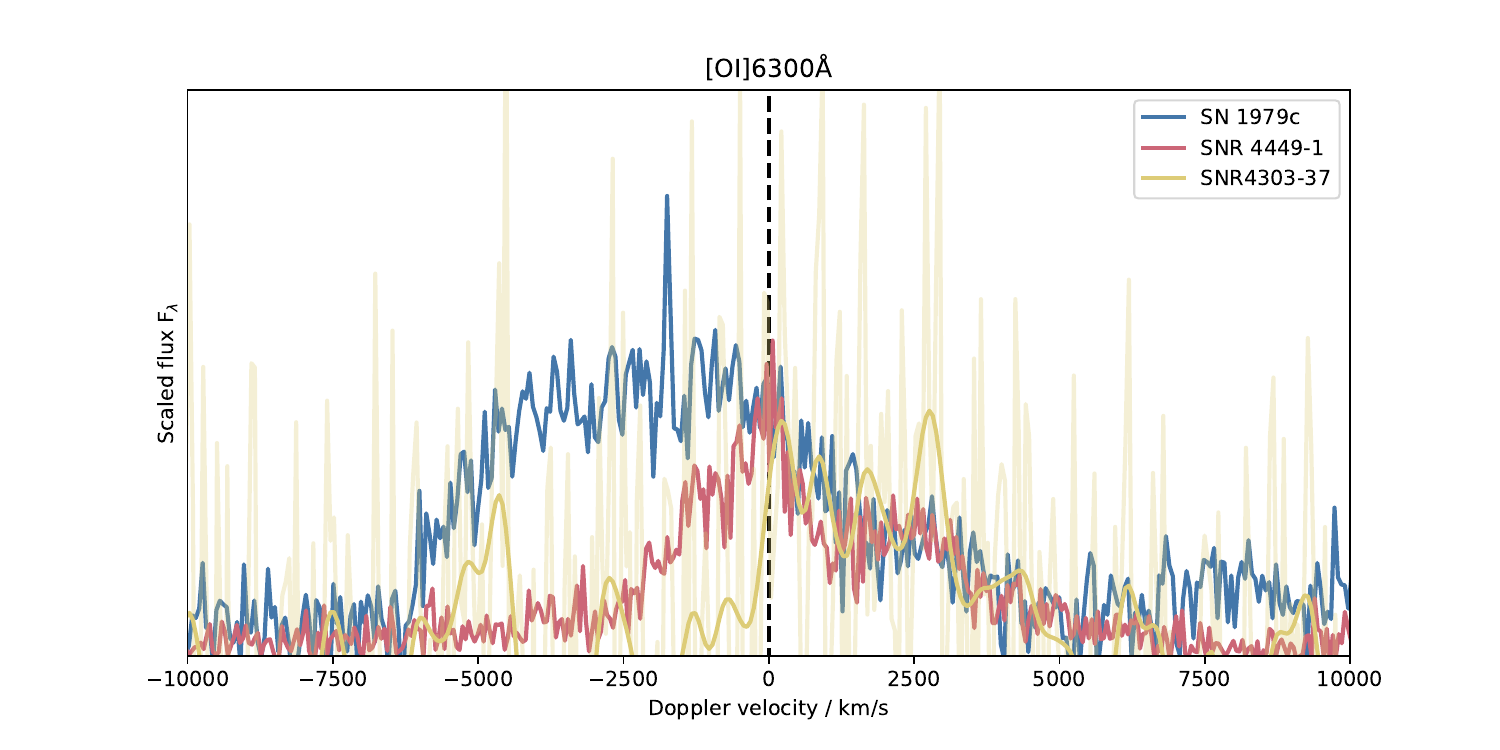}
   \includegraphics[width=\hsize]{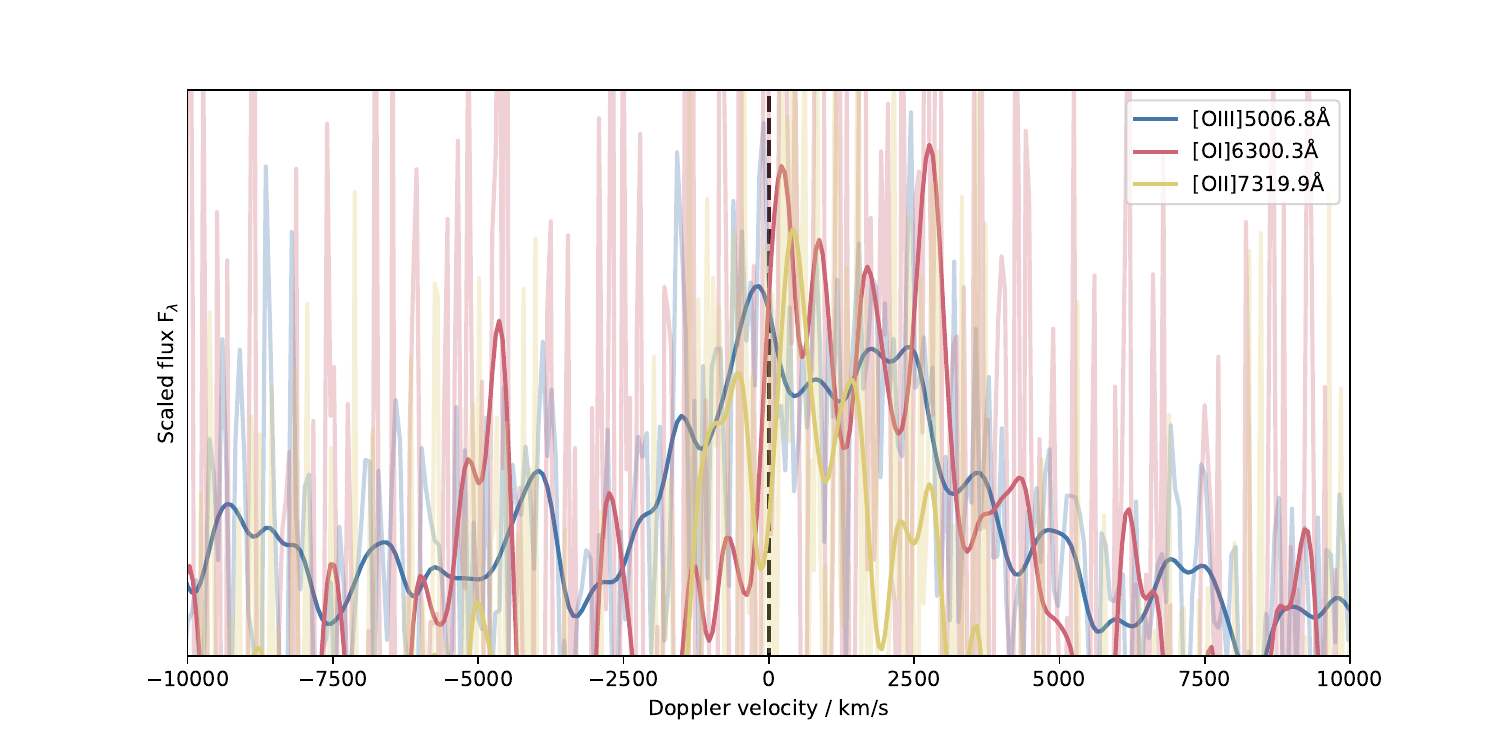}
      \caption{[O\,III]$\lambda\lambda$4959,5007 emission complex in velocity space centered around 5006.8Å (top) and same for [O\,I]$\lambda\lambda$6300,6364 centered around 6300Å (second panel). The visible light oxygen lines of SNR 4303-37 in velocity space centered on 5006.8\,\,Å, 6300.3\,\,Å and 7319.9\,\,Å (third panel). The spectra are normalized arbitrarily to have similar peak intensities.}
         \label{sourcef_etc}
   \end{figure}

   \begin{figure}[h]
   \centering
   \includegraphics[width=\hsize]{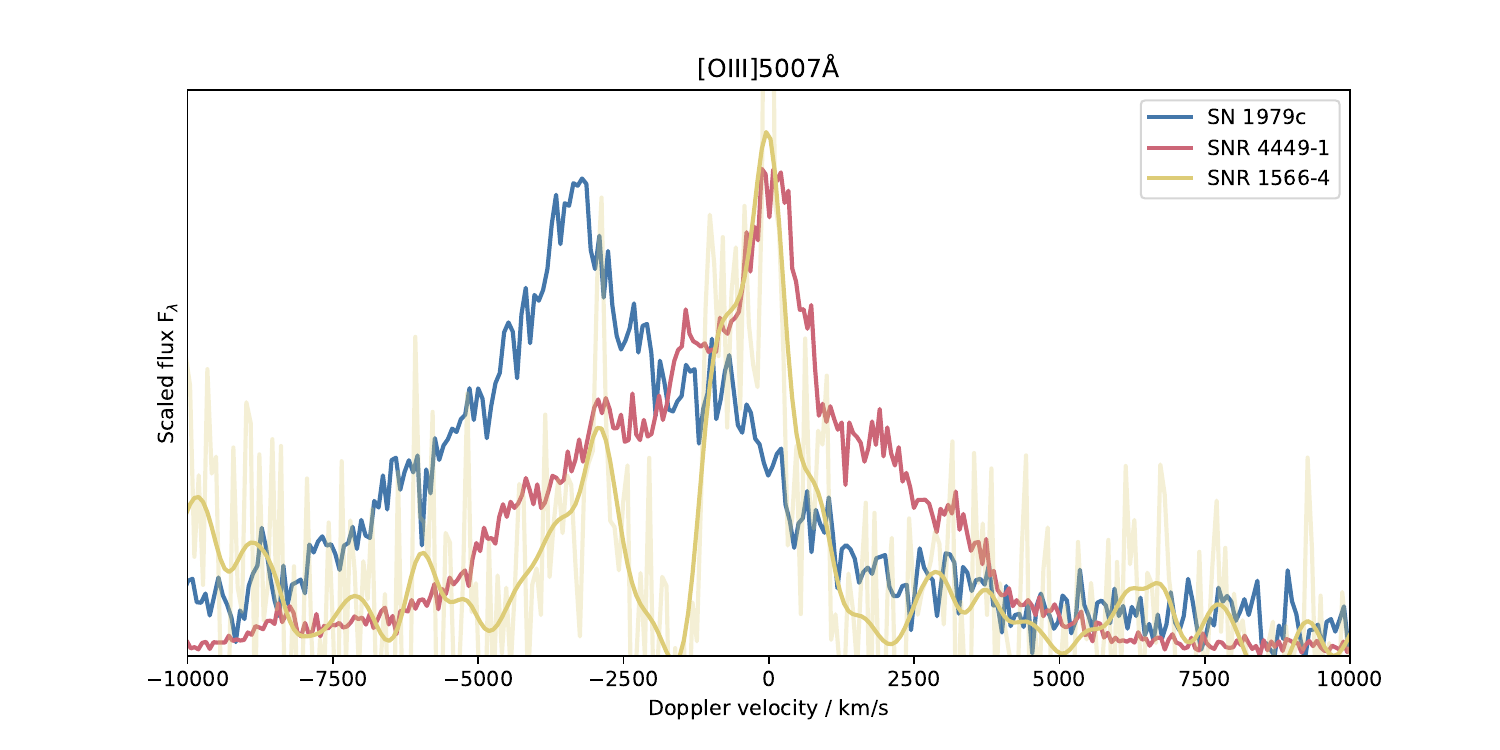}
   \includegraphics[width=\hsize]{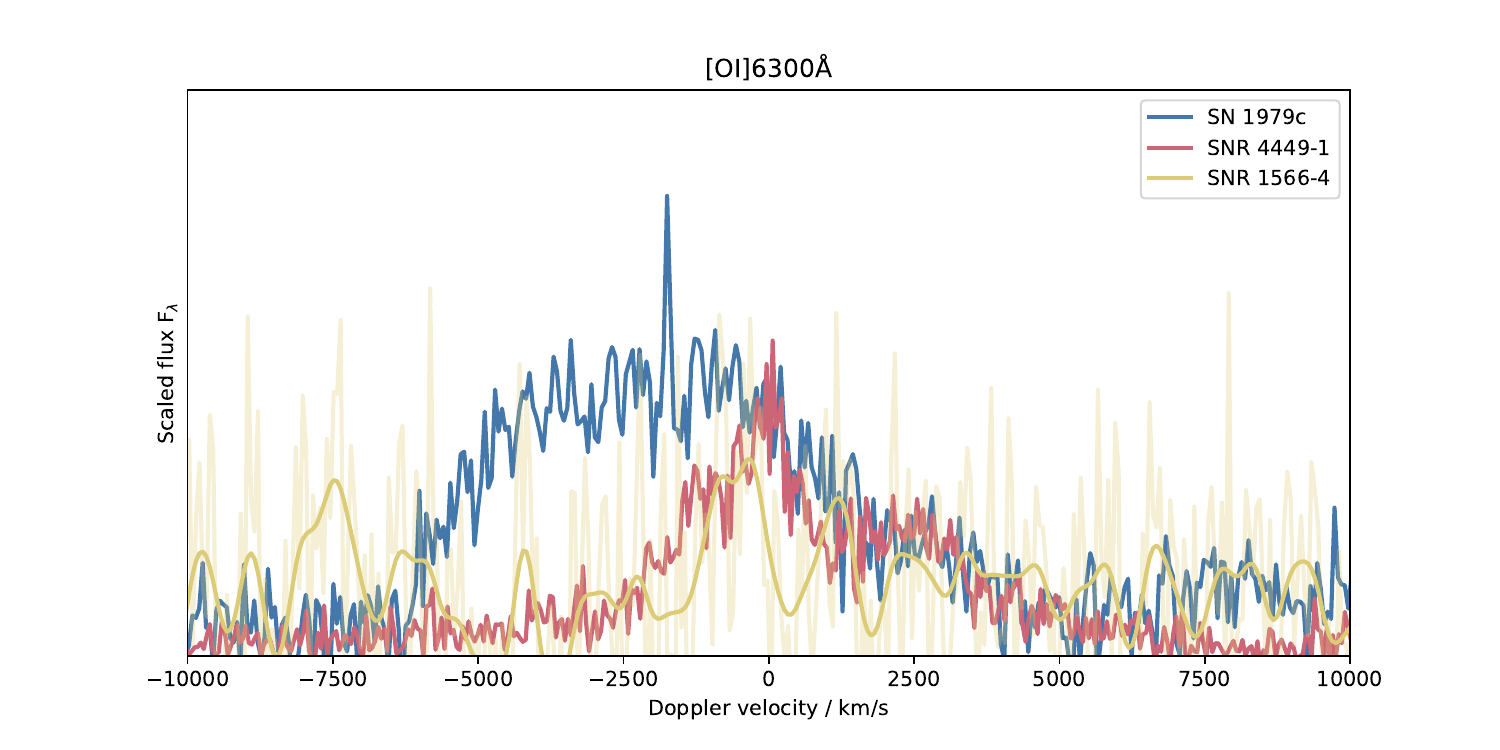}
   \includegraphics[width=\hsize]{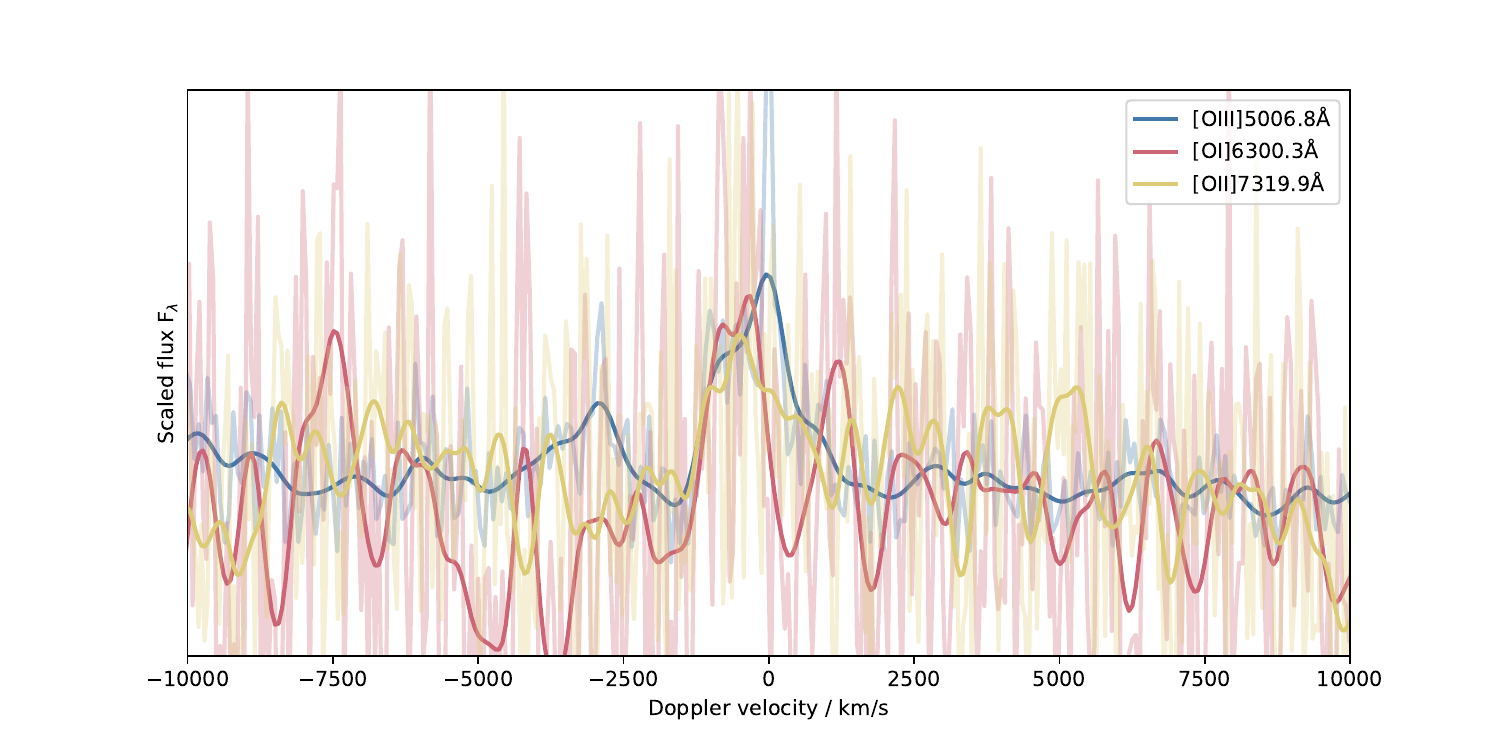}
      \caption{[O\,III]$\lambda\lambda$4959,5007 emission complex in velocity space centered around 5006.8Å (top) and same for [O\,I]$\lambda\lambda$6300,6364 centered around 6300Å (second panel). The visible light oxygen lines of SNR 1566-4 in velocity space centered on 5006.8\,\,Å, 6300.3\,\,Å and 7319.9\,\,Å (third panel). The spectra are normalized arbitrarily to have similar peak intensities.}
         \label{sourceg_etc}
   \end{figure}
    \onecolumn

\clearpage
\twocolumn
\section*{Appendix E}
    In this appendix we present a selection of HST and JWST images of several O-rich SNRs. For each figure we present the identified MUSE location and their tentatively identified counterparts in HST and JWST. The astrometric solutions between the different telescopes and instruments are better than 1", but follow-up observations are required to positively cross-match the sources.
   \begin{figure}[h]
   \centering
   \includegraphics[width=\hsize]{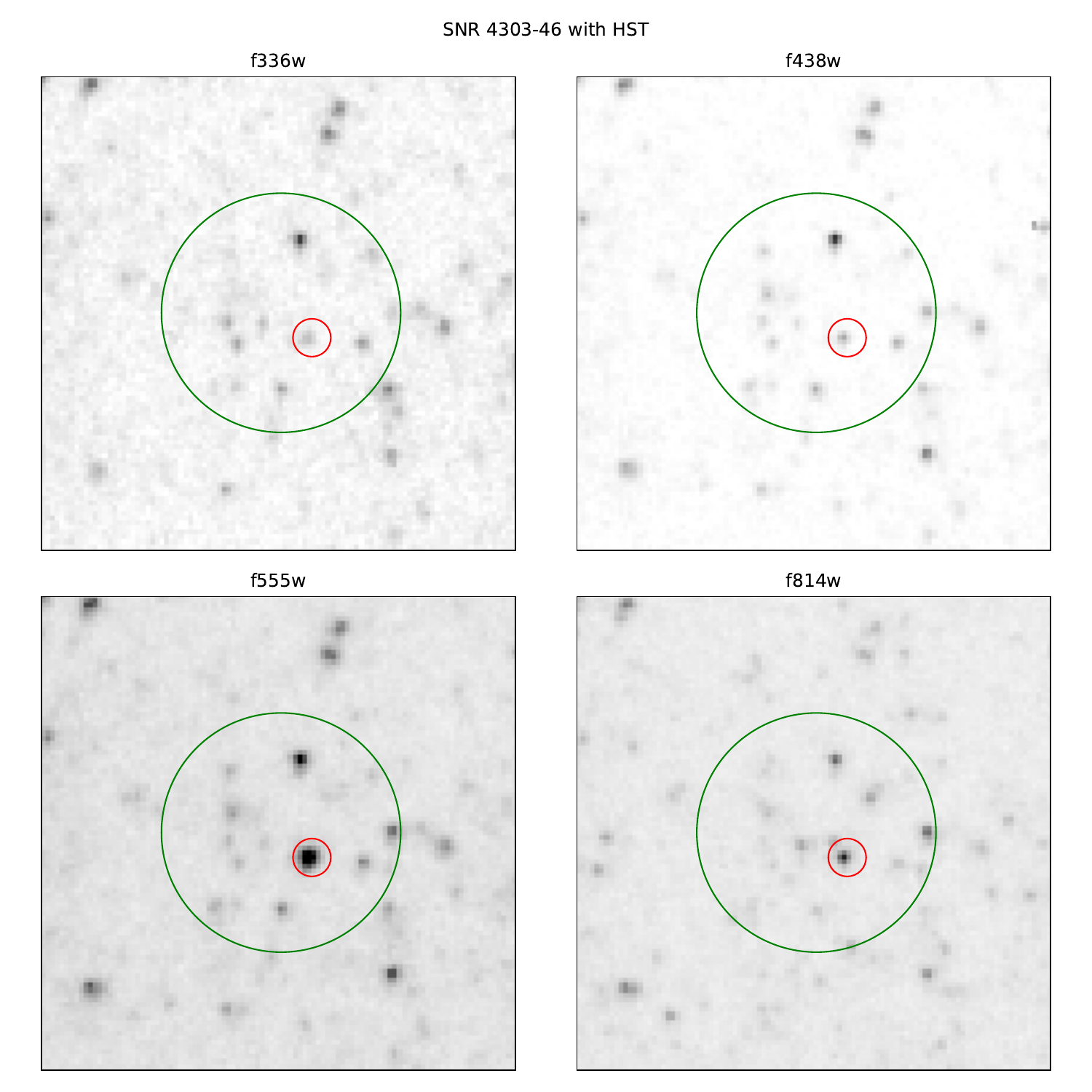}
      \caption{HST images of SNR 4303-46. Green circle is a 1" ring around MUSE coordinates. Red circle is the likely source in HST.}
         \label{fig:sourcea_hst}
   \end{figure}

   \begin{figure}[h]
   \centering
   \includegraphics[width=\hsize]{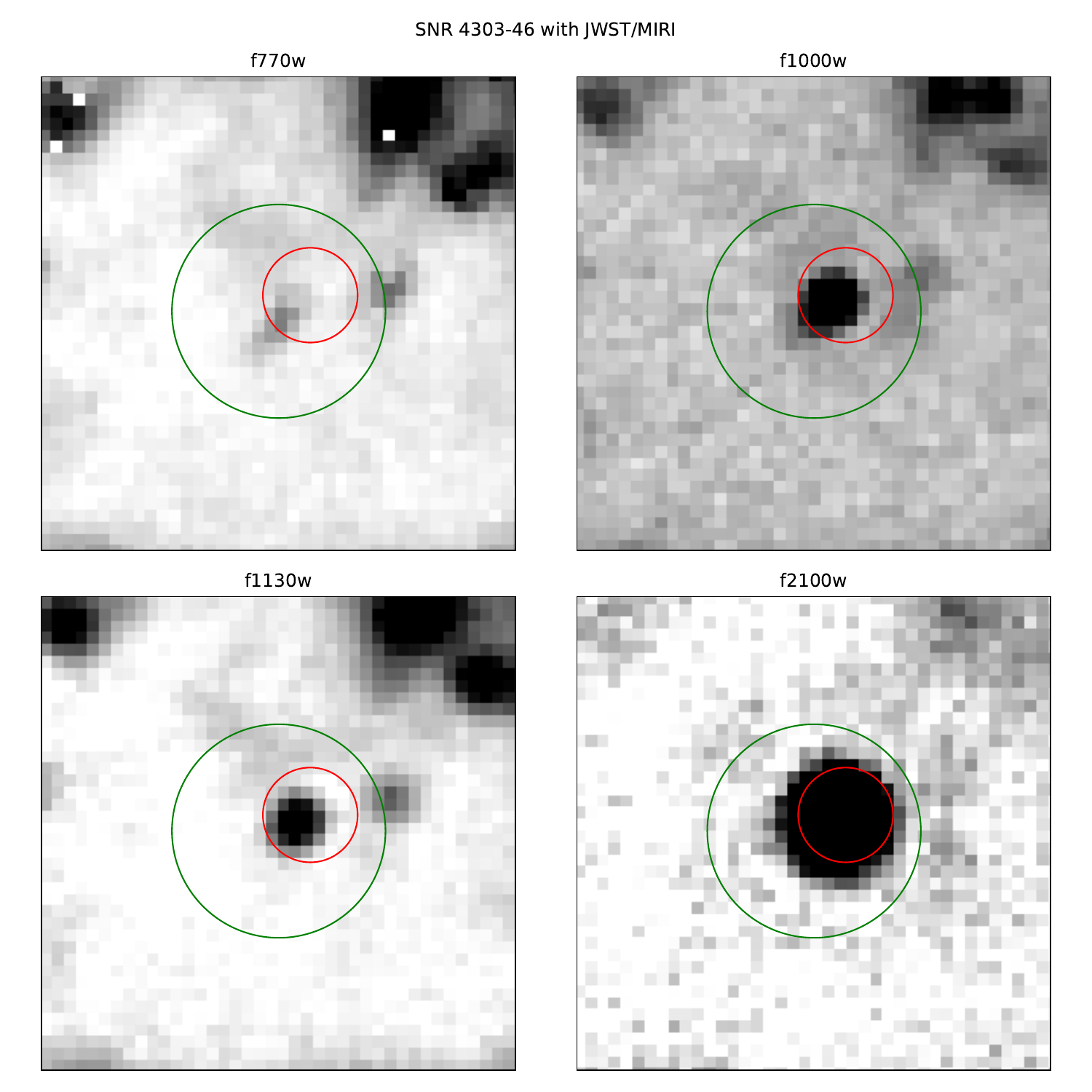}
      \caption{JWST/MIRI images of SNR 4303-46. Green circle is a 1" ring around MUSE coordinates. Red circle is the likely source in HST.}
         \label{fig:sourcea_jwst}
   \end{figure}

   \begin{figure}[h]
   \centering
   \includegraphics[width=\hsize]{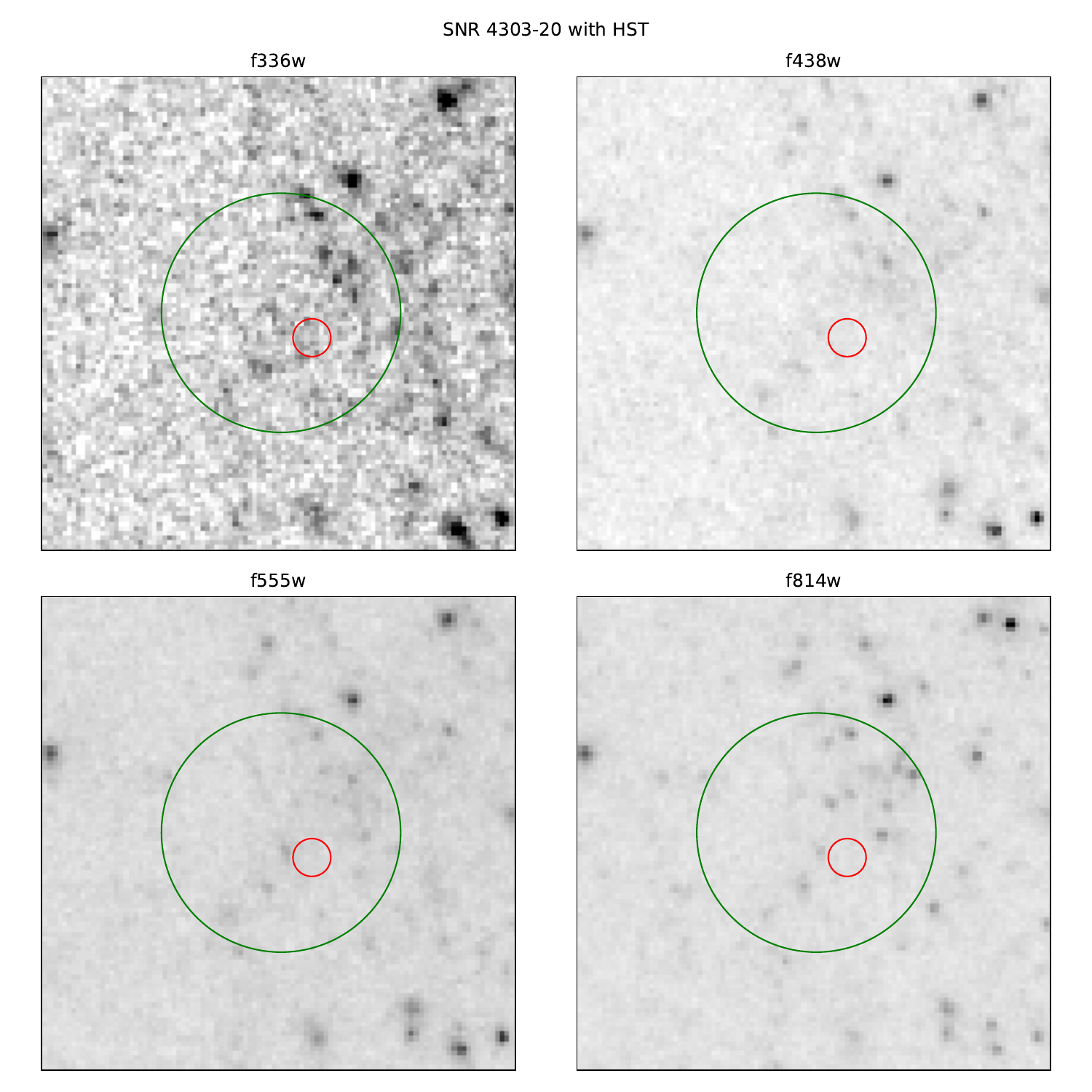}
      \caption{HST images of SNR 4303-20. Green circle is a 1" ring around MUSE coordinates. Red circle is the location of the source, if we assume that the coordinate offset between MUSE and HST are the same as with SNR 4303-46. There are no obvious sources near the MUSE coordinates.}
         \label{fig:sourceb_hst}
   \end{figure}
   
   \begin{figure}[h]
   \centering
   \includegraphics[width=\hsize]{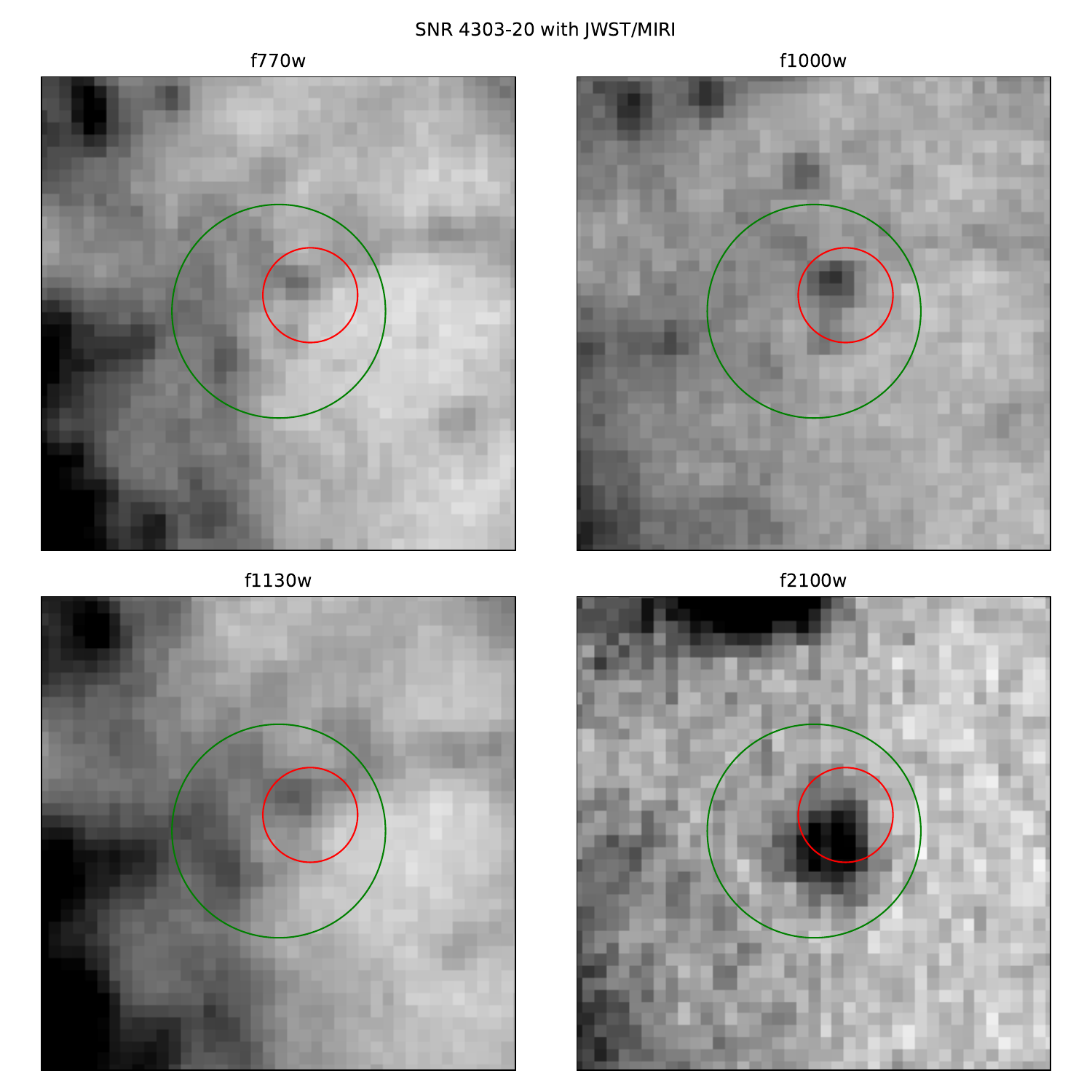}
      \caption{JWST/MIRI images of SNR 4303-20. Green circle is a 1" ring around MUSE coordinates. Red circle is the estimated location as in the previous figure. There is a clear point source in 21 $\mu$m.}
         \label{fig:sourceb_jwst}
   \end{figure}

   \begin{figure}[h]
   \centering
   \includegraphics[width=\hsize]{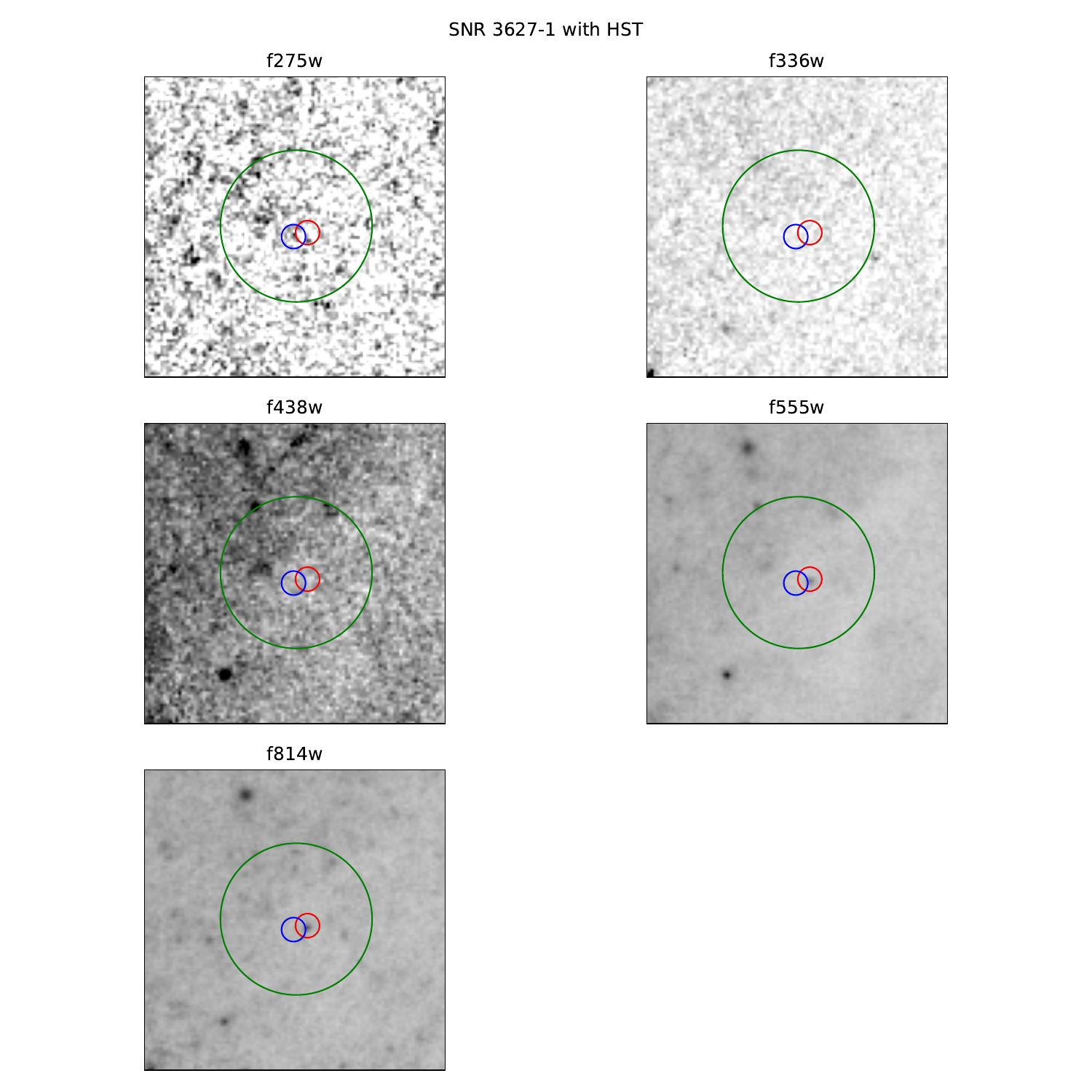}
      \caption{HST images of SNR 3627-1. Green circle is a 1" ring around MUSE coordinates. Red circle is the likely source in HST. Blue circle is the likely source location in JWST/MIRI from the next figure.}
         \label{fig:sourcec_hst}
   \end{figure}

   \begin{figure}[h]
   \centering
   \includegraphics[width=\hsize]{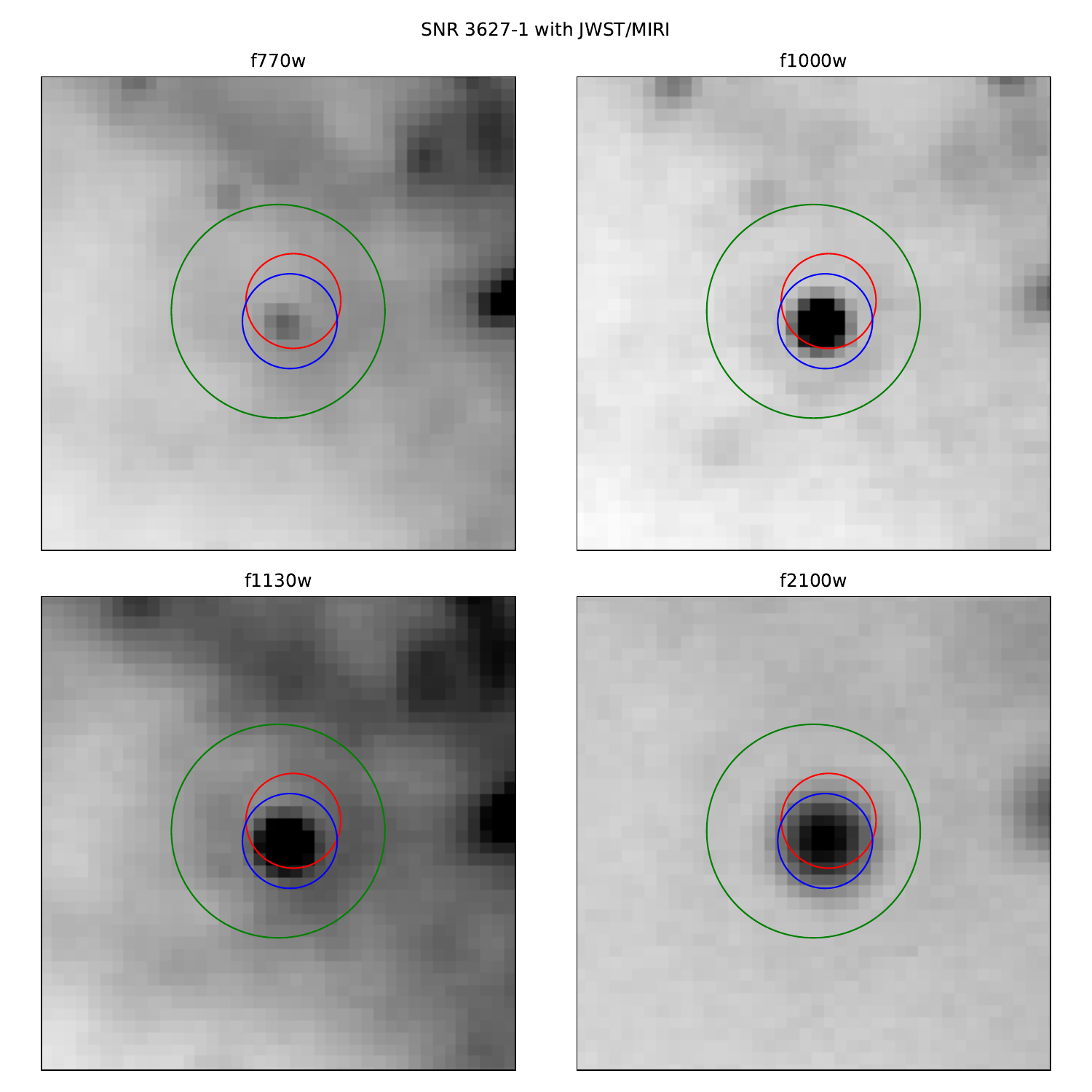}
      \caption{JWST/MIRI images of SNR 3627-1. Green circle is a 1" ring around MUSE coordinates. Red circle is the likely source in HST. Blue circle is the likely source in JWST/MIRI.}
         \label{fig:sourcec_jwst}
   \end{figure}

   \begin{figure}[h]
   \centering
   \includegraphics[width=\hsize]{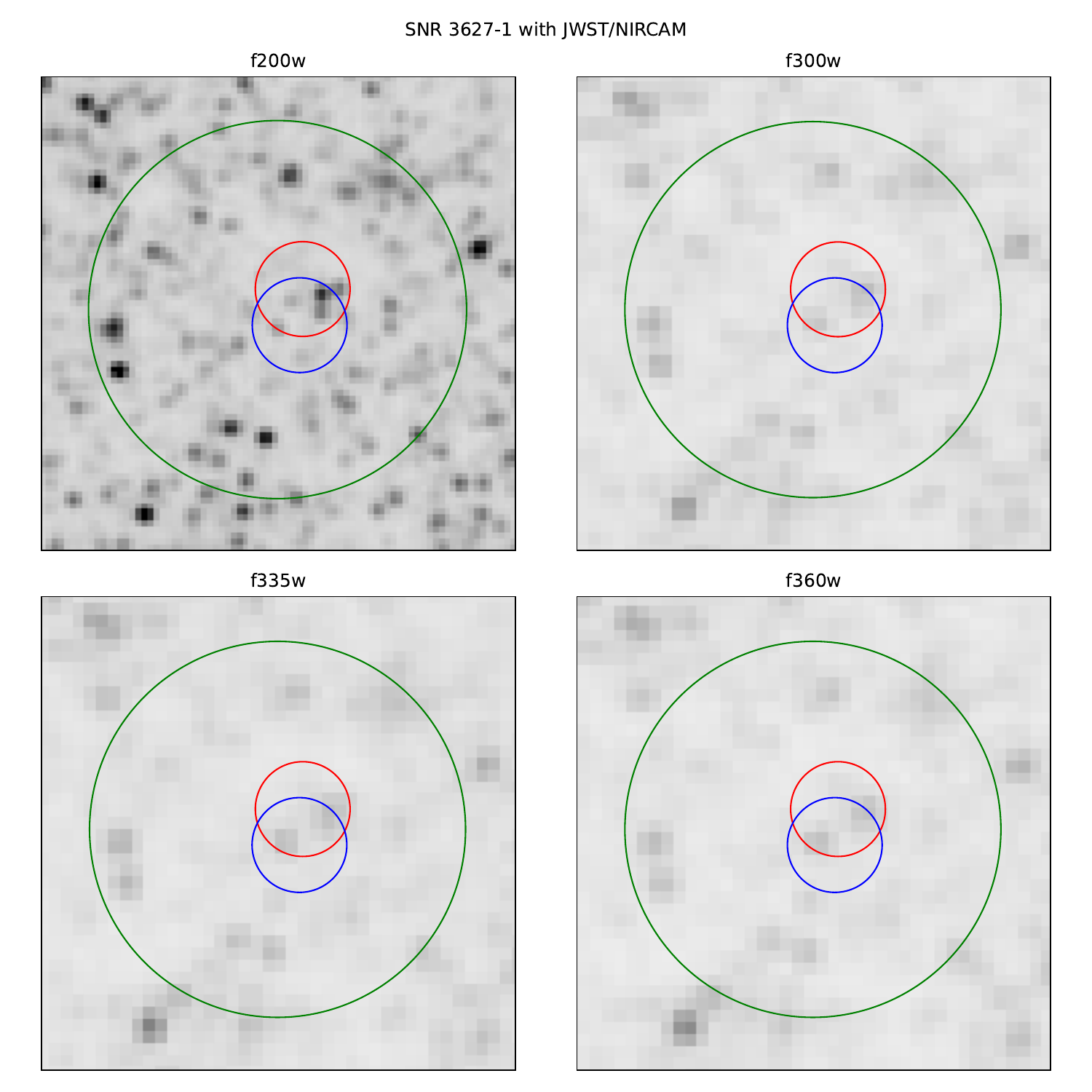}
      \caption{JWST/NIRCAM images of SNR 3627-1. Green circle is a 1" ring around MUSE coordinates. Red circle is the likely source in HST. Blue circle is the likely source in JWST/MIRI.}
         \label{fig:sourcec_jwstnircam}
   \end{figure}
   \begin{figure}
   \centering
   \includegraphics[width=\hsize]{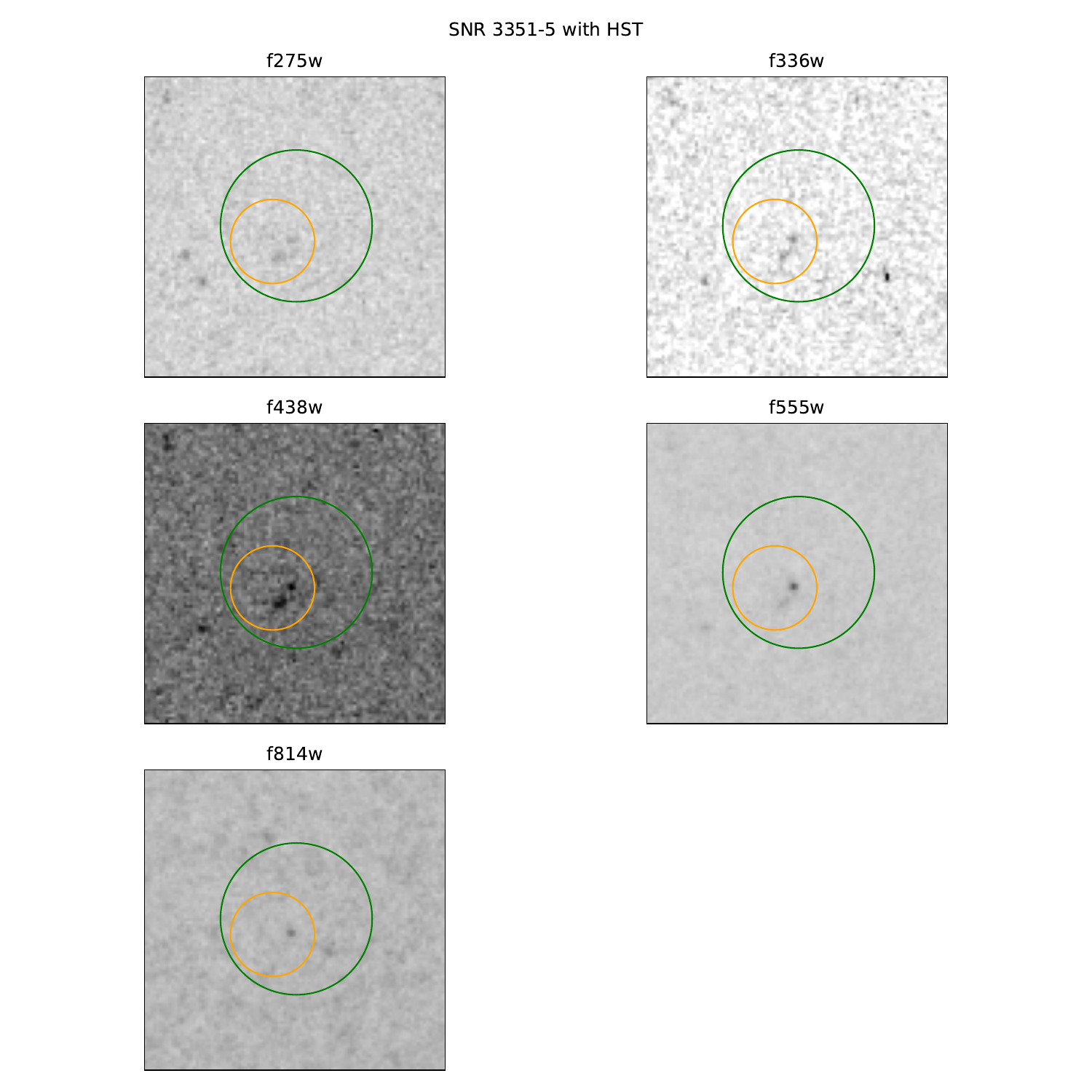}
      \caption{HST images of SNR 3351-5. Green circle is a 1" ring around MUSE coordinates. Orange circle is the location of the Chandra source. The accuracy of the coordinates of the Chandra source is about 1". There is two sources associated within both circles and it is unclear which is related to the SNR.}
         \label{fig:sourcec_hst}
   \end{figure}

\clearpage
\twocolumn
\section*{Appendix F}
In this appendix we present the discovered SNRs from this article. Each SNR is presented by a circle and in addition the remnants we identify as O-rich are marked with a gold star.
    \begin{figure}[h]
    \centering
    \includegraphics[width=\hsize]{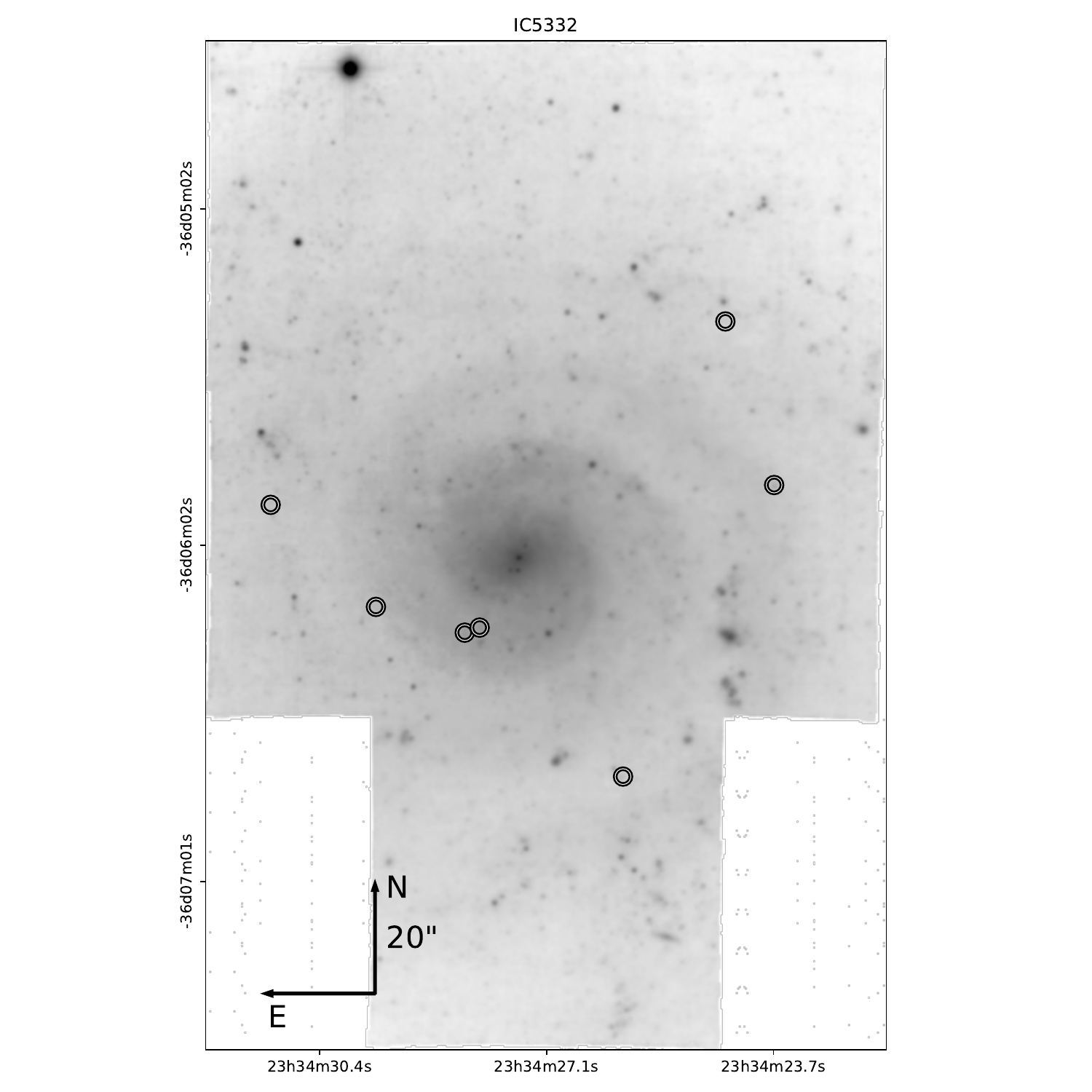}
    \end{figure}

    \begin{figure}[h]
    \centering
    \includegraphics[width=\hsize]{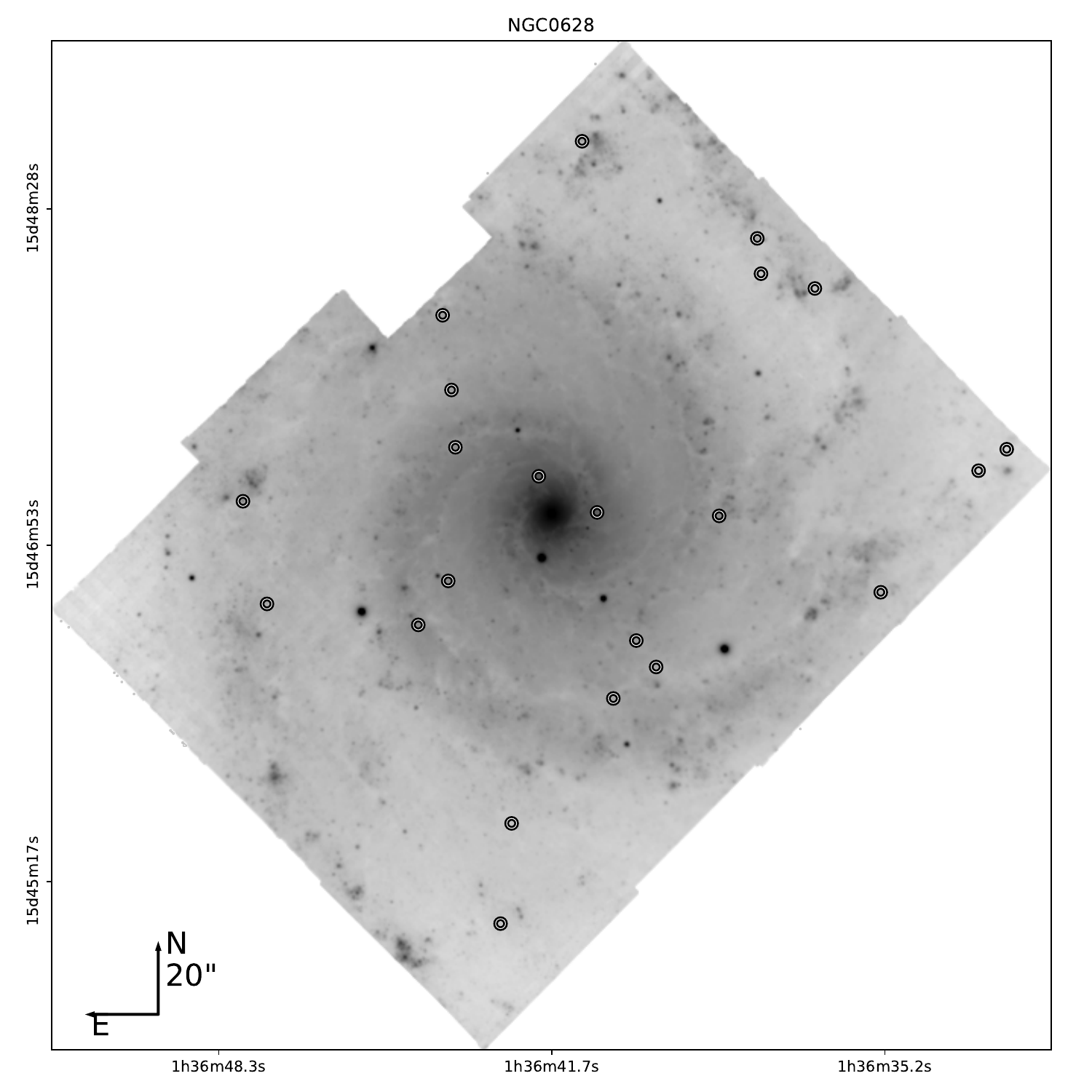}
    \end{figure}

    \begin{figure}[h]
    \centering
    \includegraphics[width=\hsize]{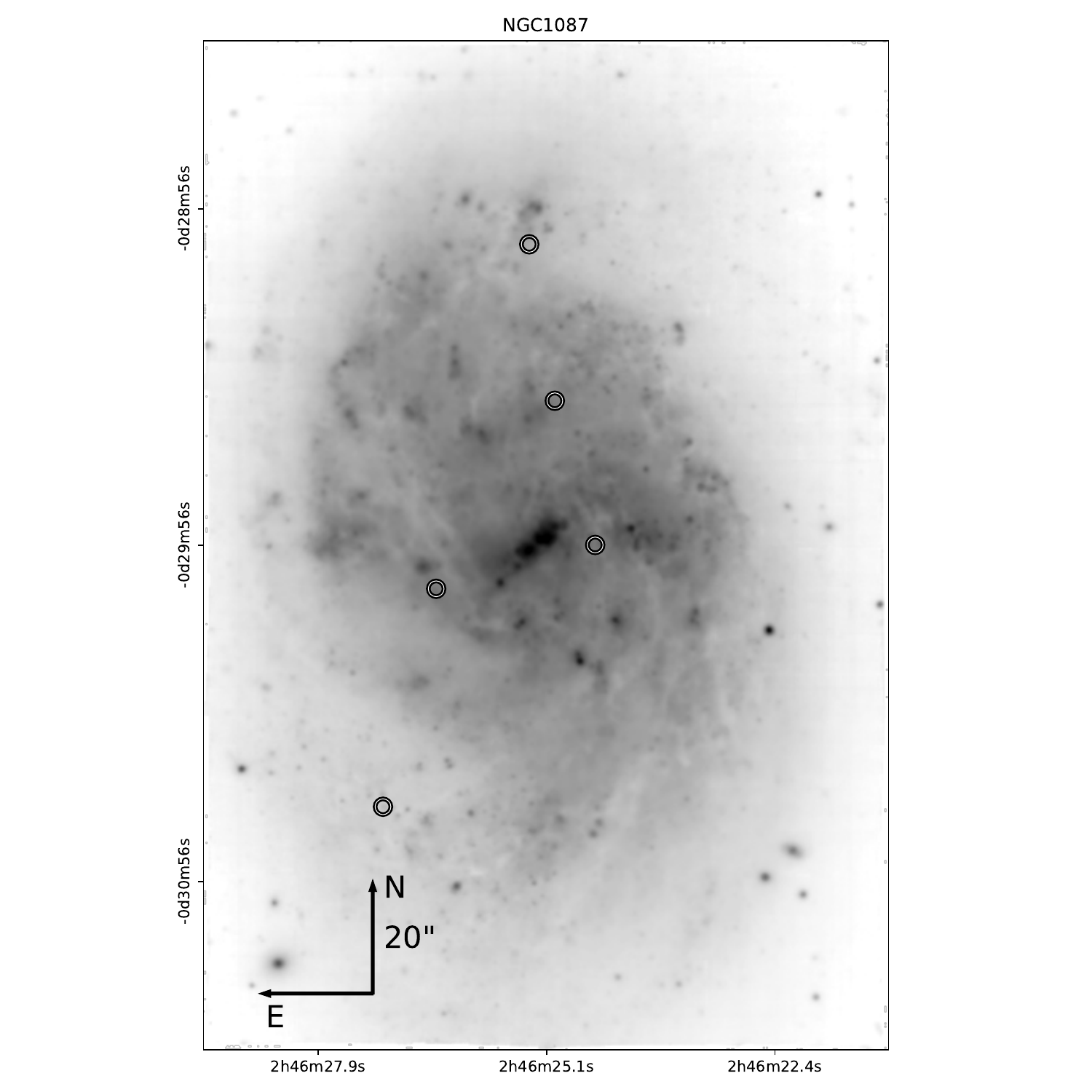}
    \end{figure}

    \begin{figure}[h]
    \centering
    \includegraphics[width=\hsize]{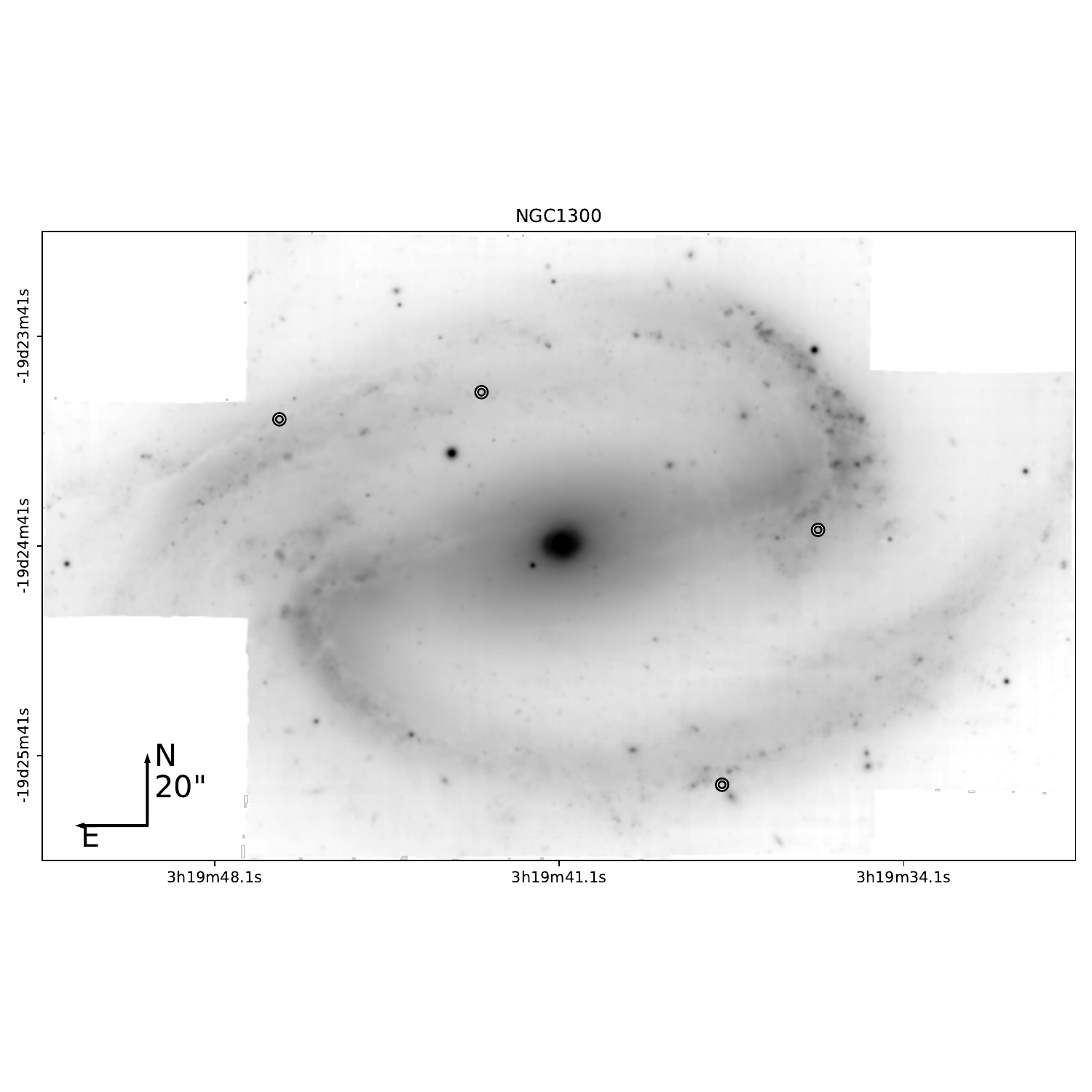}
    \end{figure}

    \begin{figure}[h]
    \centering
    \includegraphics[width=\hsize]{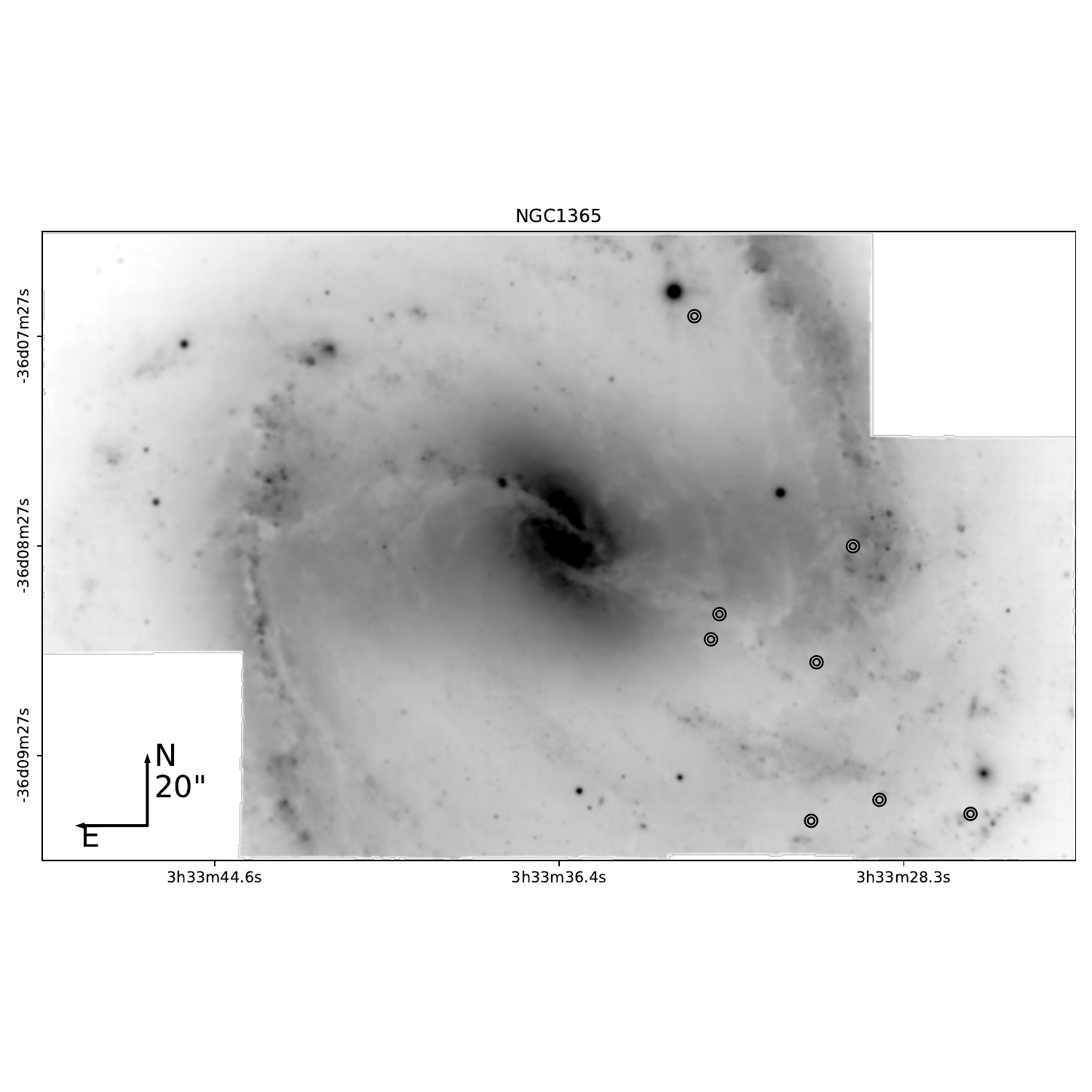}
    \end{figure}

    \begin{figure}[h]
    \centering
    \includegraphics[width=\hsize]{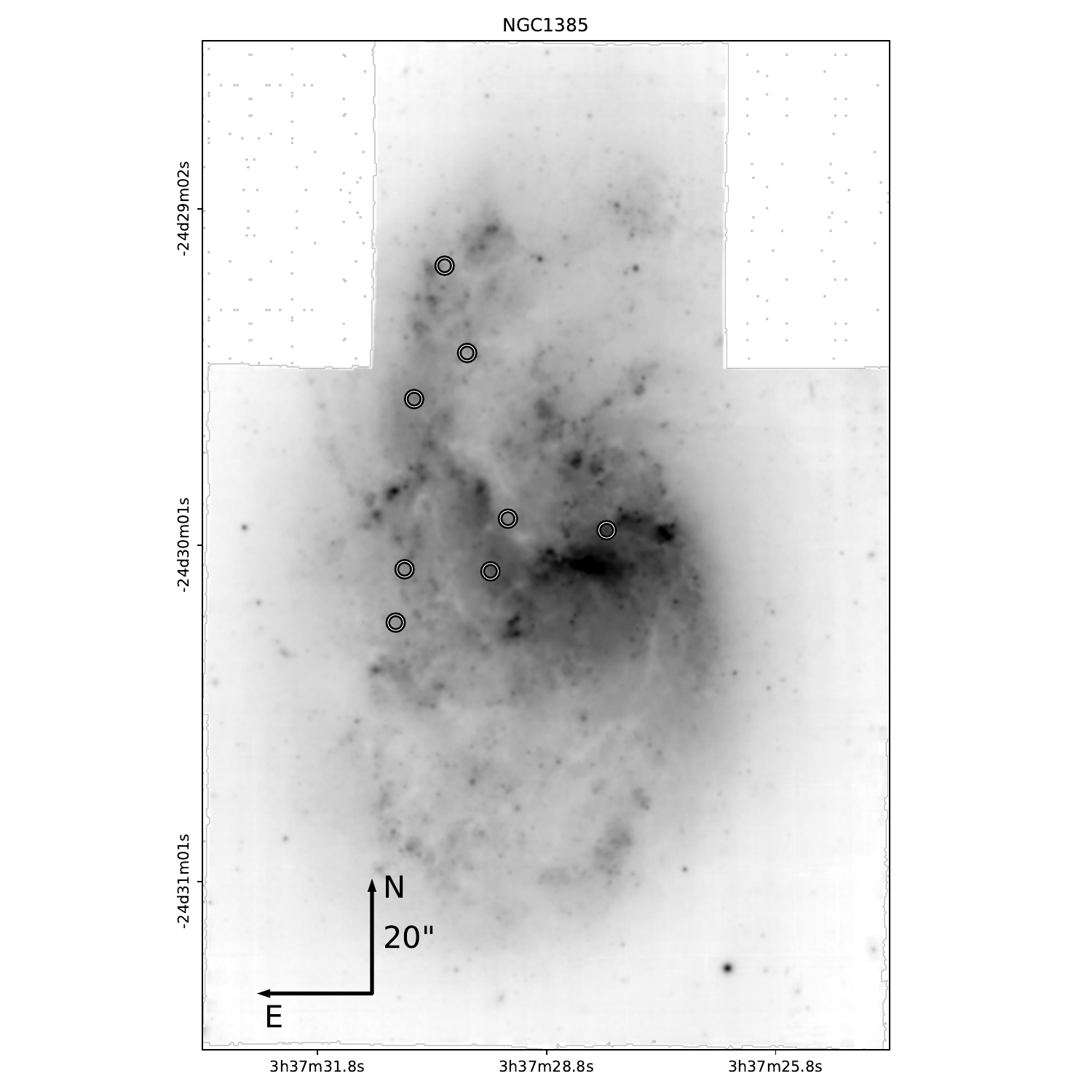}
    \end{figure}

    \begin{figure}[h]
    \centering
    \includegraphics[width=\hsize]{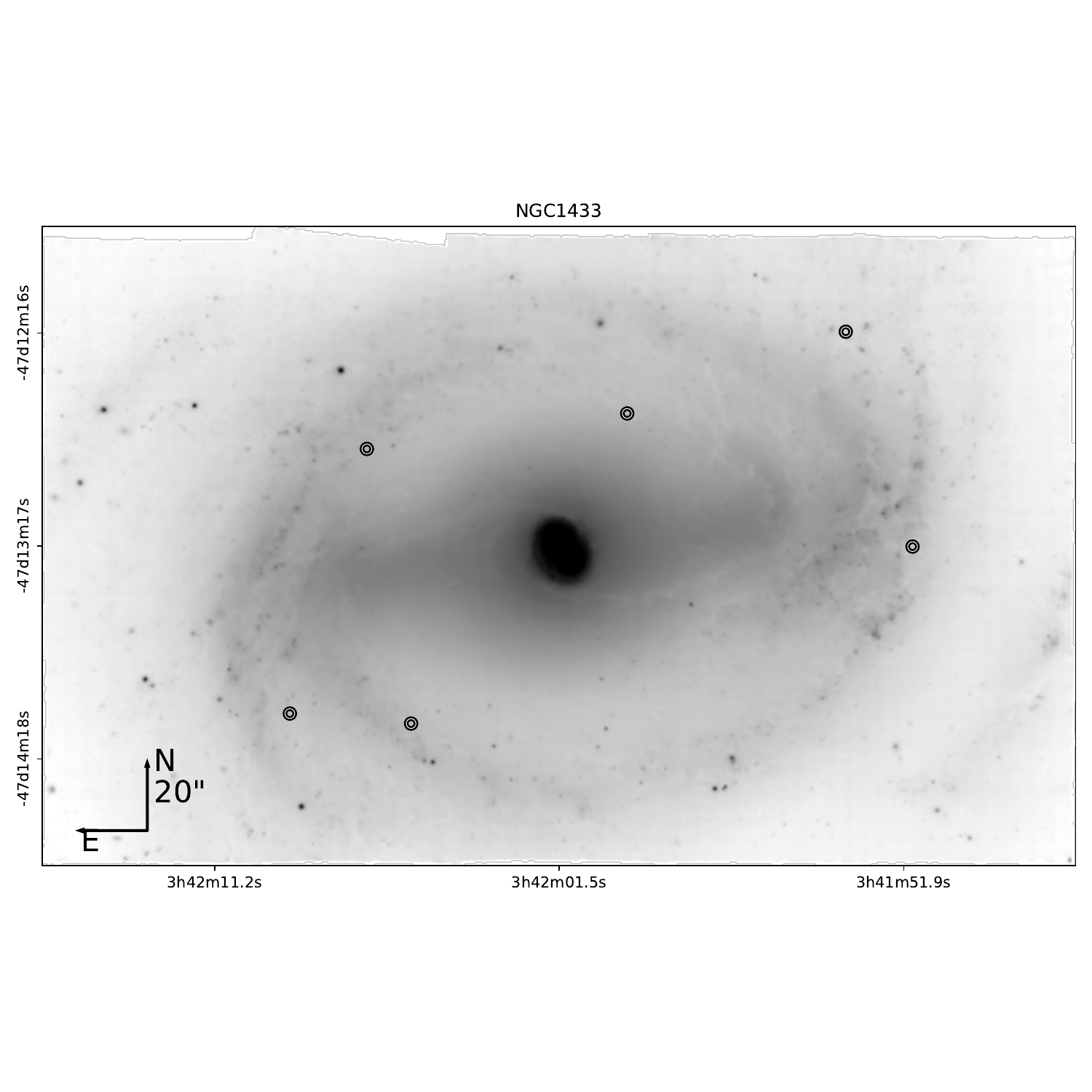}
    \end{figure}

    \begin{figure}[h]
    \centering
    \includegraphics[width=\hsize]{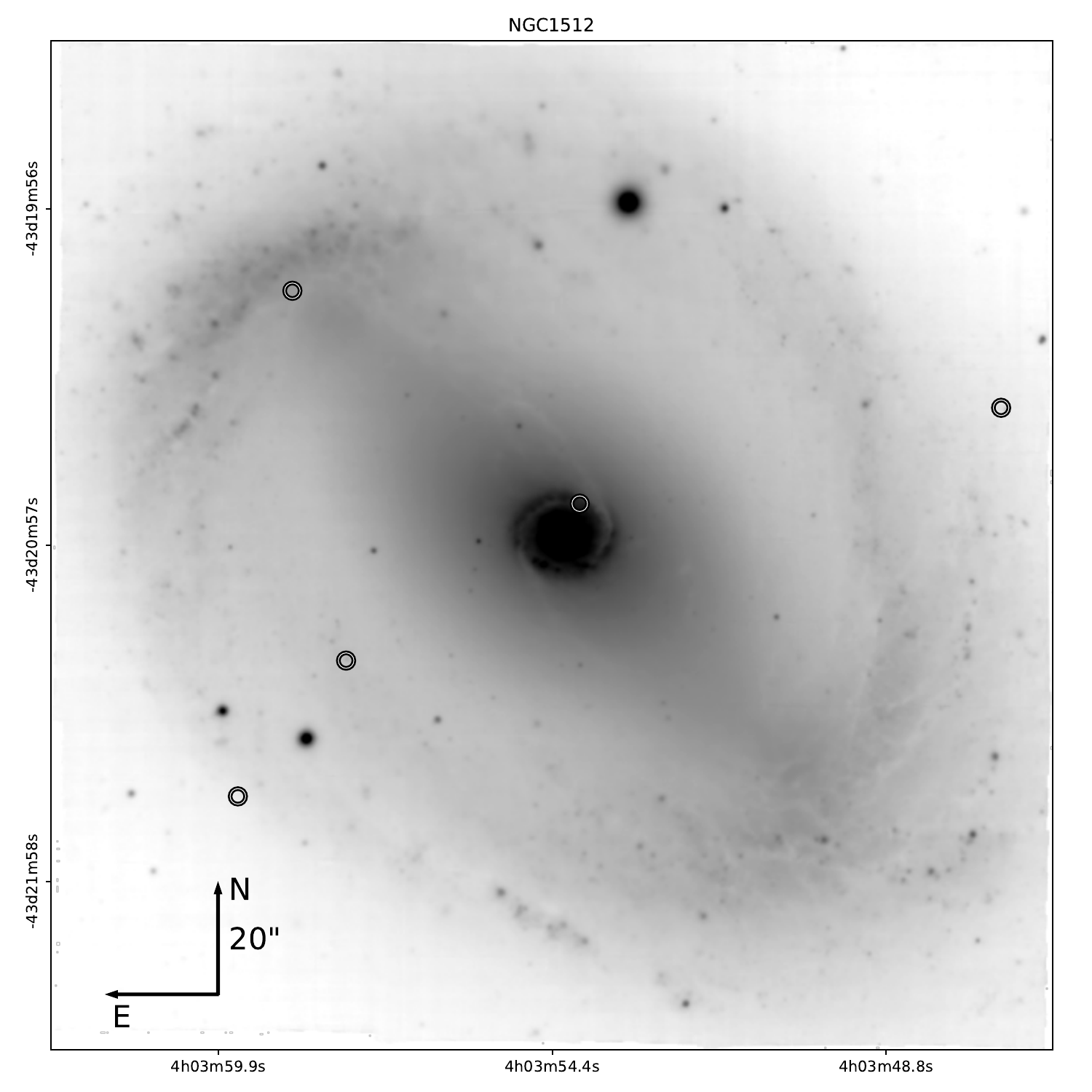}
    \end{figure}

    \begin{figure}[h]
    \centering
    \includegraphics[width=\hsize]{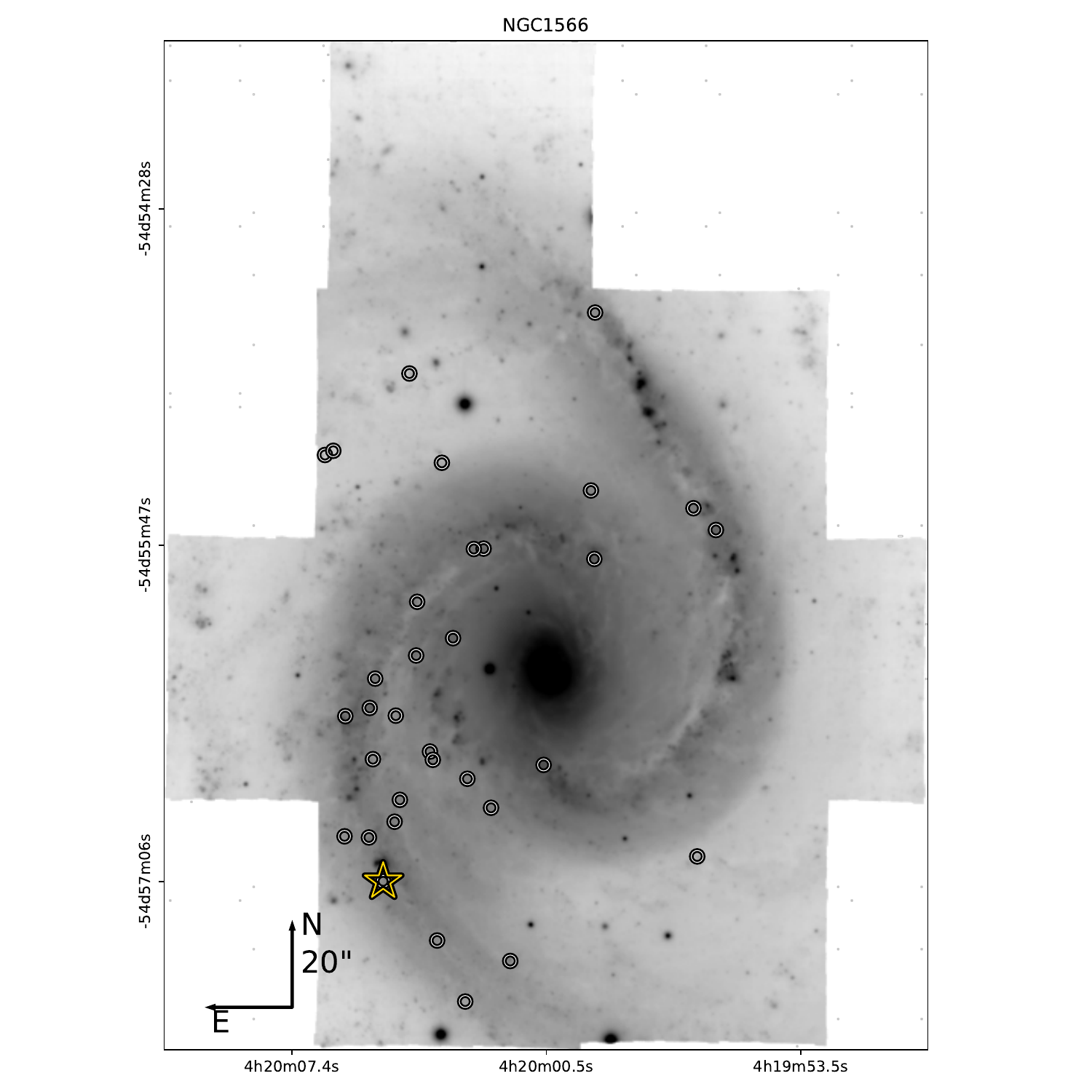}
    \end{figure}

    \begin{figure}[h]
    \centering
    \includegraphics[width=\hsize]{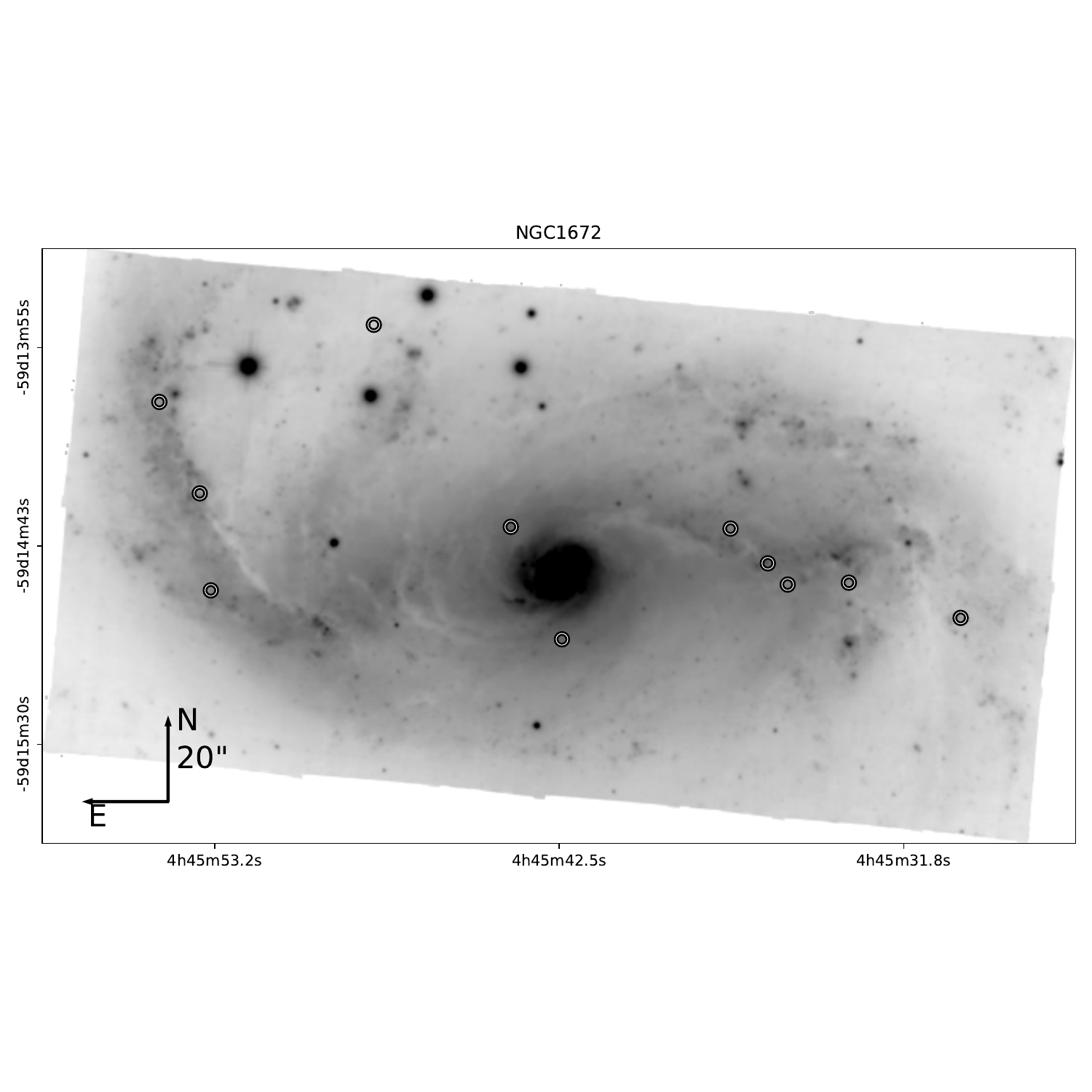}
    \end{figure}

    \begin{figure}[h]
    \centering
    \includegraphics[width=\hsize]{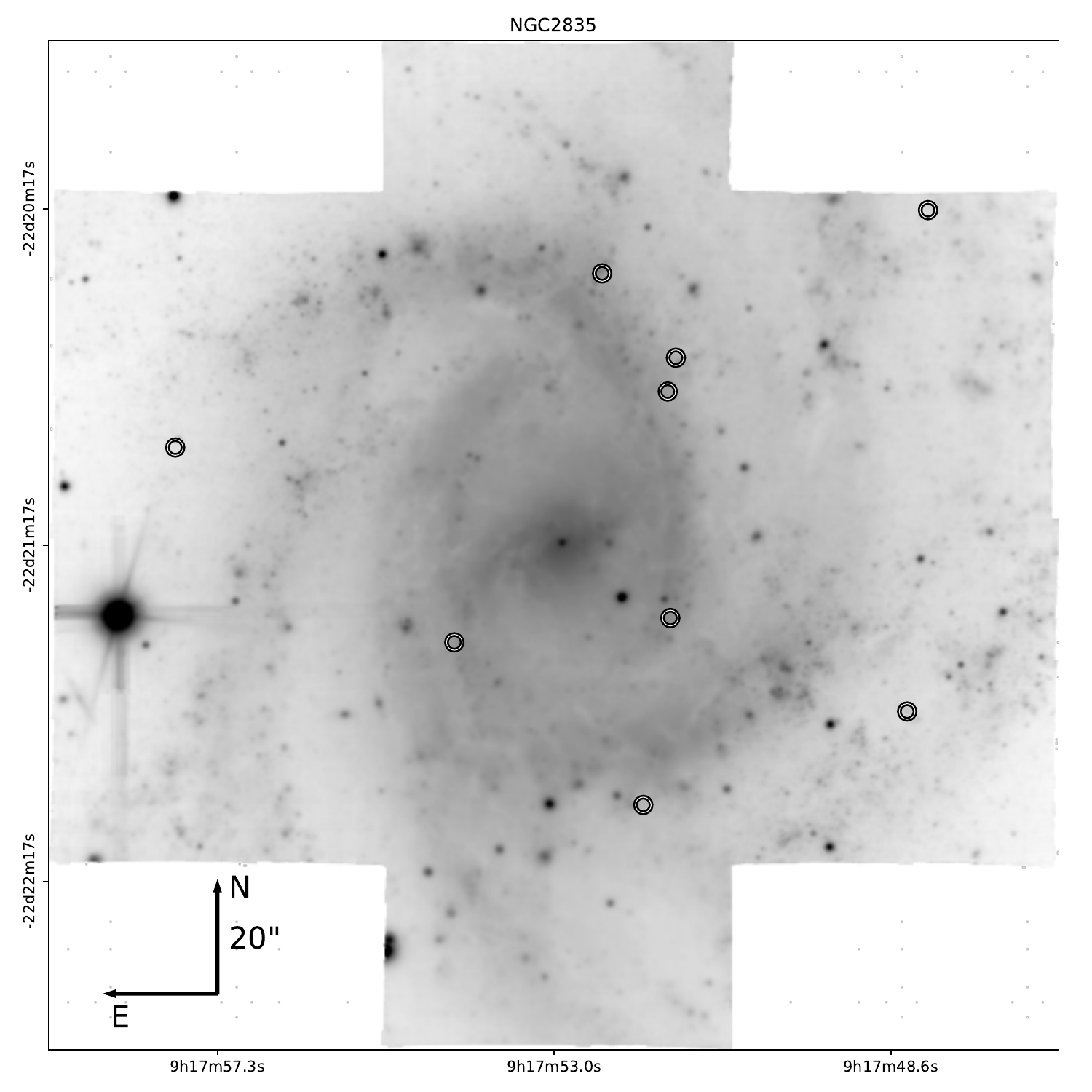}
    \end{figure}

    \begin{figure}[h]
    \centering
    \includegraphics[width=\hsize]{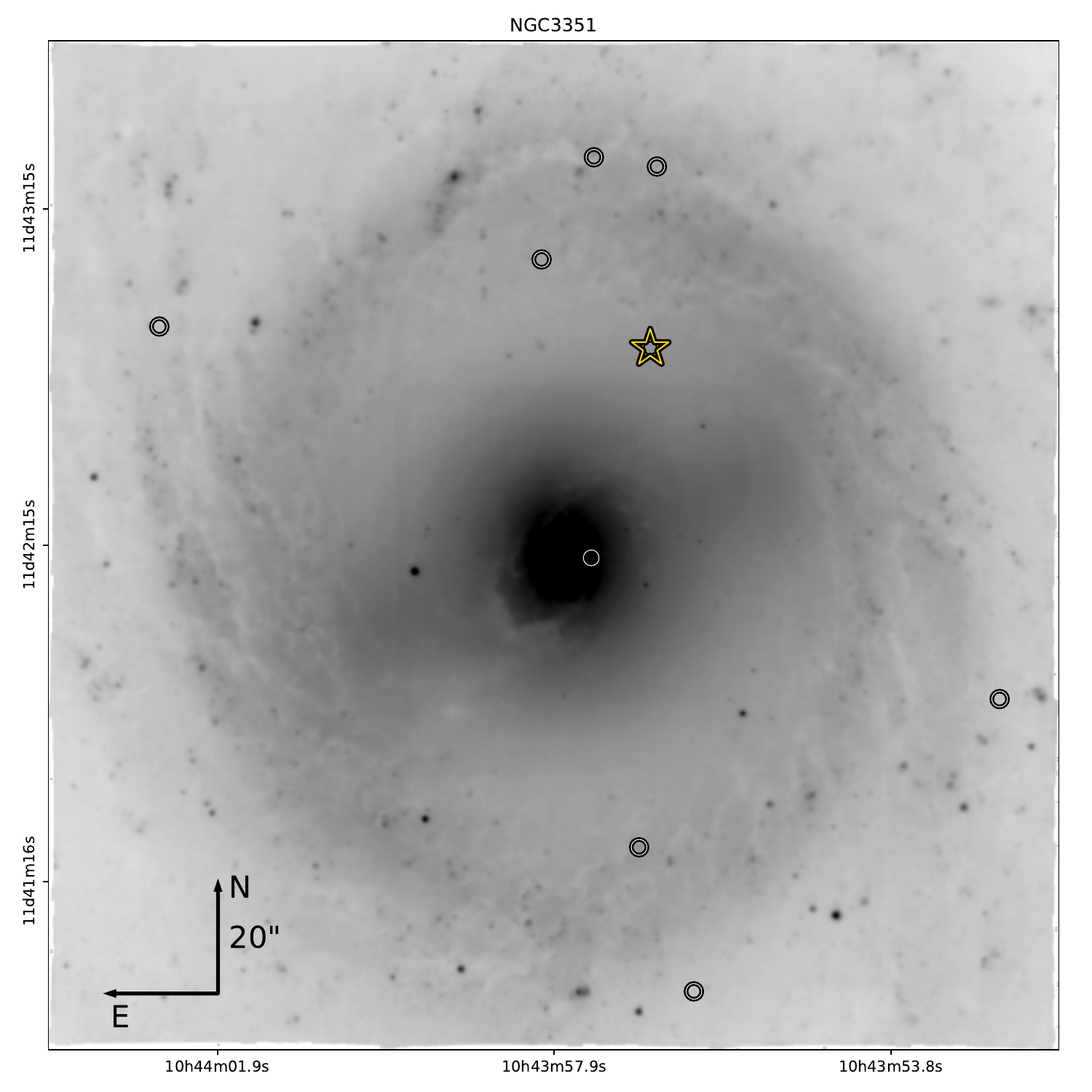}
    \end{figure}

    \begin{figure}[h]
    \centering
    \includegraphics[width=\hsize]{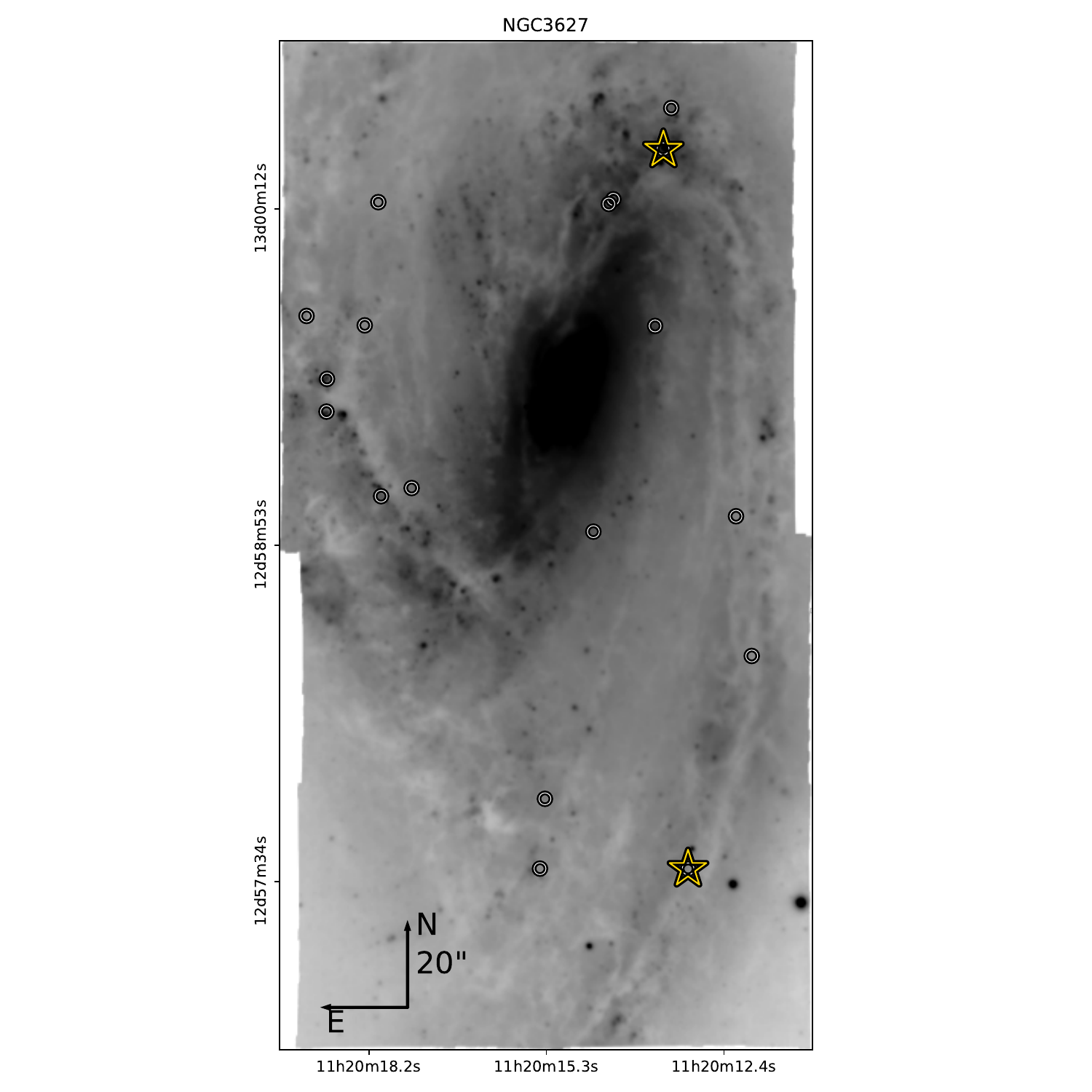}
    \end{figure}

    \begin{figure}[h]
    \centering
    \includegraphics[width=\hsize]{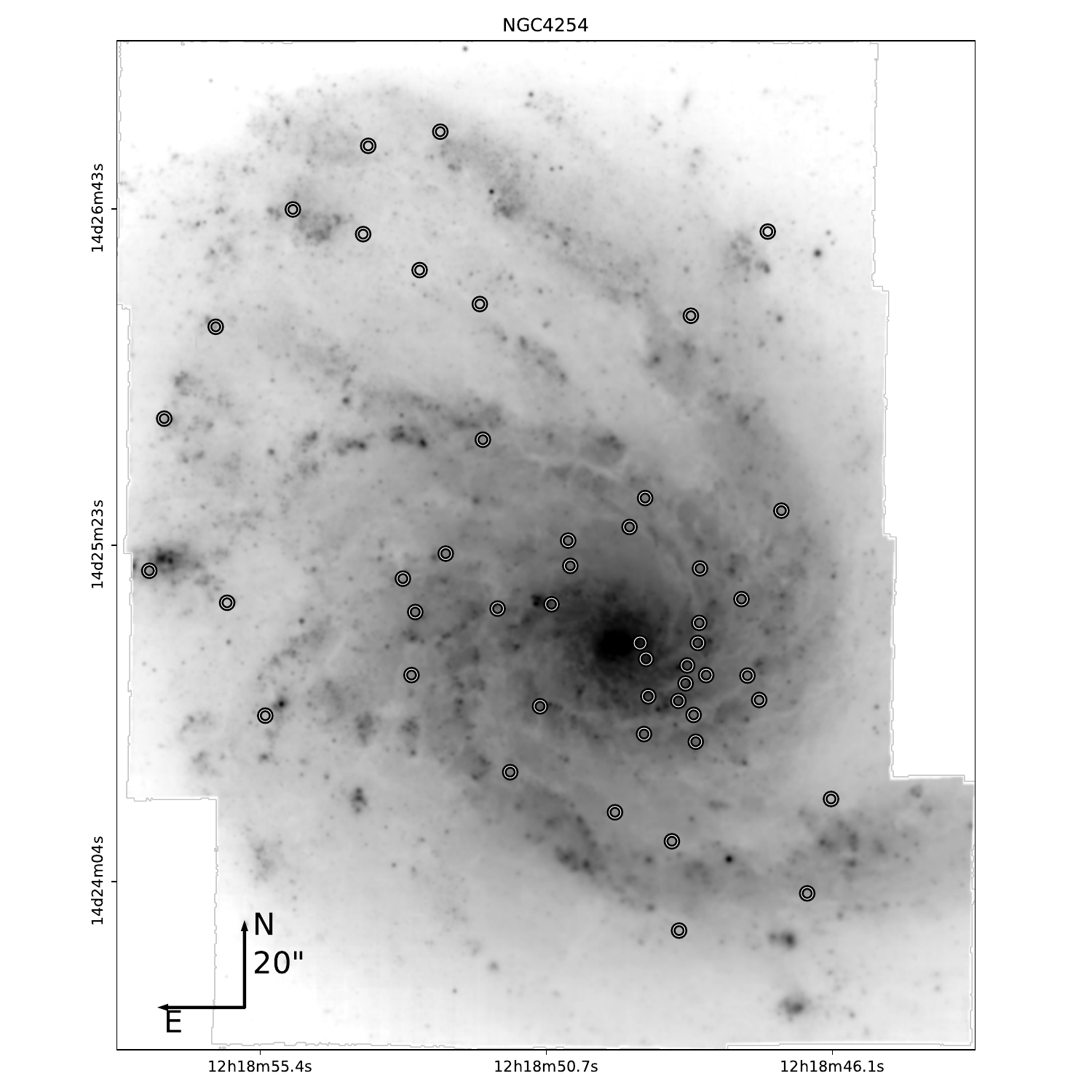}
    \end{figure}

    \begin{figure}[h]
    \centering
    \includegraphics[width=\hsize]{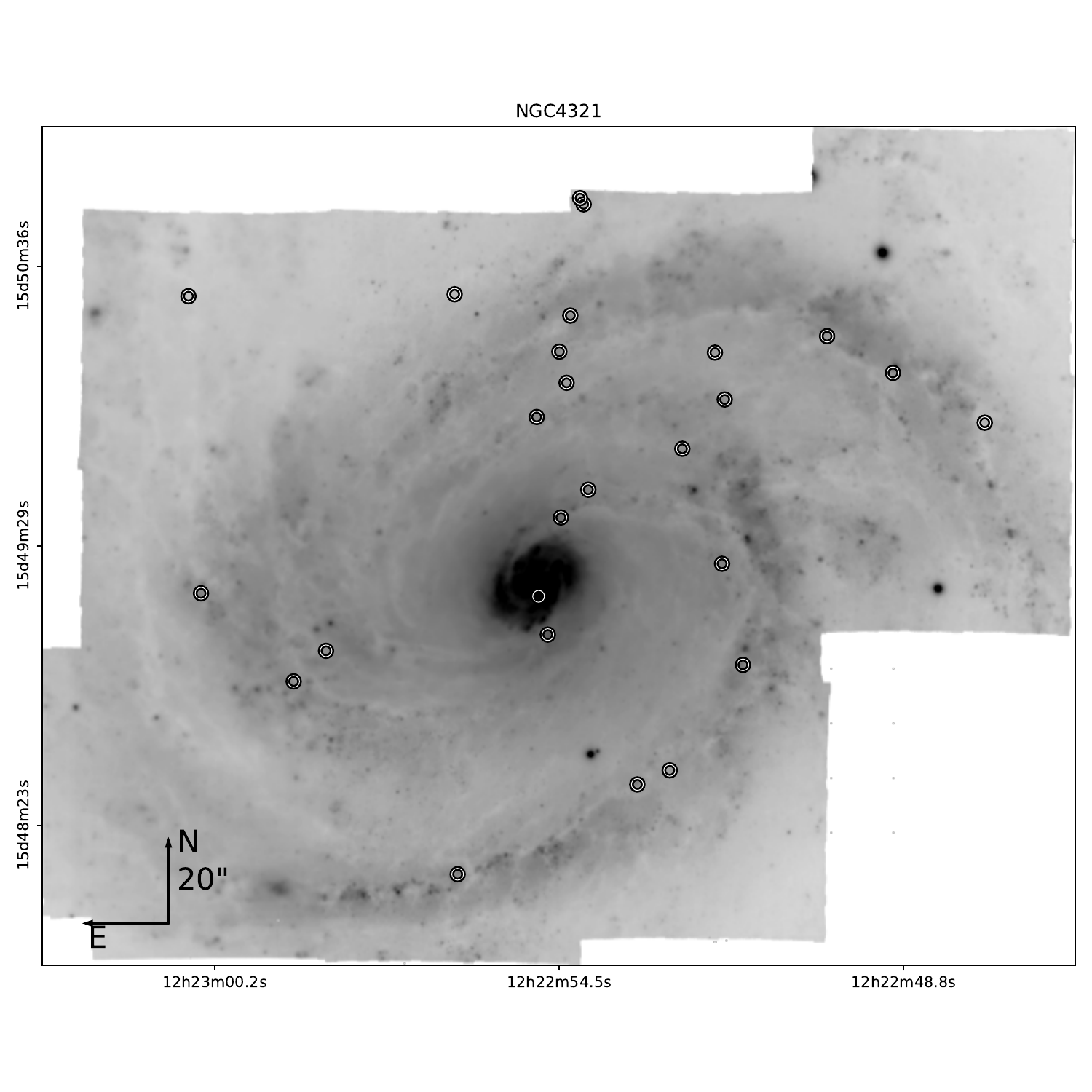}
    \end{figure}

    \begin{figure}[h]
    \centering
    \includegraphics[width=\hsize]{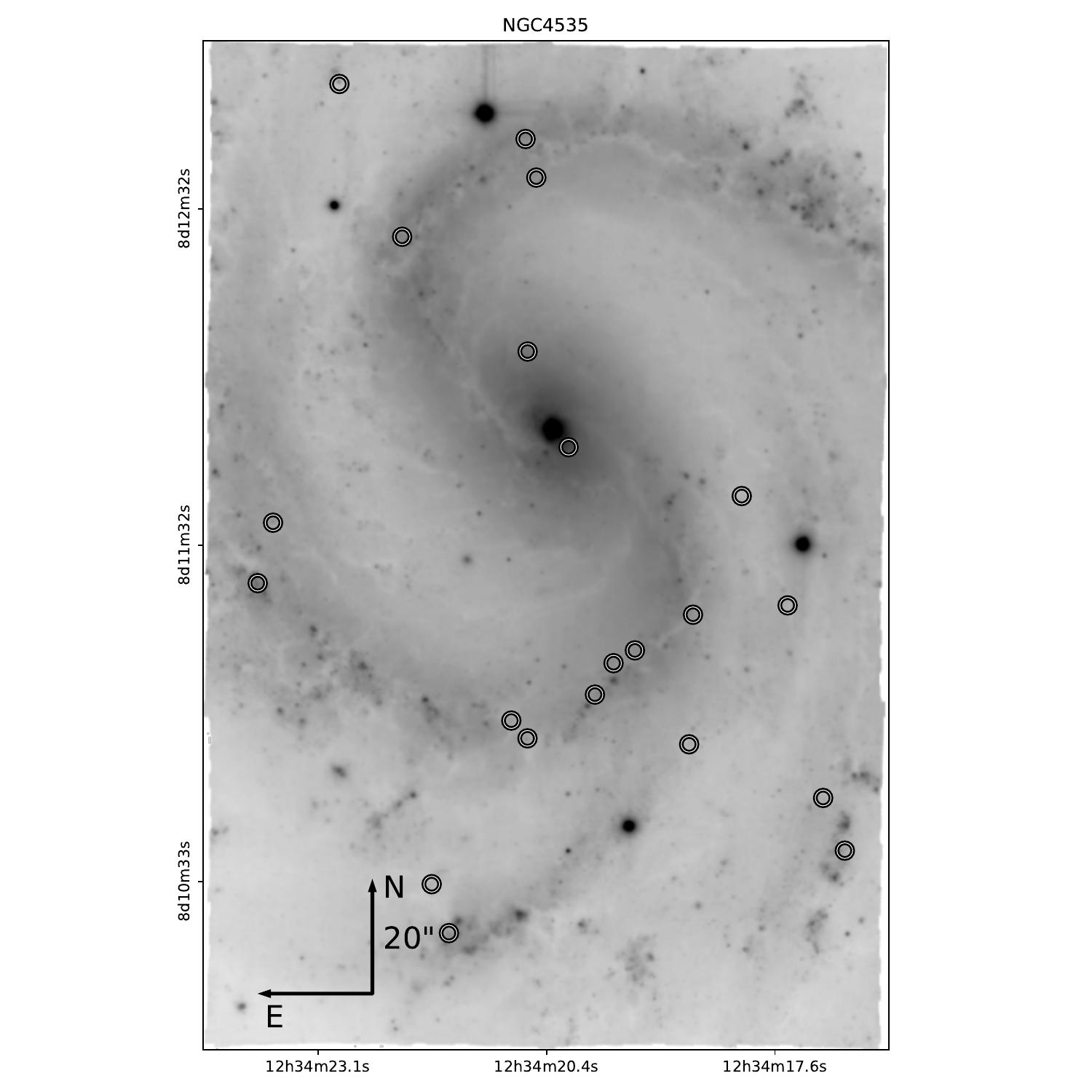}
    \end{figure}

    \begin{figure}[h]
    \centering
    \includegraphics[width=\hsize]{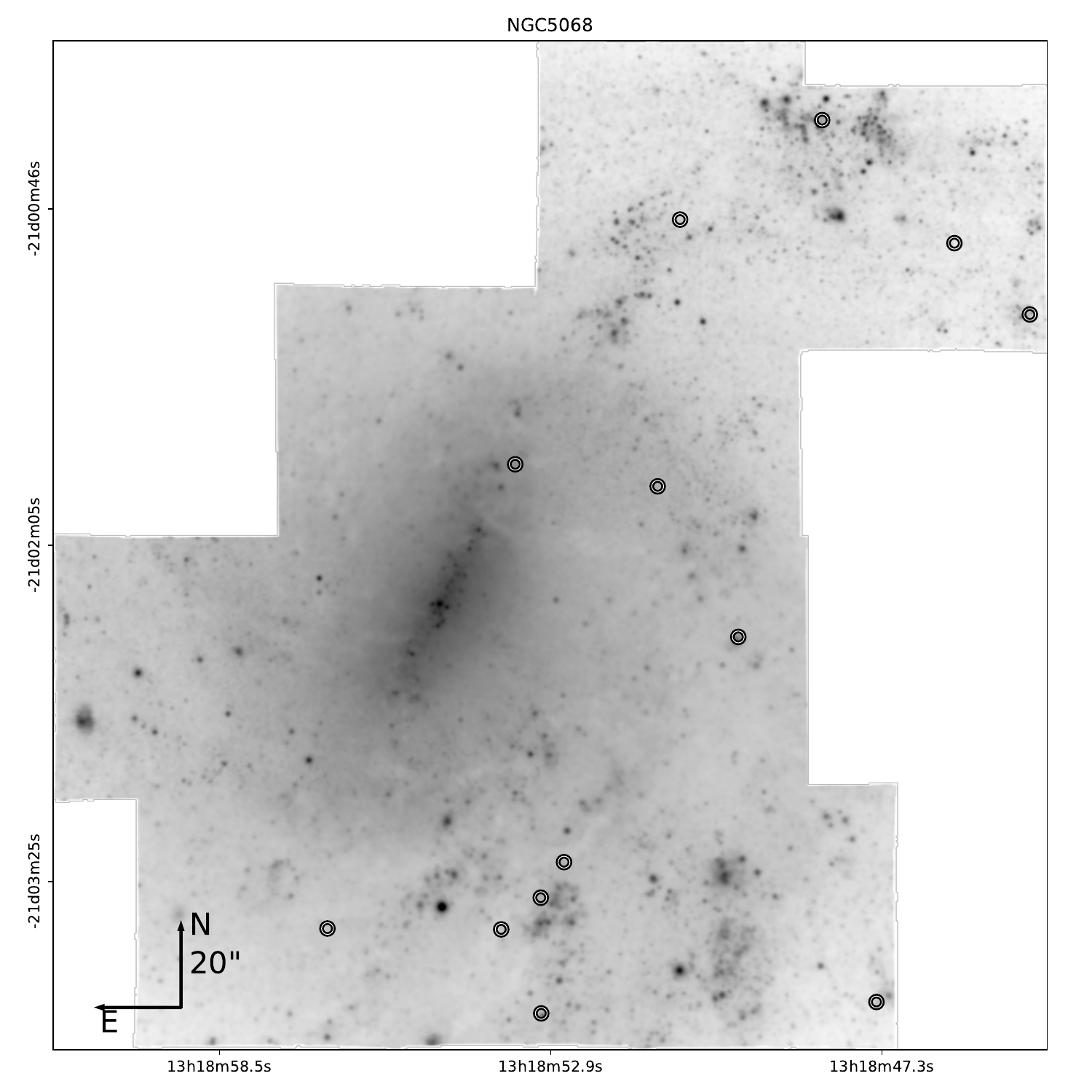}
    \end{figure}

    \begin{figure}[h]
    \centering
    \includegraphics[width=\hsize]{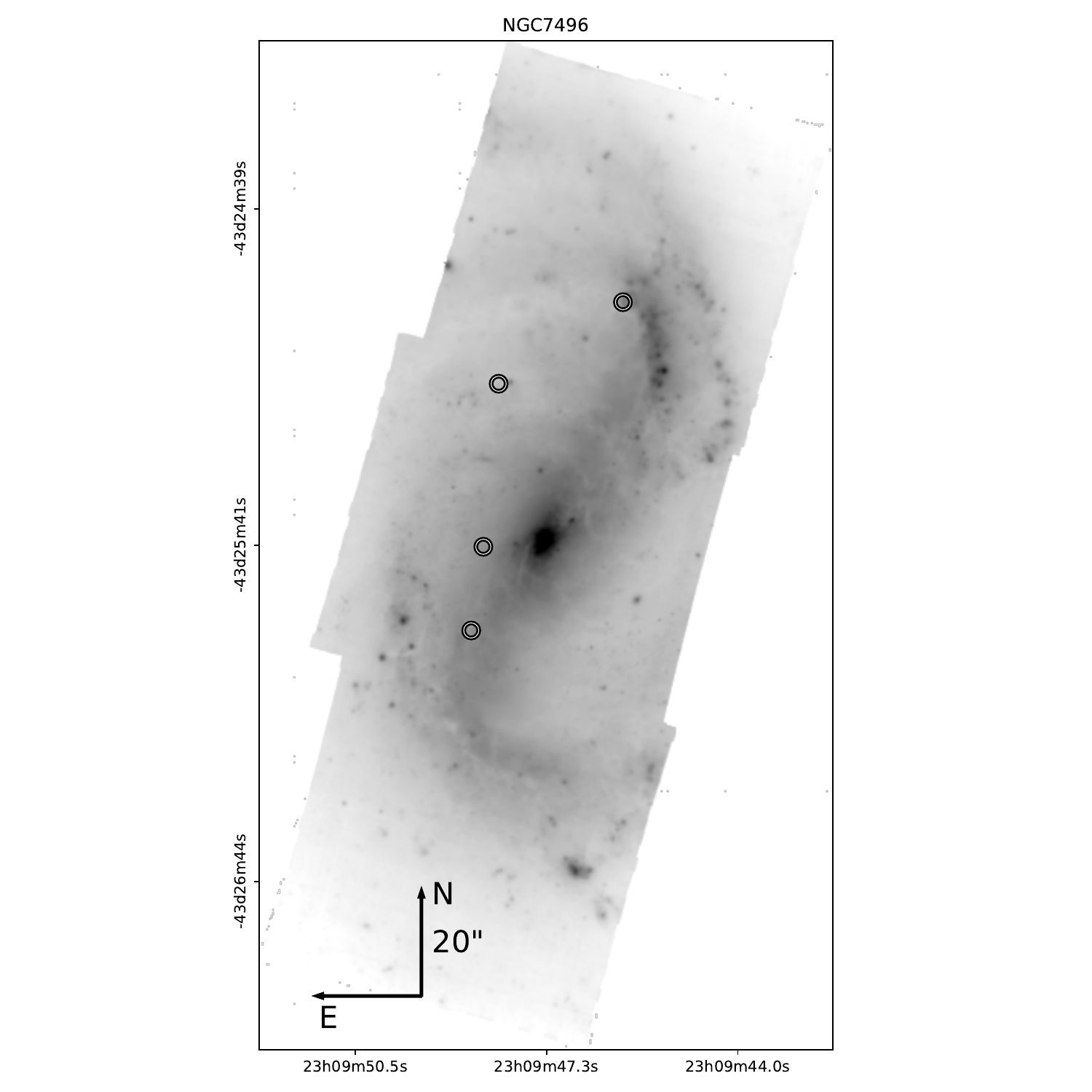}
    \end{figure}

\end{document}